\documentclass[prd,aps,twocolumn,a4paper,showkeys,nofootinbib]{revtex4-1}
\usepackage{amsmath}
\usepackage{amsfonts}
\usepackage{amssymb}	
\usepackage{graphicx}
\usepackage{bm}
\usepackage{color}
\usepackage[utf8]{inputenc}
\usepackage{mathrsfs}
\usepackage{ulem}
\usepackage{commath}
\usepackage[T1]{fontenc}
\usepackage{lmodern}

\def\p{\partial}

\def\i{{\rm i}}

\def\GMc2{G M_{\odot} c^{-2}}

\def\F{{\cal F}}
\def\lm{{\ell m}}

\def\lm{{\ell m}}
\def\p{\partial}

\def\lm{{\ell m}}

\def\ii{{\rm i}}

\def\ph{\varphi}

\def\F{{\cal F}}

\def\TEOBResumS{\texttt{TEOBResumS}}

\def\TEOBResumSlm{\texttt{TEOBResumS\_NQC\_lm}}
\def\SEOB{\texttt{SEOBNRv4HM}}

\usepackage{color}
\definecolor{cyan}{rgb}{0,0.9,0.9}
\definecolor{orange}{rgb}{0.9,0.5,0}
\definecolor{magenta}{rgb}{1,0,1}
\definecolor{purple}{rgb}{0.8,0.4,0.8}
\definecolor{gray}{rgb}{0.8242,0.8242,0.8242}
\definecolor{dodgerblue}{rgb}{0.12, 0.56, 1.0}

\newcommand{\be}{\begin{equation}}  
\newcommand{\ee}{\end{equation}}
\newcommand{\bea}{\begin{eqnarray}}           
\newcommand{\eea}{\end{eqnarray}} 
\newcommand{\beqn}{\begin{eqnarray*}}
\newcommand{\eeqn}{\end{eqnarray*}}

\newcommand{\EOB}{{\rm EOB}}

\newcommand{\virg}[1]{``#1''}
\begin{document}
\title{Waveforms and fluxes: Towards a self-consistent effective one body waveform model for nonprecessing, coalescing black-hole binaries for third generation detectors}
\author{Angelica \surname{Albertini}${}^{1,2}$}
\author{Alessandro \surname{Nagar}${}^{3,4}$}
\author{Piero \surname{Rettegno}${}^{3,5}$}
\author{Simone \surname{Albanesi}${}^{3, 5}$}
\author{Rossella \surname{Gamba}${}^{6}$}
\affiliation{${}^1$Astronomical Institute of the Czech Academy of Sciences,
Bo\v{c}n\'{i} II 1401/1a, CZ-141 00 Prague, Czech Republic}
\affiliation{${}^2$Faculty of Mathematics and Physics, Charles University in Prague, 18000 Prague, Czech Republic}
\affiliation{${}^3$INFN Sezione di Torino, Via P. Giuria 1, 10125 Torino, Italy}
\affiliation{${}^4$Institut des Hautes Etudes Scientifiques, 91440 Bures-sur-Yvette, France}
\affiliation{${}^{5}$ Dipartimento di Fisica, Universit\`a di Torino, via P. Giuria 1, 10125 Torino, Italy}
\affiliation{${}^6$Theoretisch-Physikalisches Institut, Friedrich-Schiller-Universit{\"a}t Jena, 07743, Jena, Germany}

\begin{abstract}
We present a comprehensive comparison between numerical relativity (NR) angular momentum fluxes 
at infinity and the corresponding quantity entering the radiation reaction in \TEOBResumS{}, an 
Effective-One-Body (EOB) waveform model for nonprecessing coalescing black hole binaries on 
quasi-circular orbits. This comparison prompted us to implement two changes in the model: 
(i) including Next-to-Quasi-Circular corrections in the $\ell=m$, $\ell\leq 5$ multipoles entering the radiation reaction 
and (ii) consequently updating the NR-informed spin-orbital sector of the model.
This yields a new waveform model that presents a higher self-consistency between waveform and
dynamics and an improved agreement with NR simulations. We test the model computing the EOB/NR 
unfaithfulness $\bar{F}_{\rm EOB/NR}$ over all 534 spin-aligned configurations available through the 
Simulating eXtreme Spacetime catalog, notably using the noise spectral density 
of Advanced LIGO, Einstein Telescope and Cosmic Explorer, for total mass up to $500M_\odot$. 
We find that the maximum unfaithfulness 
$\bar{F}^{\rm max}_{\rm EOB/NR}$ is mostly  between $10^{-4}$ and $10^{-3}$, and the performance
progressively worsens up to $\sim 5\times 10^{-3}$  as the effective spin of the system is increased. 
We perform similar analyses on the \SEOB{} model, that delivers $\bar{F}^{\rm max}_{\rm EOB/NR}$  values uniformly distributed 
versus effective spin and  mostly between $10^{-3}$ and $10^{-2}$.
We conclude that the improved \TEOBResumS{} model already represents a
reliable and robust first step towards the development of highly accurate waveform templates for
third generation detectors.
\end{abstract}

\maketitle

\section{Introduction}

The increasing sensitivity of gravitational-wave (GW) detectors~\cite{TheVirgo:2014hva,TheLIGOScientific:2014jea} 
and the associated compact binaries detections~\cite{LIGOScientific:2020ibl} motivate work towards physically complete, 
precise and efficient gravitational-wave models.
The effective-one-body (EOB) approach~\cite{Buonanno:1998gg,Buonanno:2000ef,Damour:2000we,Damour:2001tu,Damour:2015isa} 
is a way to deal with the general-relativistic two-body problem that, by construction,
allows the inclusion of perturbative (post-Newtonian, black hole
perturbations) and full numerical relativity (NR) results within a single theoretical framework. 
It currently represents a state-of-art approach for modeling waveforms
from binary black holes, conceptually designed to describe the entire
inspiral-merger-ringdown phenomenology of quasicircular
binaries~\cite{Nagar:2018gnk,Nagar:2018zoe,Cotesta:2018fcv,Nagar:2019wds,Nagar:2020pcj,Ossokine:2020kjp,Schmidt:2020yuu}
or even eccentric inspirals~\cite{Chiaramello:2020ehz,Nagar:2021gss,Nagar:2021xnh} 
and dynamical captures along hyperbolic orbits~\cite{Damour:2014afa,Nagar:2020xsk,Nagar:2021gss,Gamba:2021gap}.
An alternative, though less flexible, approach to generate waveforms for detection 
and parameter estimation relies on phenomenological models, whose latest avatar is 
{\tt IMRPhenomX}~\cite{Pratten:2020fqn,Garcia-Quiros:2020qpx,Pratten:2020ceb}. Note however that this 
kind of waveform models {\it does rely} on the EOB approach to accurately describe the
waveform during the long inspiral, until it is matched to (short) NR simulations.

Currently, there are two families of NR-informed EOB waveform models: 
the {\tt SEOBNR} family~\cite{Cotesta:2018fcv,Ossokine:2020kjp} and 
the {\tt TEOBResumS}~\cite{Akcay:2020qrj, Gamba:2021ydi} family. 
Both models incorporate precession and tidal effects in some form, but \TEOBResumS{}
also has spin-aligned versions that can deal with eccentric inspirals and hyperbolic encounters~\cite{Nagar:2020xsk,Nagar:2021gss}.
Although they are both EOB models, their building blocks are very different,
starting from the choice of the underlying Hamiltonians and resummation 
strategies (see e.g.~\cite{Rettegno:2019tzh}).
The quality of {\it any} waveform model (specifically, an EOB or a phenomenological one in the current context), 
is assessed by computing the unfaithfulness (or mismatch) between the waveforms 
generated by the model and the corresponding NR waveforms over the NR-covered portion
of the binary parameter space. This is an obvious procedure since the waveform is the 
crucial observable that is needed for data analysis. 
If this is the {\it only} viable procedure for phenomenological models, for EOB models
there are other quantities that might be worth considering. In particular, one has to
remember that within the EOB one has access to the {\it full relative dynamics} of
the binary and thus one can complement the waveform comparison with other,
gauge-invariant, dynamical quantities. For example, one has access to the gauge-invariant
relation between energy and angular momentum~\cite{Damour:2011fu,Nagar:2015xqa,Ossokine:2017dge}, 
to the periastron advance~\cite{LeTiec:2011bk, LeTiec:2013uey, Hinderer:2013uwa} or,
for hyperbolic encounters, to the scattering angle~\cite{Damour:2014afa}.

Together with the Hamiltonian and the waveform, the third building block of any
EOB model is the radiation reaction, i.e. the flux of angular momentum and energy
radiated via gravitational waves. Surprisingly, the only direct comparison between EOB and
NR fluxes, namely Ref.~\cite{Boyle:2008ge}, dates back to more than a decade ago.
The purpose of this paper is to update Ref.~\cite{Boyle:2008ge} focusing on spin-aligned BBHs. 
More specifically, it aims at presenting: (i) new calculations of the fluxes from (some of) the spin-aligned 
NR datasets of the Simulating eXtreme Spacetimes (SXS) catalog~\cite{Boyle:2019kee} 
and (ii) new EOB/NR comparisons between the fluxes that involve both the most recent
 version of \TEOBResumS{}~\cite{ Nagar:2019wds, Nagar:2020pcj} 
 and \SEOB{}~\cite{Cotesta:2018fcv,Ossokine:2020kjp}.
 From the EOB/NR flux comparisons with \TEOBResumS{}, we learn the importance of including
 next-to-quasi-circular (NQC) corrections also in the flux modes beyond the $\ell=m=2$
 dominant one in order to achieve a rather high level of consistency ($\lesssim 1\%$) between the 
 EOB ad NR fluxes up to merger. By contrast, the EOB/NR flux comparisons with \SEOB{} 
 show deficits of this model over the NR-covered portion of the parameter space. 
 
 While including NQC factors in the radiation reaction in \TEOBResumS{}, we eventually build an improved model,
 called \TEOBResumSlm{}, that aims at being more self-consistent and that differs from
 the standard \TEOBResumS{} also for a more precise determination of the NR-informed 
 spin-orbit dynamical parameter. By computing the unfaithfulness for the $\ell = m = 2$ mode
 over the sample of 534 nonprecessing, quasicircular simulations of the SXS catalog already considered in Ref.~\cite{Riemenschneider:2021ppj}, we find that
 both the standard model and the updated one are promising foundations in view of the requirements 
 for third generation detectors~\cite{Reitze:2021gzo, Couvares:2021ajn, Punturo:2021ryo, Katsanevas:2021fzj, 
 Kalogera:2021bya, McClelland:2021wqy}.

The paper is organized as follows. In Sec.~\ref{sec:flux} we remind the structure of the radiation reaction within
the \TEOBResumS{} model, provide a novel computation of the angular momentum flux from (a sample of)
NR simulations and compare it with the \TEOBResumS{} one. The outcome of this comparisons points to the fact
that an improved EOB model would benefit of the inclusion in the flux of NQC corrections beyond the $\ell=m=2$
ones. This improved model is constructed in  Sec.~\ref{sec:new}, notably by providing a new NR-informed fit
of the next-to-next-to-next-to-leading-order (NNNLO) effective spin-orbit parameter $c_3$ 
previously introduced in~\cite{Damour:2014sva,Nagar:2015xqa}.
In Sec.~\ref{sec:barF} we assess the accuracy of this NQC-improved model by computing the 
EOB/NR unfaithfulness 
using the PSD of advanced LIGO~\cite{aLIGODesign_PSD}, of Einstein Telescope~\cite{Hild:2009ns, Hild:2010id} and 
of Cosmic Explorer~\cite{Evans:2021gyd}. Finally, Sec.~\ref{sec:seob} provides a comprehensive comparison 
between NR, \SEOB{}~\cite{Bohe:2016gbl,Cotesta:2020qhw, Ossokine:2020kjp} and \TEOBResumS{} in its
native (i.e. non-NQC-improved) form. We gather our concluding remarks in Sec.~\ref{sec:conclusions}.

Unless otherwise specified, we use natural units with $c=G=1$.
Our notations are as follows: we denote with $(m_1,m_2)$ the individual masses,
while the mass ratio is $q\equiv m_1/m_2\geq 1$. The total mass and symmetric
mass ratio are then $M\equiv m_1+m_2$ and $\nu = m_1 m_2/M^2$.
We also use the mass fractions $X_{1,2}\equiv m_{1,2}/M$ and $X_{12}\equiv X_1-X_2=\sqrt{1-4\nu}$.
We address with $(S_1,S_2)$ the individual, dimensionful, spin components along the
direction of the orbital angular momentum. The dimensionless spin variables are
denoted as $\chi_{1,2}\equiv S_{1,2}/(m_{1,2})^2$. We also use $\tilde{a}_{1,2}\equiv X_{1,2}\chi_{1,2}$,
the effective spin $\tilde{a}_0=\tilde{a}_1+\tilde{a}_2$ and $\tilde{a}_{12}\equiv \tilde{a}_1-\tilde{a}_2$.

\section{Gravitational wave fluxes}
\label{sec:flux}

\subsection{Angular momentum fluxes from Numerical Relativity simulations}
\label{sec:cleaning}

In the systematic analysis of fluxes of Ref.~\cite{Boyle:2008ge}, performed using
NR data from the SXS collaboration, a lot of effort was devoted at the time to remove
the spurious oscillations that are present when the flux is expressed in terms of 
some, gauge-invariant, frequency parameter. The quality of SXS simulations has
hugely improved from Ref.~\cite{Boyle:2008ge}. Although SXS data has been used recently 
in the computation of the fluxes to obtain energy versus
angular momentum curves (see e.g. Refs.~\cite{Damour:2011fu,Nagar:2015xqa}), an explicit 
calculation of the flux analogous to the one presented in Ref.~\cite{Boyle:2008ge} has not been
attempted again. This is the purpose of this section.
Let us start by fixing our notations and conventions. The strain waveform is decomposed
in spin-weighted spherical harmonics as
\be
h_+ - i h_\times = \dfrac{1}{D_L}\sum_\ell \sum_{m=-\ell}^{\ell}h_\lm{}_{-2}Y_\lm(\iota,\phi)
\ee
where $D_L$ indicates the luminosity distance. The angular momentum flux radiated at infinity reads\footnote{Along the $z$-axis orthogonal to the orbital plane. Since we are considering
a nonprecessing system the components of the angular momentum along $(x,y)$ directions are zero.}
\begin{equation}
\dot{J}_{\infty} = - \frac{1}{8\pi} \sum_{\ell=2}^{\ell_{\rm max}}\sum_{m=-\ell}^{\ell} m  \Im(\dot{h}_\lm h_\lm^*).
\end{equation}
Here we will consider $\ell_{\rm max}=8$. For clarity, we work with the Newton-normalized
angular momentum flux
\be
\frac{\dot{J}_{\infty}}{\dot{J}^{\rm circ}_{\rm Newt}}	,
\ee
where the circularized Newtonian flux formally reads
\begin{equation}
\dot{J}^{\rm circ}_{\rm Newt} = \frac{32}{5}\nu^2\left(\Omega_{\rm NR}\right)^{7/3}.
\end{equation}
Here we define the NR orbital frequency $\Omega_{\rm NR}$ simply as
\begin{equation}
\Omega_{\rm NR} \equiv \frac{\omega_{22}^{\rm NR}}{2} ,
\end{equation}
where $\omega_{22}^{\rm NR}\equiv \dot{\phi}_{22}^{\rm NR}$ is the NR quadrupolar
GW frequency and $\phi_{22}^{\rm NR}$ the phase defined from $h_{22}=A_{22}^{\rm NR} e^{- \ii \phi_{22}^{\rm NR}}$.
We compute the NR fluxes out of a certain sample of SXS datasets, and choose extrapolation order\footnote{For the time-domain phasing
and unfaithfulness computations we use instead $N = 3$.}
$N=4$ to avoid systematics during the inspiral. 
\begin{figure}[t]
\includegraphics[width=0.45\textwidth]{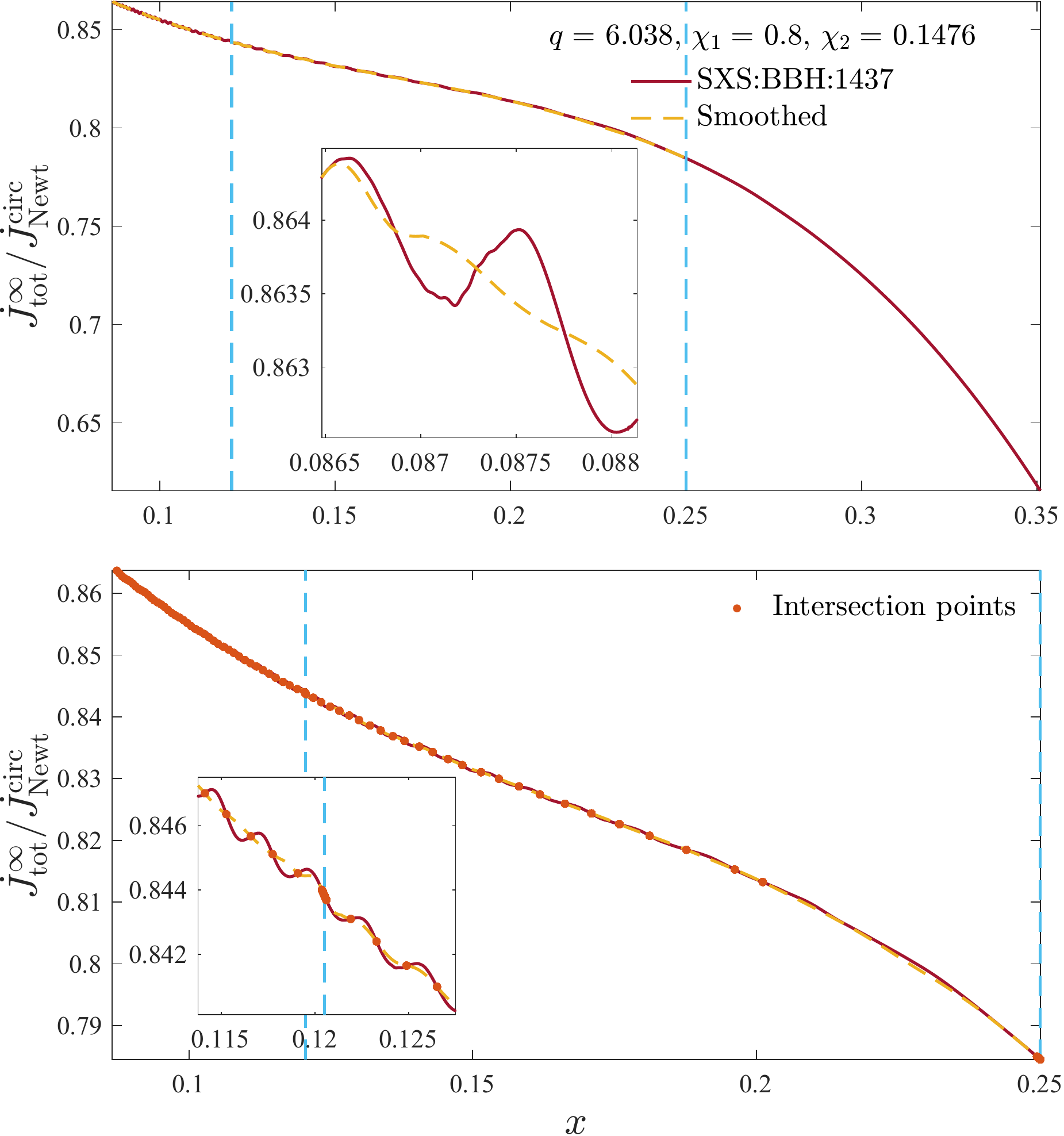} 
\caption{\label{fig:cleaning_steps} 
Intermediate steps of the cleaning procedure. The $x$-domain is separated into three different parts, 
delimited by the vertical lines (top panel). No smoothing is applied on the third region, while
the span of the moving average changes between the first and second region. The bottom panel 
focuses on the intersection points between the raw function and the smoothed
one, that are finally fitted with a polynomial. The inset highlights the behavior
around the interface between the first and second region.}
\end{figure}

When the so-computed fluxes are depicted in terms of the gauge-invariant frequency parameter
\be
\label{eq:xNR}
x_{\rm NR} \equiv  \left(\Omega_{\rm NR}\right)^{2/3}
\ee
one finds spurious oscillations. These oscillations are due to residual eccentricity (or other effects
related to the BMS symmetry being violated~\cite{Mitman:2021xkq}), and are additionally amplified when
taking the  derivatives. The amplification might be large and make the raw flux totally useless for any 
meaningful comparison with the analogous, fully nonoscillatory, EOB quantity. 
We have developed an efficient method to completely remove this oscillating behavior,
and produce a rather clean and smooth representation of the flux versus $x$.
The procedure is applied to the sample of SXS simulations reported in Table~\ref{tab:spinning_flux},
that is chosen so that the datasets distribution is approximately uniform over the NR-covered portion of 
the parameter space. We cut each flux at the NR merger, defined as the peak of $|h_{22}|$.
The procedure uses a {\tt MATLAB} function called \texttt{smooth}, i.e. a moving average whose span can 
be selected by the user\footnote{Namely, it is a lowpass filter with filter coefficients equal to the reciprocal 
of the span, meaning the higher the frequency of the oscillations to be removed, the higher the value of the chosen span.}. 
The $x$-domain on which the flux function is defined is separated into three parts: the first and the second 
ones get smoothed with different spans, as the frequency of the oscillations progressively lowers;
the third part, that is already essentially nonoscillatory, is left untouched. The three regions are optimized
manually for each dataset in Table~\ref{tab:spinning_flux}.
\begin{table}[t]
\caption{\label{tab:spinning_flux}Sample of SXS spin-aligned datasets for which we compute the
angular momentum flux. From left to right the columns display: the SXS ID; the binary parameters; the
highest and second-highest level of resolution; the average of the difference between the raw flux and the cleaned one.}
 \begin{center}
 \begin{ruledtabular}
\begin{tabular}{c c c c r}
   ID & $(q, \chi_1, \chi_2)$ & $\text{Lev}_h$ & $\text{Lev}_l$ & $\langle\Delta \dot{J}^\infty_{\rm NR - NR_{\rm clean}}\rangle$ \\ 
\hline 
BBH:1155 & $(1,0,0)$ & 3 & 2 & $1\cdot 10^{-6}$ \\ 
BBH:1222 & $(2,0,0)$ & 4 & 3 & $5.4\cdot 10^{-5}$ \\ 
BBH:1179 & $(3,0,0)$ & 5 & 4 & $1.8\cdot 10^{-5}$ \\ 
BBH:0190 & $(4.499,0,0)$ & 3 & 2 & $1.5\cdot 10^{-5}$ \\ 
BBH:0192 & $(6.58,0,0)$ & 3 & 2 & $1.3\cdot 10^{-5}$ \\ 
BBH:1107 & $(10,0,0)$ & 4 & 3 & $7.2\cdot 10^{-5}$ \\ 
\hline
BBH:1137 & $(1,-0.97,-0.97)$ & 4 & 2 & $6.3\cdot 10^{-5}$ \\ 
BBH:2084 & $(1,-0.90,0)$ & 4 & 3 & $-2\cdot 10^{-6}$ \\ 
BBH:2097 & $(1,+0.30,0)$ & 4 & 3 & $2.4\cdot 10^{-5} $\\ 
BBH:2105 & $(1,+0.90,0)$ & 4 & 3 & $2.3\cdot 10^{-5} $\\ 
BBH:1124 & $(1,+0.99,+0.99)$ & 3 & - & $2.6\cdot 10^{-5}$ \\ 
BBH:1146 & $(1.5,+0.95,+0.95)$ & 2 & 0 & $1.2\cdot 10^{-5}$ \\ 
BBH:2111 & $(2,-0.60,+0.60)$ & 4 & 3 & $-9\cdot 10^{-6} $\\ 
BBH:2124 & $(2,+0.30,0)$ & 4 & 3 & $9\cdot 10^{-6}$ \\ 
BBH:2131 & $(2,+0.85,+0.85)$ & 4 & 3 & $2\cdot 10^{-5}$ \\ 
BBH:2132 & $(2,+0.87,0)$ & 4 & 3 & $1.3\cdot 10^{-5}$ \\ 
BBH:2133 & $(3,-0.73,+0.85)$ & 4 & 3 & $2.2\cdot 10^{-5}$ \\ 
BBH:2153 & $(3,+0.30,0)$ & 4 & 3 & $3.6\cdot 10^{-5}$ \\ 
BBH:2162 & $(3,+0.60,+0.40)$ & 4 & 3 & $1.7\cdot 10^{-5}$ \\ 
BBH:1446 & $(3.154,-0.80,+0.78)$ & 3 & 2 & $9\cdot 10^{-6}$ \\ 
BBH:1936 & $(4,-0.80,-0.80)$ & 3 & 2 & $-1.8\cdot 10^{-5}$ \\ 
BBH:2040 & $(4,-0.80,-0.40)$ & 3 & 2 & $7\cdot 10^{-6}$ \\ 
BBH:1911 & $(4,0,-0.80)$ & 3 & 2 & $7\cdot 10^{-6}$ \\ 
BBH:2014 & $(4,+0.80,+0.40)$ & 3 & - & $-1\cdot 10^{-6} $\\ 
BBH:1434 & $(4.368,+0.80,+0.80)$ & 3 & - & $2.5\cdot 10^{-5}$ \\ 
BBH:1463 & $(4.978,+0.61,+0.24)$ & 3 & 2 & $1.5\cdot 10^{-5}$ \\ 
BBH:0208 & $(5,-0.90,0)$ & 3 & 2 & $9.2\cdot 10^{-5}$ \\ 
BBH:1428 & $(5.518,-0.80,-0.70)$ & 3 & 2 & $-2\cdot 10^{-6}$ \\ 
BBH:1437 & $(6.038,+0.80,+0.15)$ & 3 & 2 & $5\cdot 10^{-6} $\\ 
BBH:1436 & $(6.281,+0.009,-0.80)$ & 3 & 2 & $1\cdot 10^{-6} $\\ 
BBH:1435 & $(6.588,-0.79,+0.7)$ & 3 & 2 & $2\cdot 10^{-6}$ \\ 
BBH:1448 & $(6.944,-0.48,+0.52)$ & 3 & - &$ 2.1\cdot 10^{-5}$ \\ 
BBH:1375 & $(8,-0.90, 0)$ & 3 & - & $2.6\cdot 10^{-5}$ \\ 
BBH:1419 & $(8,-0.80,-0.80)$ & 3 & - & $-1.3\cdot 10^{-5}$ \\ 
BBH:1420 & $(8,-0.80,+0.80)$ & 3 & 2 &$ 2.2\cdot 10^{-5}$ \\ 
BBH:1455 & $(8,-0.40, 0)$ & 3 & 2 & $-3\cdot 10^{-6}$ \\ 
\end{tabular}
\end{ruledtabular}
\end{center}
\end{table}
\begin{figure}[t]
\includegraphics[width=0.44\textwidth]{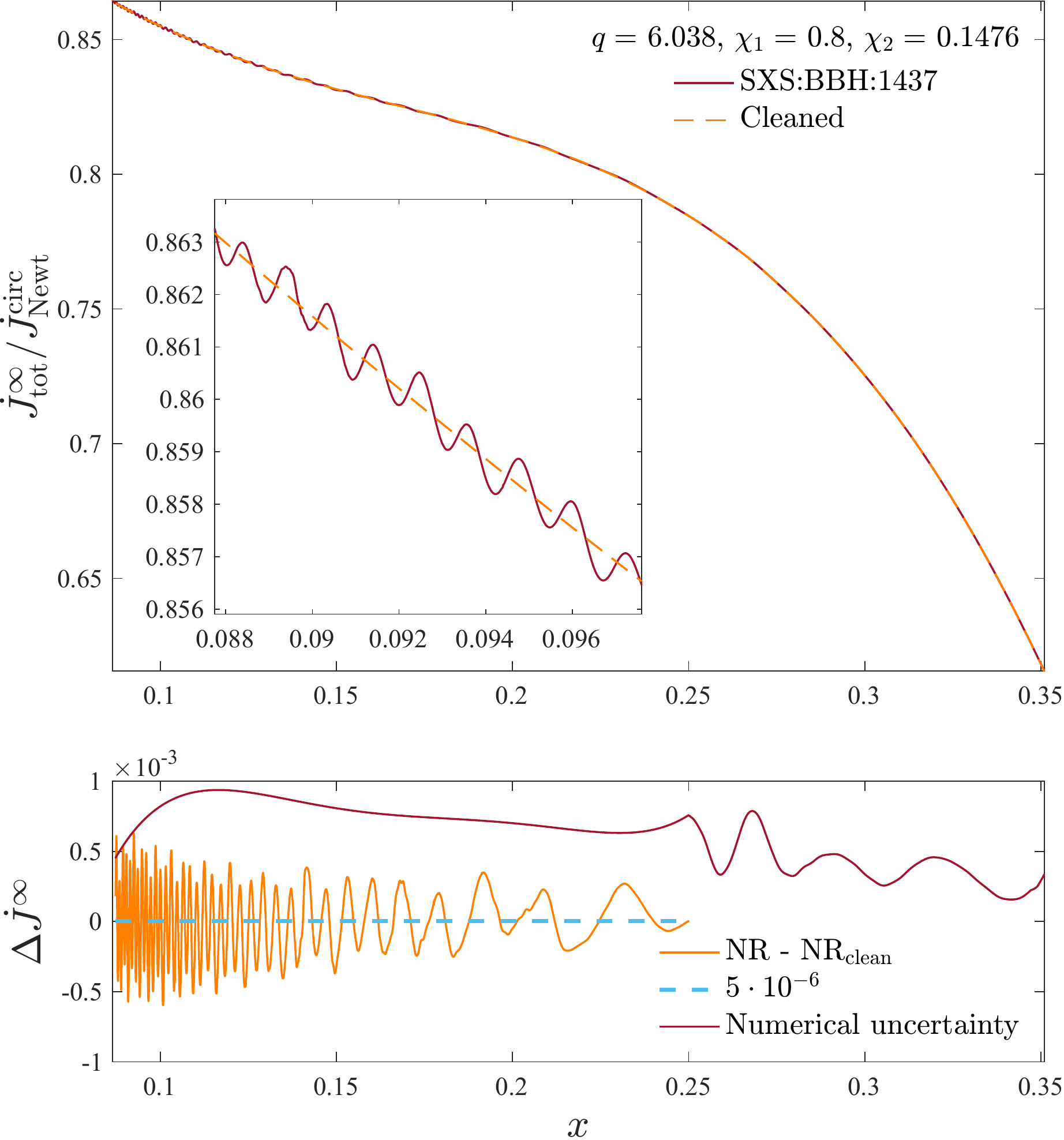} 
\caption{\label{fig:cleaning_final} 
The cleaned numerical angular momentum flux for the simulation SXS:BBH:1437 (dashed orange) is plotted against the original one (red). 
The inset in the upper panel shows how the final flux follows the original curve, averaging the oscillations.
In the lower panel we display the difference between the cleaned and the raw flux, whose mean (dashed light
blue line) is of order $10^{-5}$, hence proving the effectiveness of the procedure. Our cleaning method also 
allows to estimate numerical accuracy (red curve), that is evaluated by subtracting to the cleaned flux its 
equivalent coming from the second-highest available resolution.
}
\end{figure}
The cleaning procedure can be summarized in three steps: (i) we first apply the moving average
to reduce the amplitude of the oscillations (see inset in the upper panel of Fig.~\ref{fig:cleaning_steps});
(ii) then we find the intersection points between the raw flux and the smoothed one, 
(see markers in the inset of Fig.~\ref{fig:cleaning_steps}); (iii) as a third step, the intersection points 
 between the raw and the smoothed flux 
are fitted by a polynomial in $x$. For the datasets SXS:BBH:1155, SXS:BBH:1222, SXS:BBH:0190, 
SXS:BBH:0192 this is accomplished via a seventh
order polynomial, while it suffices a fifth order one for the others\footnote{Polynomials have been chosen
after attempting different fitting functions, but they prove to be the simplest
and more effective choice. We also found it more practical to apply a fit due to the large number of simulations taken into account.}.
The outcome of the fit is finally joined to the third part that was left unmodified. 
The final result, after some additional smoothing at the junction point, 
is shown in Fig.~\ref{fig:cleaning_final}. Its reliability can be verified by computing the 
difference with the raw data and checking that it averages zero. This is shown in the bottom panel of Fig.~\ref{fig:cleaning_final}, 
where the residual does not show any evident global trend, actually averaging at $\sim 5 \times 10^{-6}$.
To obtain a conservative estimate of the NR uncertainty on the final fluxes, we apply the cleaning procedure
to both the highest and second highest available resolution and then take the difference. This is also shown
in the bottom panel of Fig.~\ref{fig:cleaning_final}.
The procedure is found to be efficient and reliable for all configurations of Table~\ref{tab:spinning_flux},
where the quality of the cleaning procedure is indicated by the average of the difference between the raw 
flux and the cleaned one (last column of the table).
The final result is displayed in Fig.~\ref{fig:cleaned_fluxes}. The figure highlights how both the
value of the flux at merger and its global behavior have a clear dependence on the mass ratio
and the effective Kerr parameter. 
This testifies how equal-mass binaries have a more adiabatic
evolution, corresponding to slower plunges and  a lower angular momentum loss.
If the BHs have positive spins the plunge is even slower, owing to the well known effect of 
spin-orbit coupling (or hang-up effect)~\cite{Damour:2001tu, Campanelli:2006uy}.
Conversely for high mass ratio binaries (nearer to the test-mass limit) and negative spins, 
the fact that the system is progressively more and more nonadiabatic implies larger angular 
momentum losses, and the evolution ends at lower frequencies. 
\begin{figure}[t]
\includegraphics[width=0.43\textwidth]{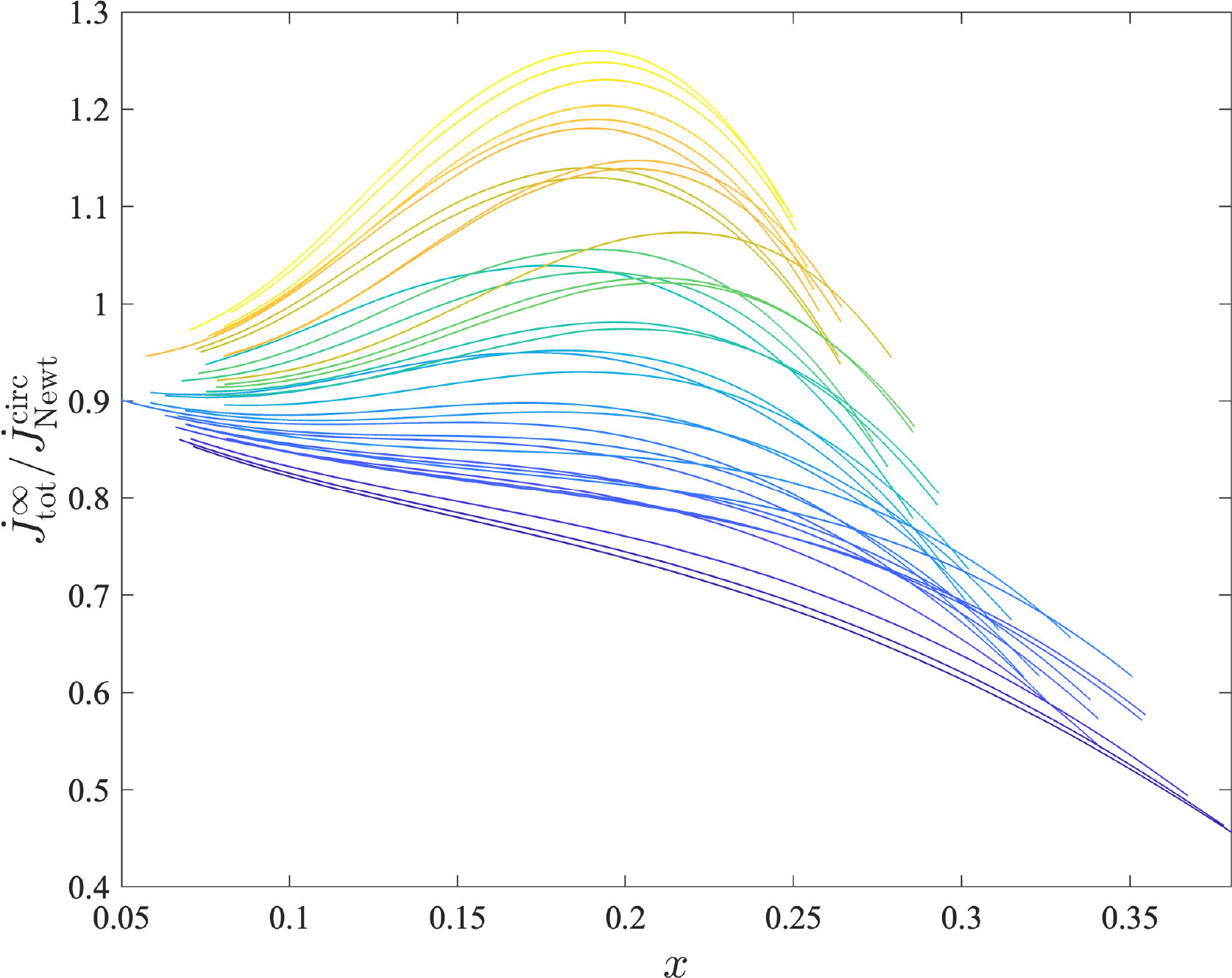} \\
\vspace{2mm}
\includegraphics[width=0.43\textwidth]{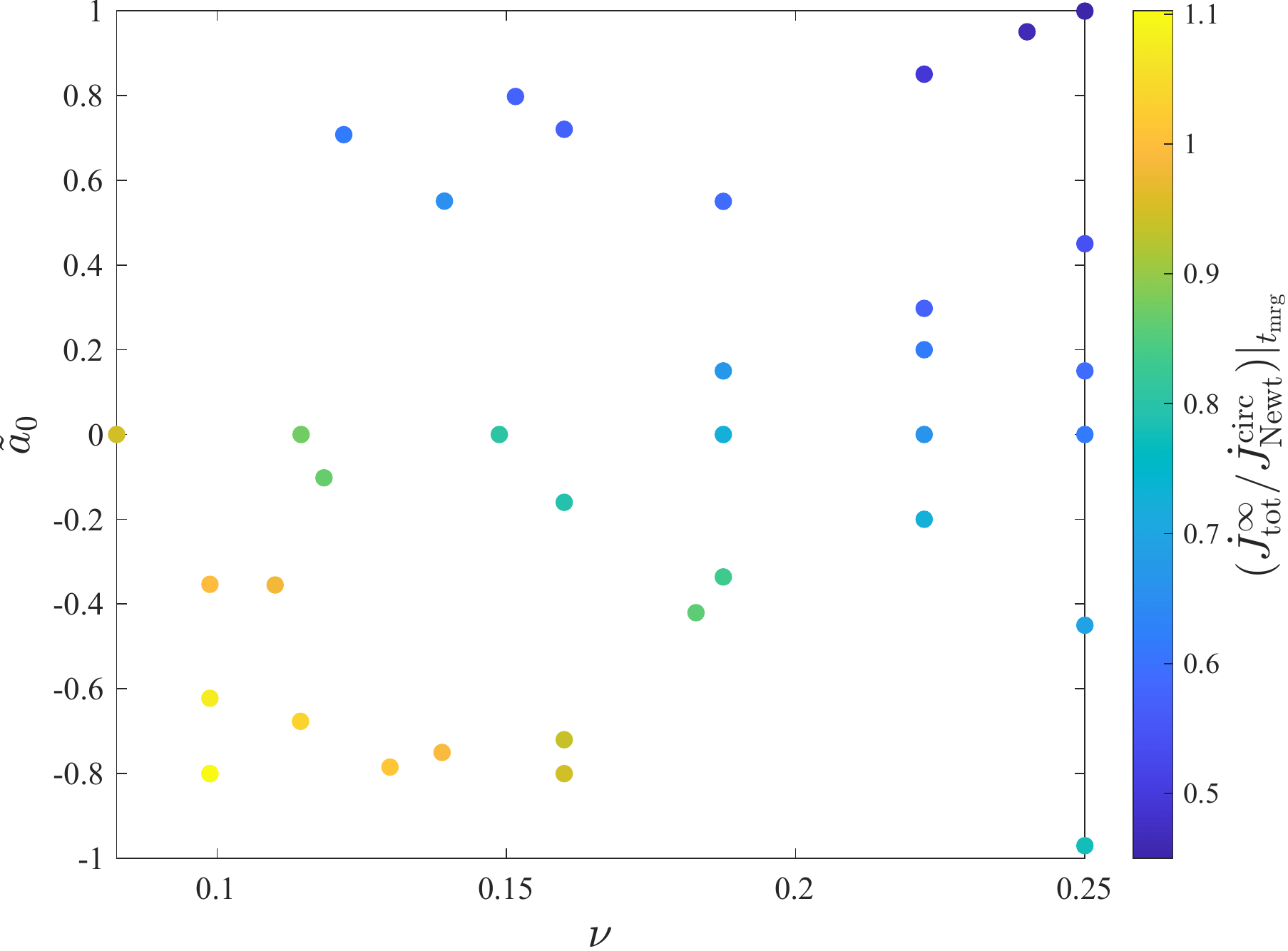} 
\caption{\label{fig:cleaned_fluxes} The top panel shows the final outcome of the Newton-normalized
angular momentum flux calculation for the NR datasets of Table~\ref{tab:spinning_flux},
with $x$ given by Eq.~\eqref{eq:xNR}, shown up to merger. The color code
is chosen depending on the final value of the flux, displayed in the lower panel.  
The merger values show a clear dependence on $\nu$ and $\tilde{a}_0$,
mirroring whether the dynamics is more or less adiabatic: larger emissions correspond to
faster plunges (with $\tilde{a}_0<0$).}
\end{figure}

\subsection{Angular momentum flux and radiation reaction within EOB}
\label{sec:EOB_flux}
Let us now turn to discuss EOB fluxes within \TEOBResumS{}.
To do so, we start by reviewing the analytical elements of \TEOBResumS{}
that will be useful for our discussion.
We use mass-reduced phase-space variables $(r,\varphi,p_\varphi,p_{r_*})$,  
related to the physical ones by $r=R/M$ (relative separation), $p_{r_*}=P_{R_*}/\mu$ 
(radial momentum), $\varphi$ (orbital phase), 
$p_\varphi=P_\varphi/(\mu M)$ (angular momentum) and $t=T/M$ (time).
The \virg{tortoise} radial momentum is 
$p_{r_*}\equiv (A/B)^{1/2}p_r$, where $A$ and $B$ are the EOB potentials 
(with included spin-spin interactions~\cite{Damour:2014sva}).
The Hamilton's equations for the relative dynamics read
\begin{align}
\dot{\ph} &= \Omega = \p_{p_\ph} \hat{H}_\EOB, \\
\dot{r} &= \left( \frac{A}{B} \right)^{1/2} \p_{p_{r_*}} \hat{H}_\EOB, \\
\dot{p}_\ph &= \hat{\F}_\ph , \nonumber \\
\dot{p}_{r_*} &= - \left( \frac{A}{B} \right)^{1/2} \p_{r} \hat{H}_\EOB,
\end{align}
where $\hat{H}_\EOB$ is the EOB Hamiltonian~\cite{Nagar:2018zoe}, 
$\Omega$ is the orbital frequency 
and $\hat{\F}_\ph$ is the radiation  reaction force accounting for mechanical 
angular momentum losses due to GW emission. Note that within this context 
we are assuming that the radial force $\hat{\F}_r=0$, that is equivalent to a 
gauge choice for circular orbits~\cite{Buonanno:2000ef}.
For a balance argument, the system angular momentum loss should be equal to the 
sum of the GW flux emitted at infinity, $\dot{J}_{\infty}$, and absorbed by the event horizons
of the two black holes, $\dot{J}_{\rm H_{1,2}}$, that is
\begin{equation}
\dot{J}_{\rm system} = \hat{\F}_\ph = - \dot{J}_{\infty} - \dot{J}_{\rm H_1} - \dot{J}_{\rm H_2}.
\end{equation}
In general, within this equation there should be an additional term accounting for
Schott contributions, that are due to the interactions between the radiation and the field.
However, it is always possible to choose a gauge such that there is no Schott 
contribution to the angular momentum~\cite{Bini:2012ji} and this is the choice 
made here (on top of neglecting $\hat{\cal{F}}_r$). 
The azimuthal radiation reaction force is hence written as
\begin{equation}
	\hat{\F}_\ph = \hat{\F}_\ph^\infty + \hat{\F}_\ph^{\rm H},
\end{equation}
where $\hat{\F}_\ph^{\rm H}$ is the horizon flux contribution~\cite{Damour:2014sva}. 
The asymptotic term reads
\begin{equation}
\label{eq:RR}
\hat{\F}_\ph^\infty  = -\frac{32}{5} \nu r_{\omega}^4 \Omega^5 \hat{f}^{\infty}(v_\varphi^2;\nu),
\end{equation}
where $\hat{f}^{\infty}(v_\varphi^2;\nu)$ is the reduced (i.e., Newton-normalized) flux function,
$v_\ph^2 \equiv (r_{\omega}\Omega)^2$ and $r_{\omega}$ is a modified radial separation 
defined in such a way that $1 = \Omega^2 r_{\omega}^3$ is valid during the plunge, 
fulfilling a modified Kepler's law that accounts for non-circularity~\cite{Damour:2006tr,Damour:2007xr}.  
The reduced flux function is defined by normalizing the resummed circularized energy flux 
as $\hat{f} \equiv (\F_{22}^{\rm Newt})^{-1} \sum \F_\lm$, with all multipoles (except $m = 0$ modes)
up to $\ell = 8$. The Newtonian term reads $\F_{22}^{\rm Newt} = (32/5) \nu^2 x^5$ and 
the multipolar terms $\F_\lm$ are factorized and 
resummed analogously to what is done for the waveform~\cite{Damour:2012ky}. 
Explicitly, building upon Ref.~\cite{Damour:2008gu}, the structure of each flux multipole is
\begin{equation}
\label{eq:flux_multipoles}
\F_\lm = \F_\lm^{\rm Newt} |\hat{h}_\lm |^2 \F_\lm^{\rm NQC}.
\end{equation}
This is related to the correction entering the factorization of the waveform multipoles
\be
h_\lm = h_\lm^{\rm Newt} \, \hat{h}_\lm  \, \hat{h}_\lm^{\rm NQC}
\ee
where $h_\lm^{\rm Newt}$ is the Newtonian 
prefactor\footnote{As pointed out in Ref.~\cite{Nagar:2019wds}, the standard Newtonian 
prefactors proportional to some power of $v_\varphi$ are replaced in some multipoles by suitable 
powers of $v_\varphi v_\Omega$, with $v_\Omega=\Omega^{1/3}$. 
This is  a practical solution to ease the action of the NR-informed NQC amplitude corrections 
and allow them to correctly capture the peak amplitude of each multipole. When including 
NQC corrections also in the higher mode contribution to the flux, this choice will eventually 
yield a partial inconsistency between the waveform and the flux. 
In Appendix~\ref{sec:hlm_Newt} we show that by using the 
standard Newtonian prefactors in the waveform we generically improve the EOB/NR flux
agreement for positive spins, but get inconsistent results for negative spins.}, $\hat{h}_\lm$ is the resummed PN
correction and $\hat{h}_\lm^{\rm NQC}$ is the next-to-quasi-circular factor.
The latter is described in more detail in 
Refs.~\cite{Damour:2014sva,Nagar:2017jdw,Nagar:2019wds,Riemenschneider:2021ppj}
(see in particular Sec.~IIID of~\cite{Nagar:2019wds}).
For each flux mode we have
\begin{equation}
\F_\lm^{\rm NQC} = \left|\hat{h}_\lm^{\rm NQC}\right| ^2 = \left(1 + a_1^\lm n_1 ^\lm + a_2^\lm n_2 ^\lm\right)^2 
\end{equation}
where $(n_1^\lm, n_2^\lm)$ are functions of the radial momentum and 
of the radial acceleration (and a priori depend on the mode); $(a_1^\lm, a_2^\lm)$ are 
numerical coefficients that are informed by NR simulations~\cite{Damour:2014sva,Nagar:2017jdw}
via an iterative procedure~\cite{Damour:2009kr}.
NQC corrections can, and actually should, be applied to each 
waveform (and thus flux) mode since they complete the analytical waveform,
that is quasicircular  by construction. In practice,
within \TEOBResumS{} we add NQC corrections {\it only} in the $(2,2)$ flux mode,
while the waveform is NQC-completed up to $\ell=m=5$~\cite{Nagar:2019wds}.

Finally, we remind that \TEOBResumS{} is NR-informed via two different parameters,
$a_6^c(\nu)$ and $c_3(\nu, \tilde{a}_1, \tilde{a}_2)$, respectively tuning the $A$ potential and the
spin-orbit sector of the model. Details on these functions can be found in Sec.~IIC of Ref.~\cite{Nagar:2020pcj}.

For most of the analyses carried out in the following, we make use of the private {\tt MATLAB} version
of \TEOBResumS{}, in which we implement the changes for \TEOBResumSlm{}. 
The publicly available $C$ version is used in the unfaithfulness calculation for the standard \TEOBResumS{}.

\subsection{Comparing NR and EOB fluxes}
\label{sec:comparison}
\begin{figure}[t]
\includegraphics[width=0.44\textwidth]{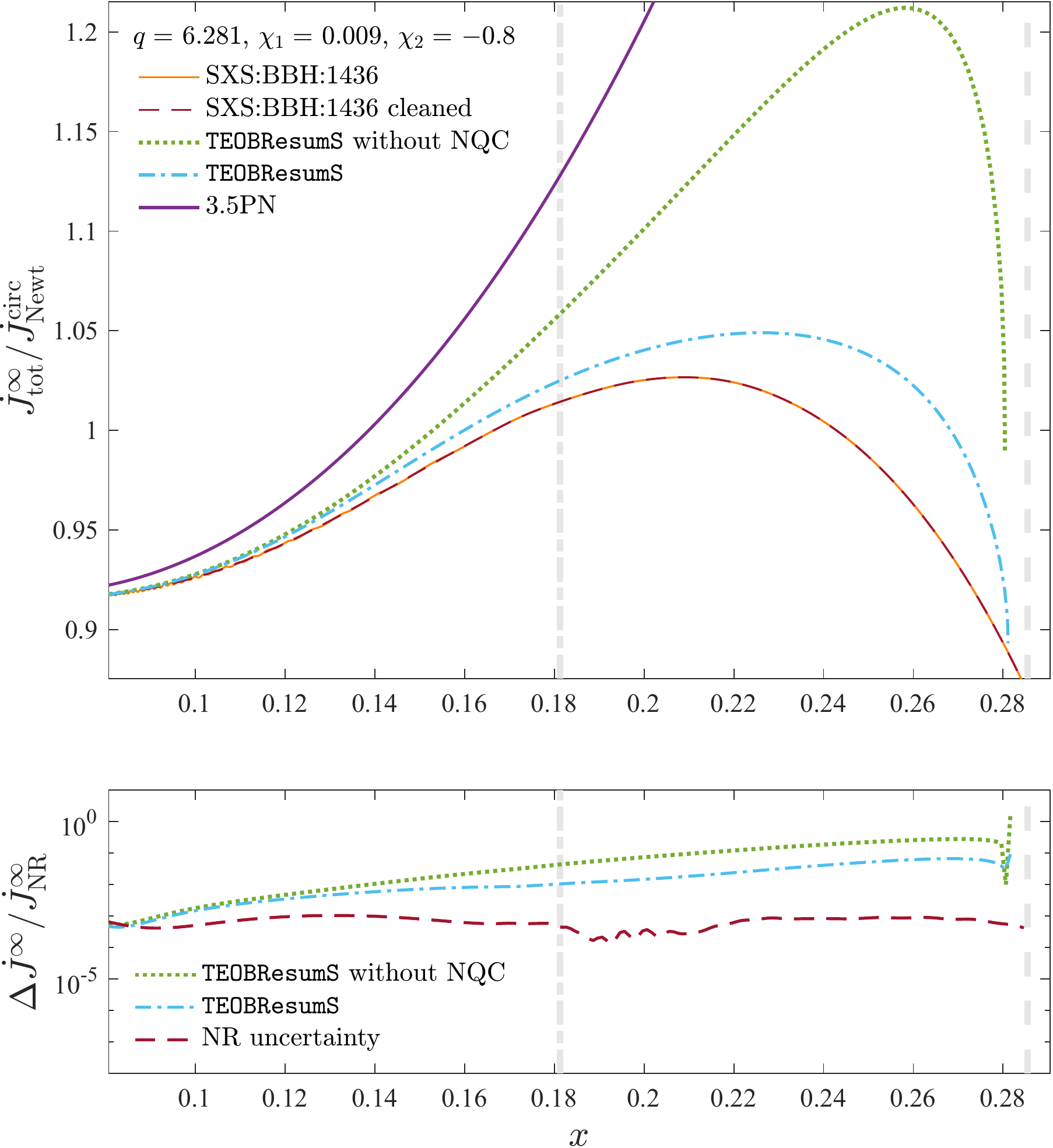} 
\caption{\label{fig:1436old} Comparing Newton normalized total angular 
momentum fluxes summed up to $\ell_{\rm max}=8$.
The upper panel shows:
(i) the raw numerical flux (orange) and as its cleaned version (dashed red);
(ii) the EOB flux with $\ell=m=2$ NQC corrections (dash-dotted
light blue) and without (dotted green); 
(iii) the 3.5PN flux (purple). From left to right, the vertical lines indicate the EOB LSO and 
the NR merger respectively. Fractional differences are shown in the bottom panel, 
together with the NR uncertainty. NQC corrections are essential to reduce 
the  gap between the EOB and NR curves.}
\end{figure}
\begin{figure}[t]
\includegraphics[width=0.44\textwidth]{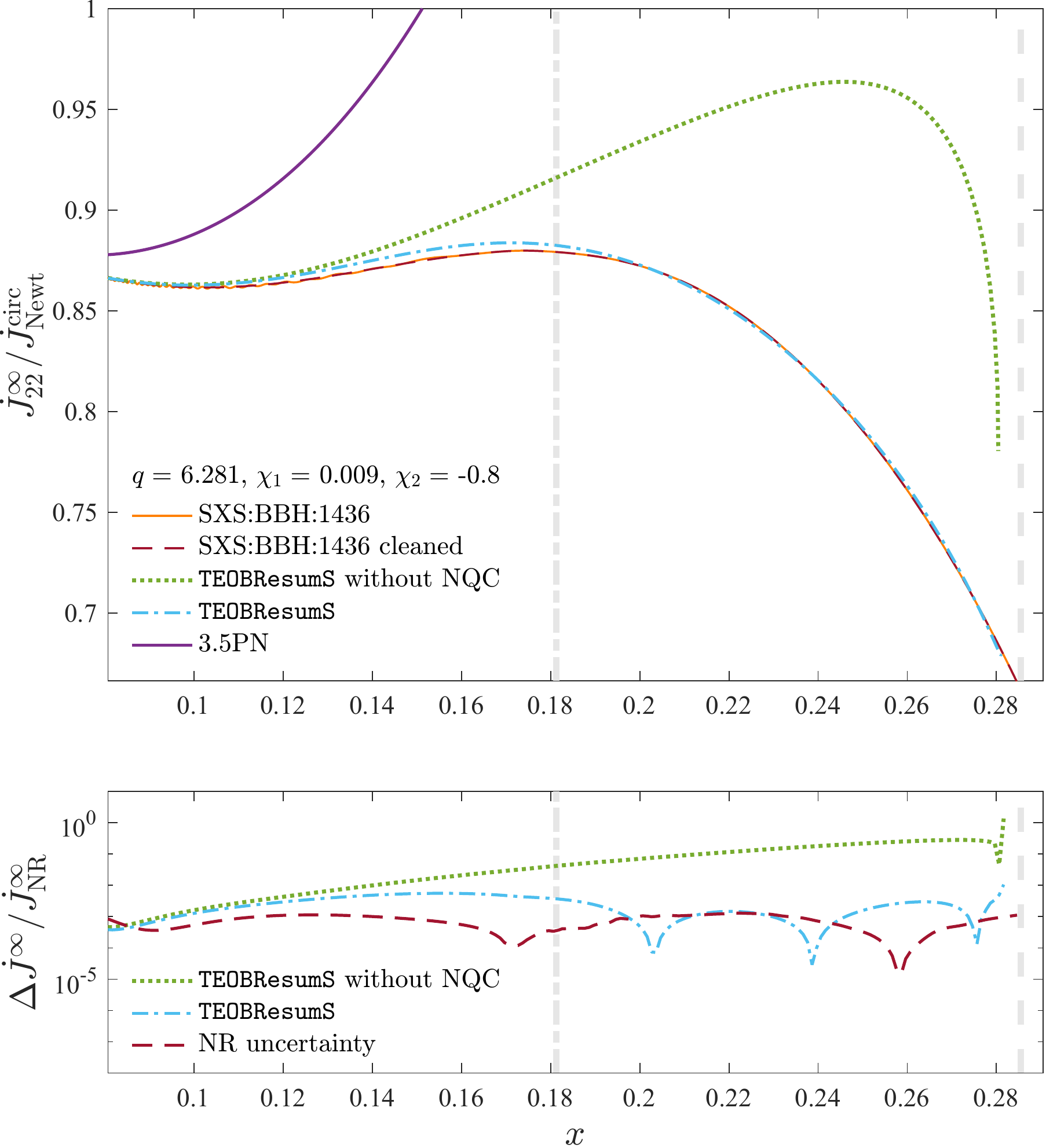} \\
\vspace{1.5mm}
\includegraphics[width=0.44\textwidth]{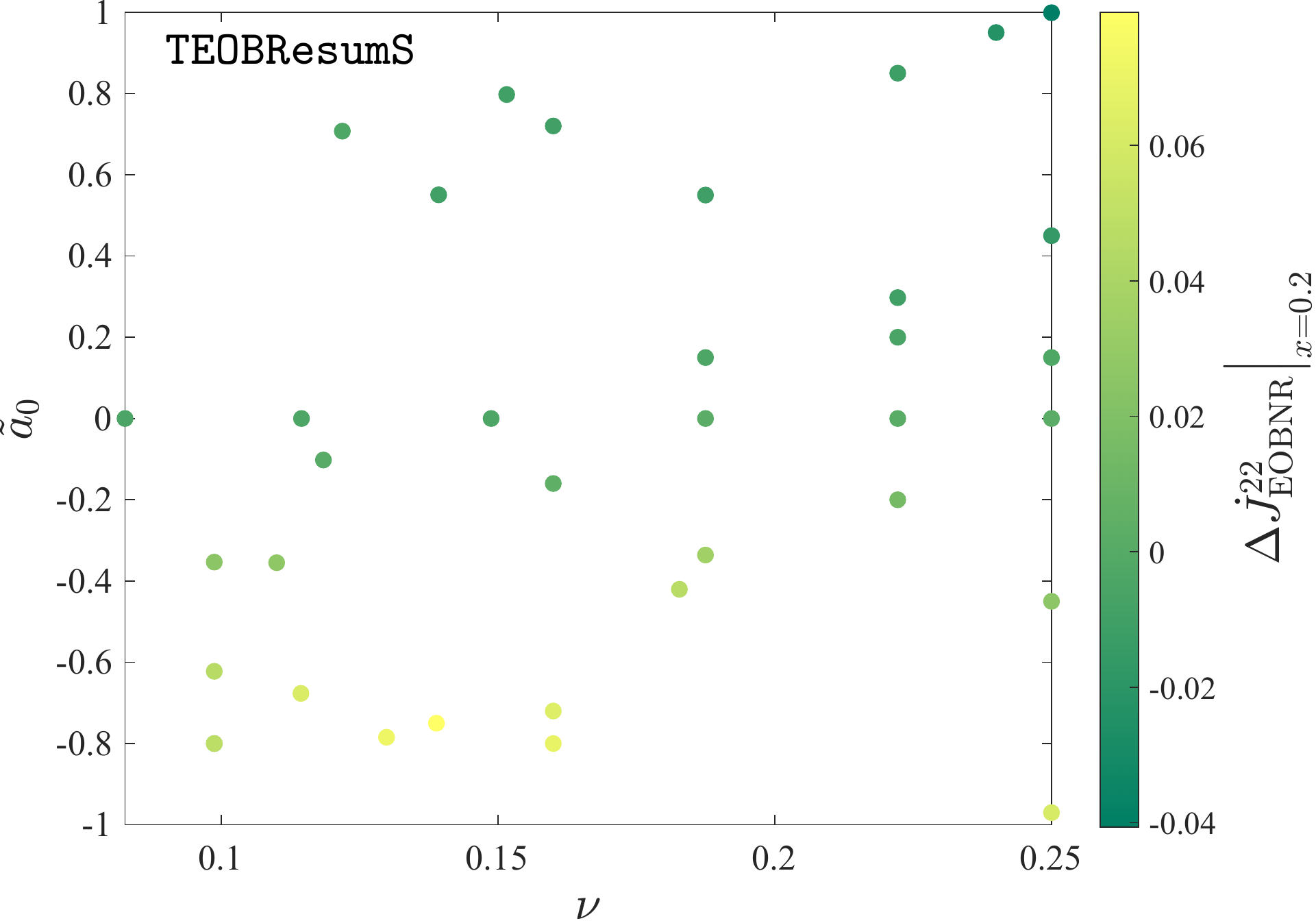}
\caption{\label{fig:1436old_22} {\it Top}: 
Comparing Newton-normalized $\ell = m = 2$ angular 
momentum fluxes, including again NR, the EOB fluxes with and without NQC corrections,
and the 3.5PN result. 
Remarkably, the fractional difference with NR for the NQC-corrected EOB curve
is of order $10^{-3}$ up to merger. The vertical lines indicate the LSO and the merger point. 
The $\ell = m = 2$ numerical flux has been cleaned separately from the total
one, and the difference between the raw flux and the final fit averages to $-2 \cdot 10^{-5}$.
{\it Bottom}: Fractional differences for the EOB/NR $\ell = m = 2$ fluxes at $x = 0.2$ 
for all configurations of Table~\ref{tab:spinning_flux}. The largest differences occur
when  $\tilde{a}_0 < 0$, where $x=0.2$ approximately corresponds to the plunge regime.}
\end{figure}

Let us now move to compare EOB and NR fluxes. The Newton-normalized EOB flux is
expressed versus $x_{\rm EOB}=\Omega^{2/3}$, while the NR curve is expressed versus
$x_{\rm NR}=\Omega_{\rm NR}^{2/3}$ as defined above.  To simplify the notation, in the
figure we will simply use $x$ for the horizontal axis, but it is intended that $x=x_{\rm NR}$
when dealing with the NR curve and $x=x_{\rm EOB}$ for the EOB curve.
As an illustrative configuration we choose SXS:BBH:1436, corresponding to parameters $(q, \chi_1, \chi_2) = (6.281, 0.009, -0.8)$.
The Newton-normalized, total, angular momentum flux, summed up to $\ell_{\rm max}=8$ is displayed in
Fig.~\ref{fig:1436old}. In particular, the figure shows:  (i) the raw and cleaned NR fluxes, that are effectively indistinguishable
on this scale; (ii) two  EOB fluxes, one with the $\ell=m=2$ NQC correction in the flux and another without it;
(iii) the 3.5PN flux. The EOB fluxes prove both the power of resummation techniques and 
the effectiveness of NQC corrections in achieving a good  agreement with the NR quantities.
The upper panel in Figure~\ref{fig:1436old_22} is analogous to Fig.~\ref{fig:1436old}, but only focuses on 
the $\ell=m=2$ contribution. The most interesting fact inferred by the plot is that the 
NQC factor is crucial to yield a fractional difference $\sim 10^{-3}$ up to merger.
The lower panel of the same figure shows the distribution of the EOB/NR fractional difference at 
$x = 0.2$ over the parameter space. This seems to point out to a decreased agreement for configurations 
having a negative $\tilde{a}_0$, but one has to note however that, as can be seen in Fig.~\ref{fig:cleaned_fluxes},
the fluxes for these datasets end at lower frequencies and hence $x = 0.2$ corresponds to the late plunge.

\begin{figure}[t]
\center
\includegraphics[width=0.45\textwidth]{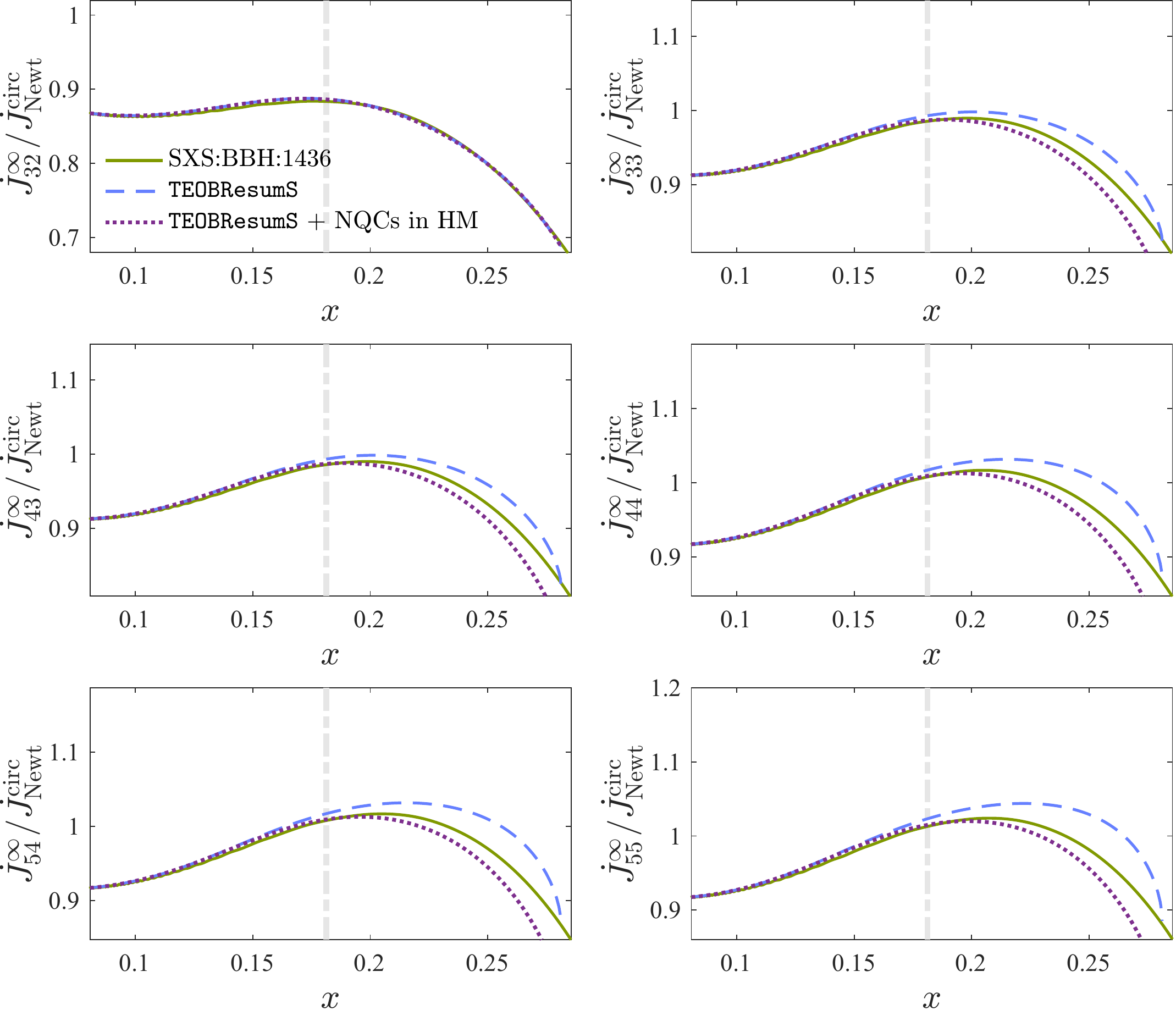} 
\caption{\label{fig:1436_multipoles} Exploring the importance of higher modes using the
SXS:BBH:1436 dataset. In each panel the flux is summed up to the indicated $(\ell,m)$ mode.
NQC corrections are included either in the $\ell=m=2$ EOB mode only (dashed blue line) or
in all $\ell=m$ modes up to $\ell=5$ (dotted purple line). The NR-informed NQC corrections in higher 
modes are essential to improve the EOB/NR agreement beyond plunge (the vertical line indicates 
the LSO frequency) and up to merger.}
\end{figure}
 \begin{table}[t]
   \caption{\label{tab:c3}Binary configurations, first-guess values of $c_3$
     used to inform the global interpolating fit given in Eq.~\eqref{eq:c3fit},
     and the corresponding $c_3^{\rm fit}$ values.}
   \begin{center}
 \begin{ruledtabular}
   \begin{tabular}{lllc|cc}
     $\#$ & ID & $(q,\chi_1,\chi_2)$ & $\tilde{a}_0$ &$c_3^{\rm first\;guess}$ & $c_3^{\rm fit}$\\
     \hline
1 & BBH:1137 & $(   1, -0.   97, -0.   97)$ & $-0.97$ & 89.7 & 89.33 \\ 
2 & BBH:0156 & $(   1, -0.9498, -0.9498)$ & $-0.95$ & 88.5 & 88.33 \\ 
3 & BBH:0159 & $(   1, -0.   90, -0.   90)$ & $-0.90$ & 84.5 & 85.86 \\ 
4 & BBH:2086 & $(   1, -0.   80, -0.   80)$ & $-0.80$ & 82 & 80.93 \\ 
5 & BBH:2089 & $(   1, -0.   60, -0.   60)$ & $-0.60$ & 71 & 71.19 \\ 
6 & BBH:0150 & $(   1, +0.   20, +0.   20)$ & $+0.20$ & 35.5 & 35.73 \\ 
7 & BBH:2102 & $(   1, +0.   60, +0.   60)$ & $+0.60$ & 22.2 & 21.67 \\ 
8 & BBH:2104 & $(   1, +0.   80, +0.   80)$ & $+0.80$ & 15.9 & 16.31 \\ 
9 & BBH:0153 & $(   1, +0.   85, +0.   85)$ & $+0.85$ & 15.05 & 15.29 \\ 
10 & BBH:0160 & $(   1, +0.   90, +0.   90)$ & $+0.90$ & 14.7 & 14.5 \\ 
11 & BBH:0157 & $(   1, +0.   95, +0.   95)$ & $+0.95$ & 14.3 & 14.1 \\ 
12 & BBH:0177 & $(   1, +0.   99, +0.   99)$ & $+0.99$ & 14.2 & 14.29 \\ 
13 & BBH:0004 & $(   1, -0.   50,  0.    0)$ & $-0.25$ & 55.5 & 54.44 \\ 
14 & BBH:0005 & $(   1, +0.   50,  0.    0)$ & $+0.25$ & 35 & 34.17 \\ 
15 & BBH:2105 & $(   1, +0.   90,  0.    0)$ & $+0.45$ & 27.7 & 27.21 \\ 
16 & BBH:2106 & $(   1, +0.   90, +0.   50)$ & $+0.70$ & 19.1 & 19.09 \\ 
17 & BBH:0016 & $( 1.5, -0.   50,  0.    0)$ & $-0.30$ & 56.2 & 56.14 \\ 
18 & BBH:1146 & $( 1.5, +0.   95, +0.   95)$ & $+0.95$ & 14.35 & 13.98 \\ 
19 & BBH:2129 & $(   2, +0.   60,  0.    0)$ & $+0.40$ & 29.5 & 29.31 \\ 
20 & BBH:2130 & $(   2, +0.   60, +0.   60)$ & $+0.60$ & 23 & 22.41 \\ 
21 & BBH:2131 & $(   2, +0.   85, +0.   85)$ & $+0.85$ & 16.2 & 15.73 \\ 
22 & BBH:2139 & $(   3, -0.   50, -0.   50)$ & $-0.50$ & 65.3 & 62.45 \\ 
23 & BBH:0036 & $(   3, -0.   50,  0.    0)$ & $-0.38$ & 58.3 & 57.62 \\ 
24 & BBH:0174 & $(   3, +0.   50,  0.    0)$ & $+0.37$ & 28.5 & 30.87 \\ 
25 & BBH:2158 & $(   3, +0.   50, +0.   50)$ & $+0.50$ & 27.1 & 26.64 \\ 
26 & BBH:2163 & $(   3, +0.   60, +0.   60)$ & $+0.60$ & 24.3 & 23.56 \\ 
27 & BBH:0293 & $(   3, +0.   85, +0.   85)$ & $+0.85$ & 17.1 & 17.05 \\ 
28 & BBH:1447 & $(3.16, +0.7398, +0.   80)$ & $+0.75$ & 19.2 & 19.46 \\ 
29 & BBH:2014 & $(   4, +0.   80, +0.   40)$ & $+0.72$ & 21.5 & 21.52 \\ 
30 & BBH:1434 & $(4.37, +0.7977, +0.7959)$ & $+0.80$ & 19.8 & 20.05 \\ 
31 & BBH:0111 & $(   5, -0.   50,  0.    0)$ & $-0.42$ & 54 & 57.18 \\ 
32 & BBH:0110 & $(   5, +0.   50,  0.    0)$ & $+0.42$ & 32 & 30.98 \\ 
33 & BBH:1432 & $(5.84, +0.6577, +0. 793)$ & $+0.68$ & 25 & 24.42 \\ 
34 & BBH:1375 & $(   8, -0.   90,  0.    0)$ & $-0.80$ & 64.5 & 65.12 \\ 
35 & BBH:0114 & $(   8, -0.   50,  0.    0)$ & $-0.44$ & 57 & 56.07 \\ 
36 & BBH:0065 & $(   8, +0.   50,  0.    0)$ & $+0.44$ & 29.5 & 31.78 \\ 
37 & BBH:1426 & $(   8, +0.4838, +0.7484)$ & $+0.51$ & 30.3 & 29.98 \\ 
 \end{tabular}
 \end{ruledtabular}
 \end{center}
 \end{table}

\looseness=-2
The cumulative importance of higher modes with respect to the $\ell=m=2$ one is studied
in Fig.~\ref{fig:1436_multipoles} for the same SXS:BBH:1436 configuration.
The figure contrasts the EOB flux with the NR one, where both
functions incorporate modes summed up to the indicated $(\ell,m)$ value. 
The plot shows that for the standard $\TEOBResumS{}$ the EOB/NR agreement progressively worsens during the
late inspiral up to merger, due to the lack of the NR-informed NQC corrections
beyond the $\ell=m=2$ ones. Including NQC corrections in 
the flux in all the $\ell=m$ modes up to $\ell=5$ yields a 
closer agreement between the analytical and numerical fluxes up to merger.
The NQC parameters are determined with the usual iteration procedure, 
although we maintain the same values of the NR-informed parameters $(a_6^5,c_3)$
determined with the standard $\ell=m=2$ NQC correction.
The effect is very evident for this specific dataset, but it is a feature that 
is always present, also for other configurations.
This exercise indicates that to increase the physical completeness and 
NR-consistency of \TEOBResumS{}  it would be needed to include NQC
corrections {\it at least} in the $\ell=m$ multipoles in the flux. Evidently, this operation 
will eventually imply the need of constructing new NR-informed $(a_6^c,c_3)$ 
functions that are consistent with the new choice of radiation 
reaction\footnote{Note that part of the residual difference cannot be totally removed
because the Newtonian prefactors in the waveform are not consistent with those
in the flux for $\ell=m>2$, as pointed out above. See Appendix~\ref{sec:hlm_Newt} for other details.}. 


\begin{figure}[t]
\begin{center}
\includegraphics[width=0.45\textwidth]{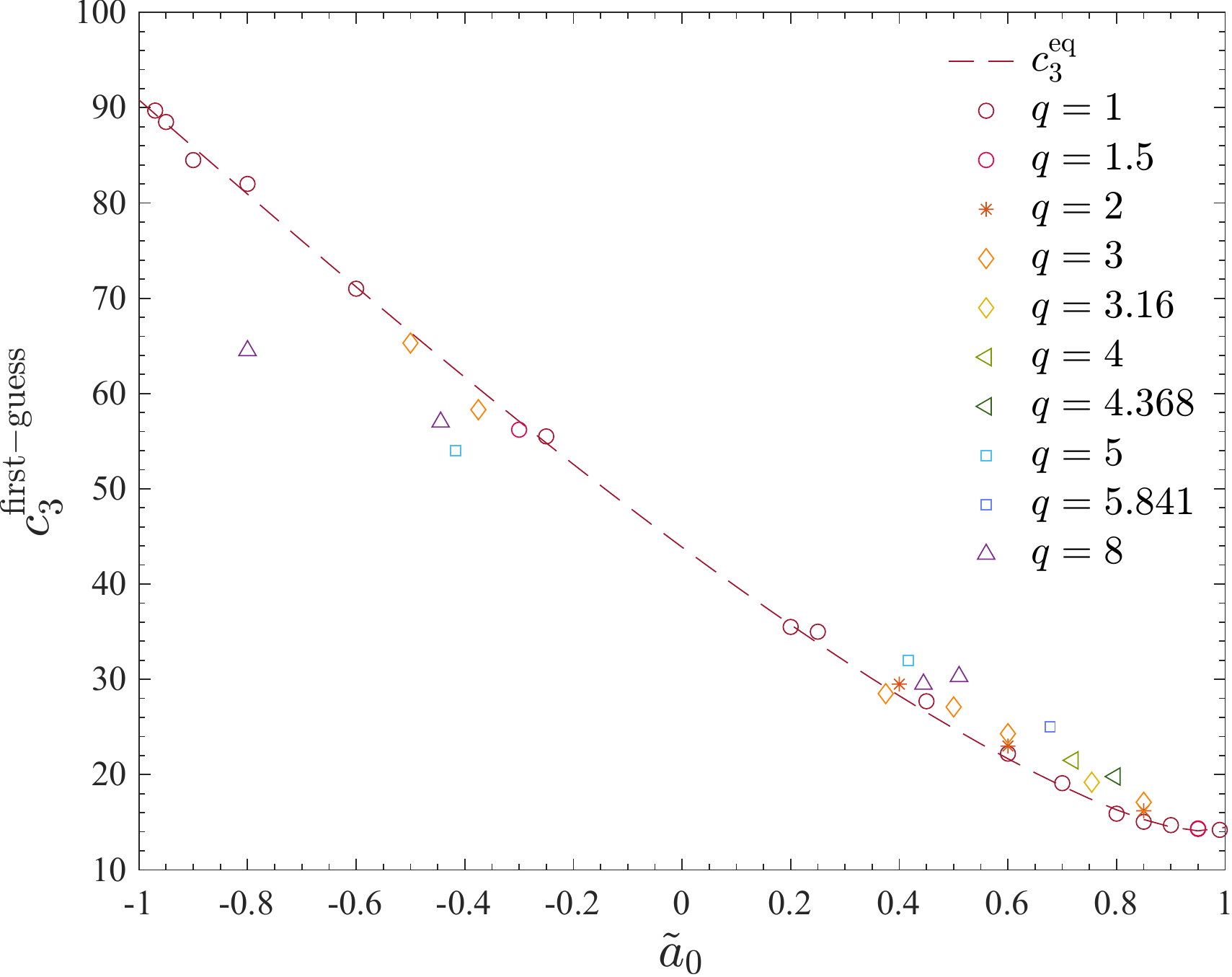} 
\caption{The first-guess $c_3$ values of Table~\ref{tab:c3} versus
  the spin variable $\tilde{a}_0$.
  The unequal-spin and unequal-mass points can be essentially
  seen as a correction to the equal-mass, equal-spin values. The latter
  are fitted to obtain the first part of the fit, $c_3^{\rm eq}$ (dashed red).}
\label{fig:c3}
\end{center}
\end{figure}

\section{Improving the consistency between waveform and flux of \TEOBResumS{}}
\label{sec:new}
Let us construct a modified \TEOBResumS{} model that incorporates
iterated NQC corrections in all $\ell=m$ modes in the flux up to $\ell=5$.
Since we are modifying the radiation reaction, this choice in principle 
calls for a new determination of both the $a_6^c$ and $c_3$ functions. 
However, we have verified that the improvements brought by a newly 
tuned $a_6^c(\nu)$ are marginal, so that, for the sake of simplicity, 
we keep its standard expression that we quote here for completeness as
\be
a_6^c=n_0\dfrac{1+n_1\nu + n_2\nu^2+n_3\nu^3}{1+d_1\nu},
\ee
where
\begin{align}
  n_0 &= \;\;\;5.9951,\\
  n_1 &=-34.4844,\\
  n_2 &=-79.2997,\\
  n_3 &=\;\;\;713.4451,\\
  d_1 &=-3.167.
\end{align}
By contrast, we look for a new NR-informed representation of $c_3$. We follow our usual
procedure, that is described for example in Sec.~IIB.2 of Ref.~\cite{Nagar:2018zoe}.
Typically, for each NR dataset one determines a value of $c_3$ so that the EOB/NR
accumulated phase difference up to merger is within (or compatible with) the NR phase uncertainty 
at NR merger. This leaves a certain flexibility and arbitrariness in the choice of $c_3$ and, in
previous attempts, we were typically accepting EOB/NR phase differences of the order of
0.1-0.2~rad at merger. Here, on the understanding that the NR phase uncertainty might
be overestimated by taking the difference between the two highest resolutions, 
we attempt to ask more, requiring that the EOB/NR phase difference is {\it as flat as possible} 
through inspiral, merger and ringdown when the two waveforms are aligned during the 
early inspiral. As a cross check, we also align the two waveforms
during the late plunge, just before merger, to verify that the phase difference keeps remaining flat. 
This further proves that the $c_3$ determination, that mostly affects the plunge phase, is 
done robustly. To exploit at best current NR information, we consider a sample of 37 SXS configurations,
most of which were already taken into account in the previous determinations of $c_3$. 
Here we replaced some datasets used in Ref.~\cite{Nagar:2018zoe} with newer ones with improved 
accuracy and included a few more simulations so as to cover the parameter space
more efficiently. Table~\ref{tab:c3} reports the SXS configurations, the corresponding 
values of $\tilde{a}_0$ , the first-guess values of $c_3$ obtained 
with the procedure explained above as well as the corresponding ones obtained 
after a global fit.
Specifically, the $c_3^{\rm first-guess}$ data of Table~\ref{tab:c3} 
are fitted with a global function as $c_3(\nu,\tilde{a}_0,\tilde{a}_{12})$ 
that reads
\begin{align}
  \label{eq:c3fit}
c_3(\nu,\tilde{a}_0,\tilde{a}_{12})=\,
&p_0\dfrac{1 + n_1\tilde{a}_0 + n_2\tilde{a}_0^2 + n_3\tilde{a}_0^3 + n_4\tilde{a}_0^4}{1 + d_1\tilde{a}_0}\nonumber\\
+ &p_1 \tilde{a}_0\sqrt{1-4\nu} + p_2\tilde{a}_0^2\sqrt{1-4\nu} \nonumber\\
+ &p_3 \tilde{a}_0\nu\sqrt{1-4\nu} + p_4  \tilde{a}_{12}\nu^2,
\end{align}
where the fitted parameters are
\begin{align}
p_0&=\;\;\;43.872788,\\
n_1&=-1.849495,\\
n_2&=\;\;\;1.011208,\\
n_3&=-0.086453,\\
n_4&=-0.038378,\\
d_1&=-0.888154, \\
p_1&= \;\;\;26.553,\\
p_2&= -8.65836, \\
p_3&= -84.7473, \\
p_4&= \;\;\;24.0418 \ .
\end{align}
Figure~\ref{fig:c3} highlights that the span of the ``best'' (first-guess)
values of $c_3$ is rather limited (especially for spins aligned with the 
orbital angular momentum) around the equal-mass, equal-spin case.
As in previous work, the fitting procedure consists of two steps.
First, one fits the equal-mass, equal-spin data with
a quasi-linear function of $\tilde{a}_0=\tilde{a}_1+\tilde{a}_2$
with $\tilde{a}_1=\tilde{a}_2$. This delivers the six
parameters $(p_0,n_1,n_2,n_3,n_4,d_1)$. The corresponding fit $c_3^{\rm eq}$
is shown as a dashed red curve in Fig.~\ref{fig:c3}.
Note that the
analytical structure of the fitting function was chosen in order
to accurately capture the nonlinear behavior of $c_3$ for $\tilde{a}_0\to 1$.
In the second step one subtracts this fit from the corresponding $c_3^{\rm first-guess}$ values and
fits the residual. This determines the parameters $(p_1,p_2,p_3,p_4)$.
The novelty with respect to previous work is that the functional form chosen
for the unequal-mass, unequal-spin fit
is more effective in capturing the first-guess values all over the SXS sample
considered.

\begin{figure*}[t]
\includegraphics[width=0.45\textwidth]{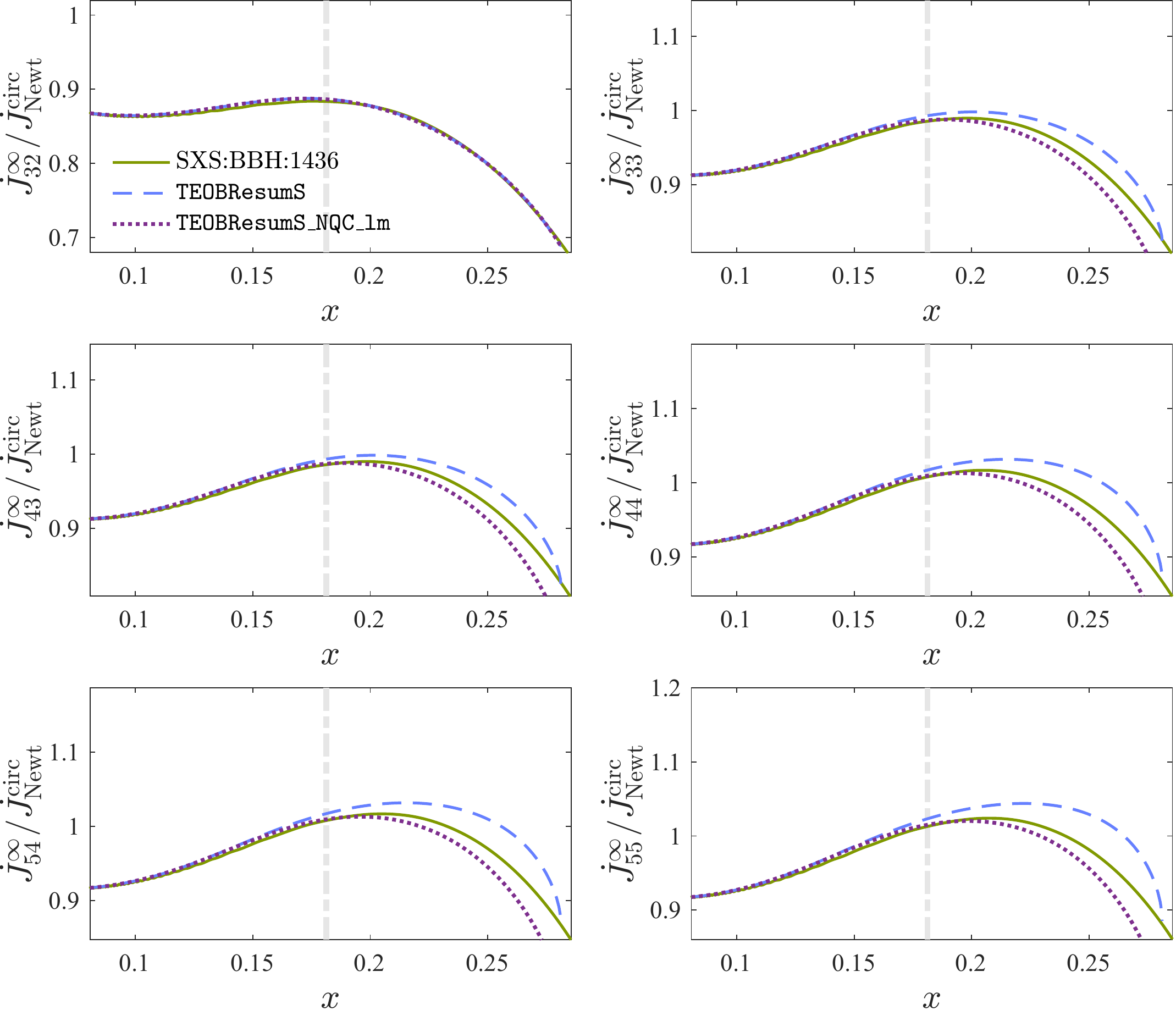} 
\hspace{1.5mm}
\includegraphics[width=0.45\textwidth]{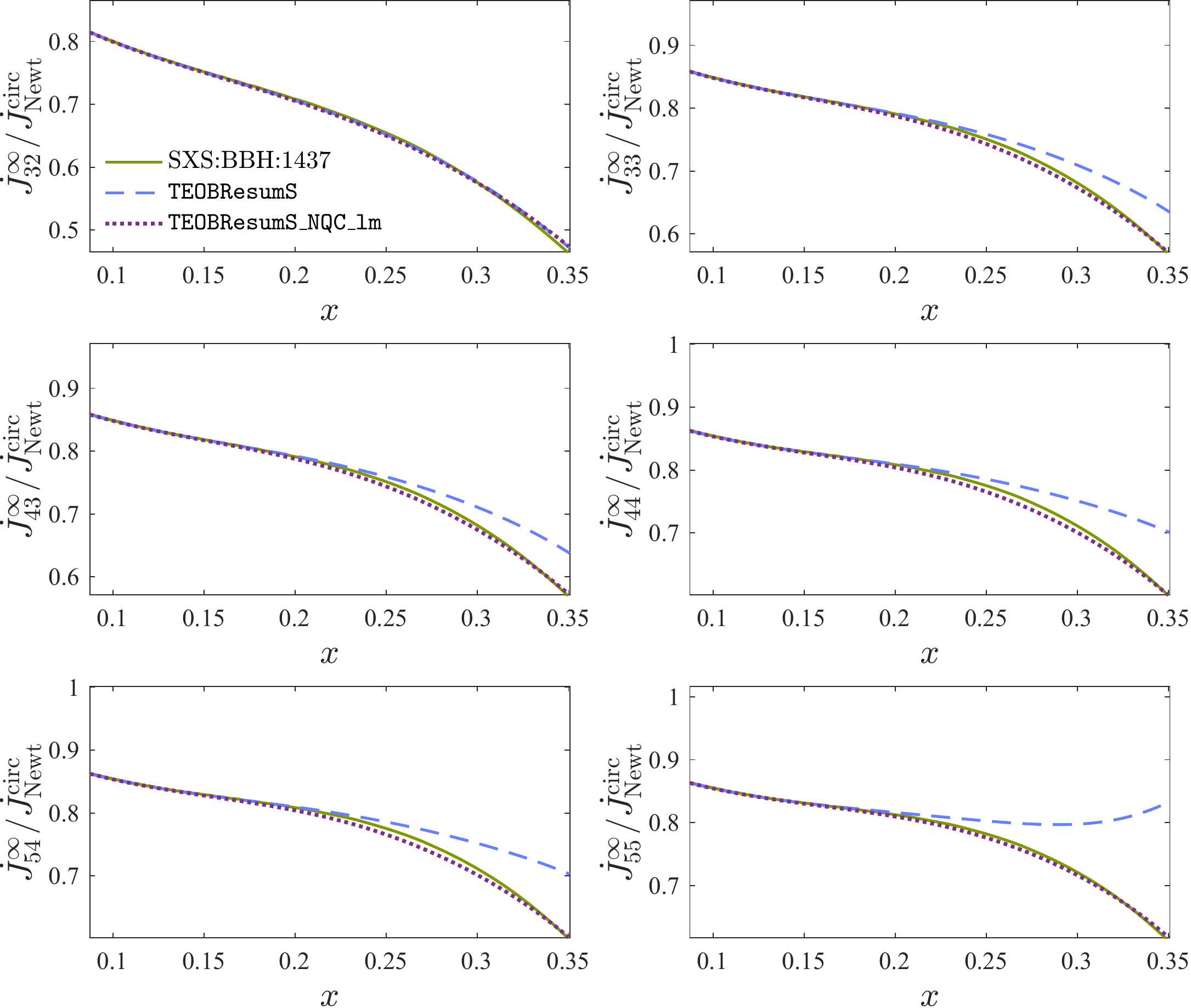}
\caption{\label{fig:new_multipoles} {\it Left}: Analogous of Fig.~\ref{fig:1436_multipoles} obtained using the new model,
showing that the change in $c_3$ does not affect the behavior of the flux. 
When summing up to $\ell = 8$ as done in Fig.~\ref{fig:1436old}, the EOB/NR fractional difference for \TEOBResumSlm{} 
remains below $10^{-2}$ for most of the evolution, even beyond the LSO.
{\it Right}: Contrasting the performance of \TEOBResumS{} and \TEOBResumSlm{} for the dataset 
SXS:BBH:1437 with $(q, \chi_1, \chi_2) = (6.038, 0.8, 0.1476)$. The behavior of the flux up to $\ell = 8$
progressively gets less robust and is discussed in Appendix~\ref{sec:NQCissues}.}
\end{figure*}

To give a flavor of the improved EOB/NR agreement that can be obtained with
the new $c_3$ and with the new radiation reaction, let us report a few
examples. From now on we will refer to the improved 
model as  \TEOBResumSlm{}, to easily distinguish it from \TEOBResumS{}.
Figure~\ref{fig:new_multipoles} shows  the updated flux comparison for SXS:BBH:1436,
and also includes the dataset SXS:BBH:1437 with $(q, \chi_1, \chi_2) = (6.038, 0.8, 0.1476)$.
The addition of NQC corrections to $\ell = m$ modes up to $\ell=5$ of the radiation reaction
is essential to improve the behavior of the analytic flux towards merger. 
For \TEOBResumSlm{} the fractional difference between EOB/NR total fluxes 
for the configuration SXS:BBH:1436 remains below $10^{-2}$ until 
$x \sim 0.26$. By contrast, in Fig.~\ref{fig:1436old}, the fractional difference 
for \TEOBResumS{} already reached $10^{-2}$ at the LSO and kept growing until merger. 

\begin{figure}[t]
\includegraphics[width=0.43\textwidth]{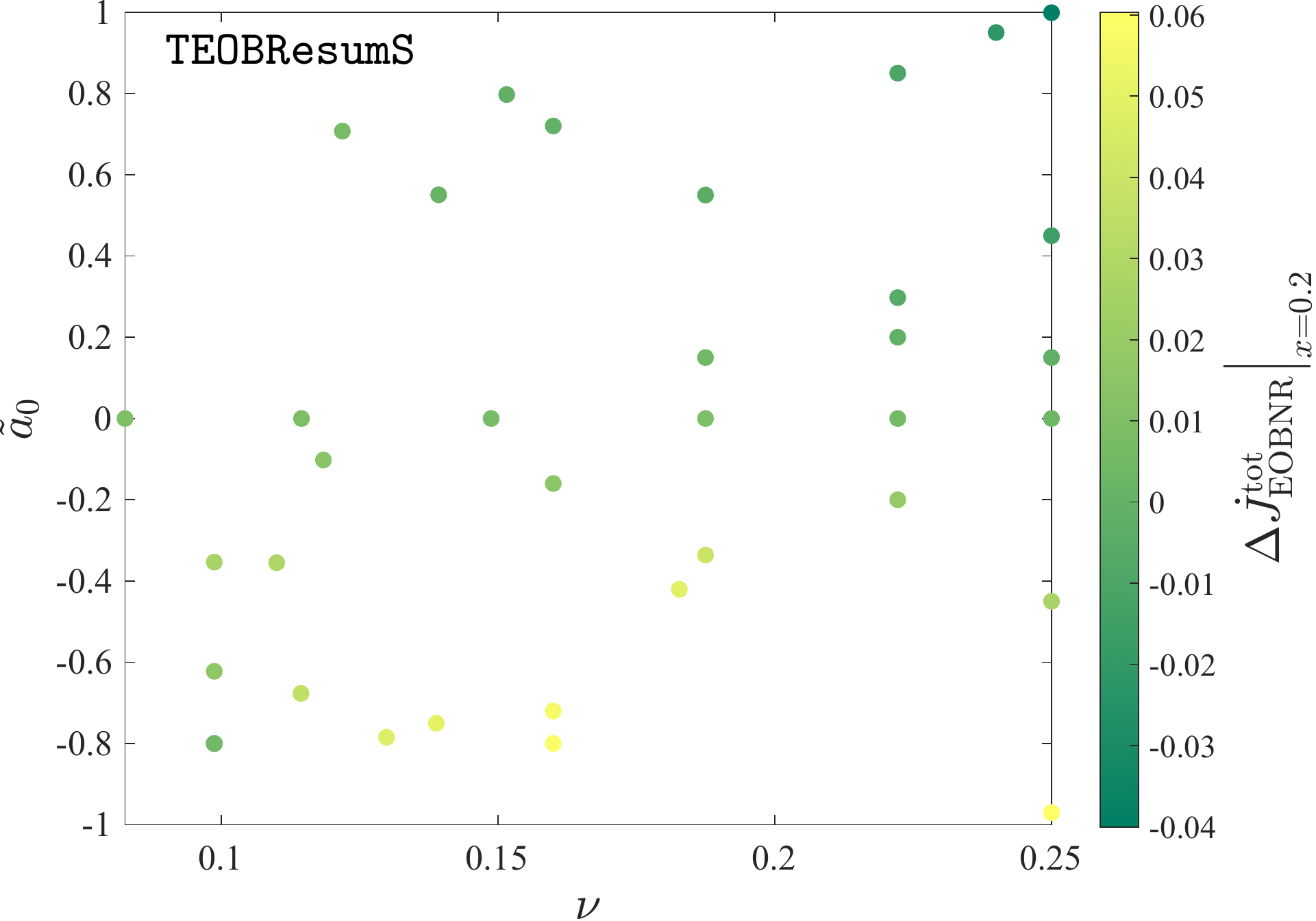} \\
\vspace{1mm}
\includegraphics[width=0.43\textwidth]{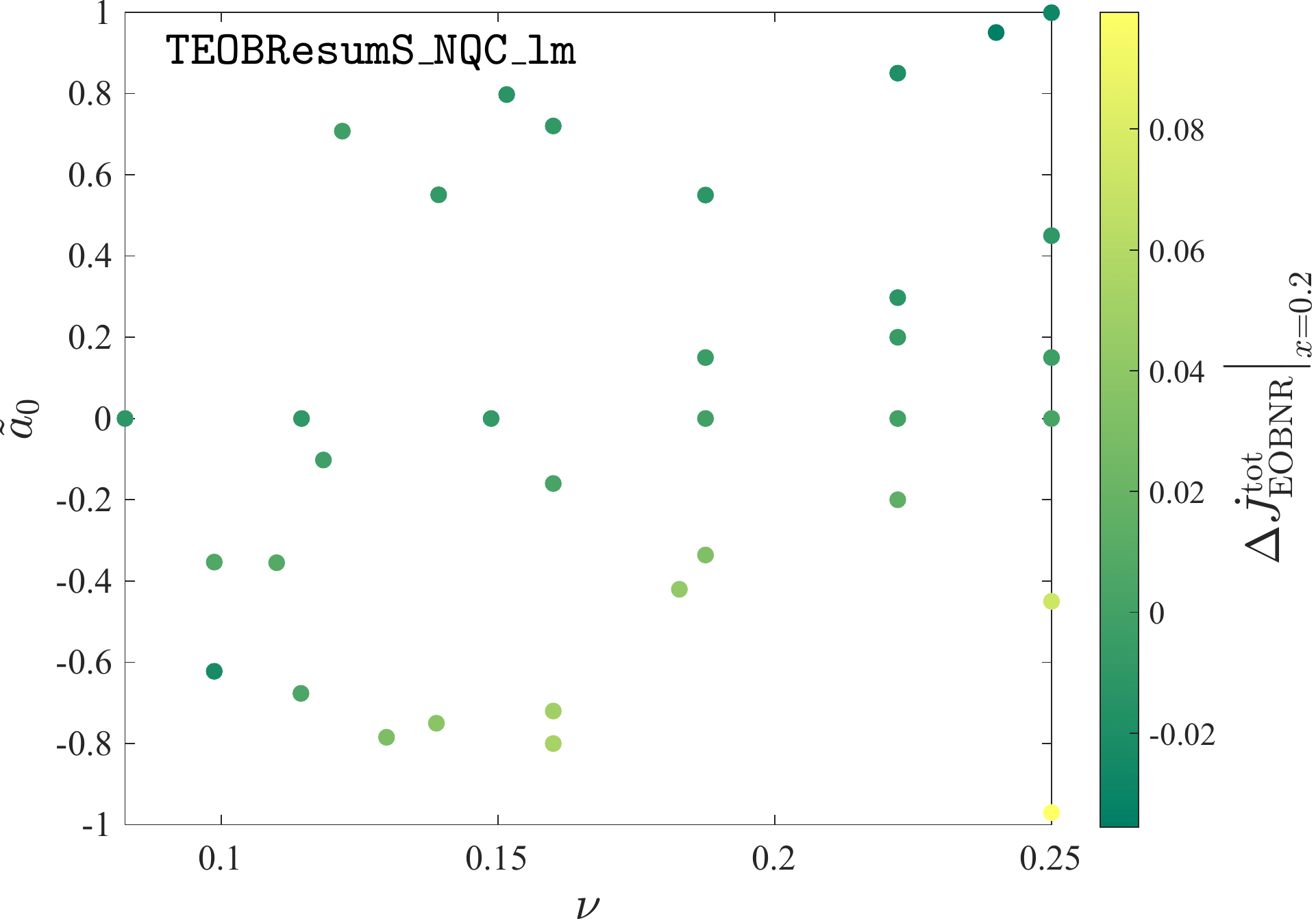}
\caption{\label{fig:flux_diff} Fractional EOB/NR flux differences at \mbox{$x = 0.2$}
for \TEOBResumS{} (top) and \TEOBResumSlm{} (bottom) evaluated
for the sample of SXS data of Table~\ref{tab:spinning_flux}.
For \TEOBResumSlm{} we are excluding from the points
two configurations, corresponding to datasets SXS:BBH:1419 and SXS:BBH:1375,
where the contribution of modes with \mbox{$\ell_{\rm max}>5$} becomes important 
towards merger. These will be discussed in Appendix~\ref{sec:NQCissues}.}
\end{figure}

We finally test the performance of the model over all datasets of Table~\ref{tab:spinning_flux},
by computing the fractional difference between EOB and NR (total) fluxes at $x = 0.2$ for both
\TEOBResumS{} and \TEOBResumSlm{}, as shown respectively in the top and bottom panel of
Fig.~\ref{fig:flux_diff}. Here one can see an evident improvement for larger mass ratios and negative 
values of the effective Kerr parameter.

\begin{figure}[t]
\includegraphics[width=0.23\textwidth]{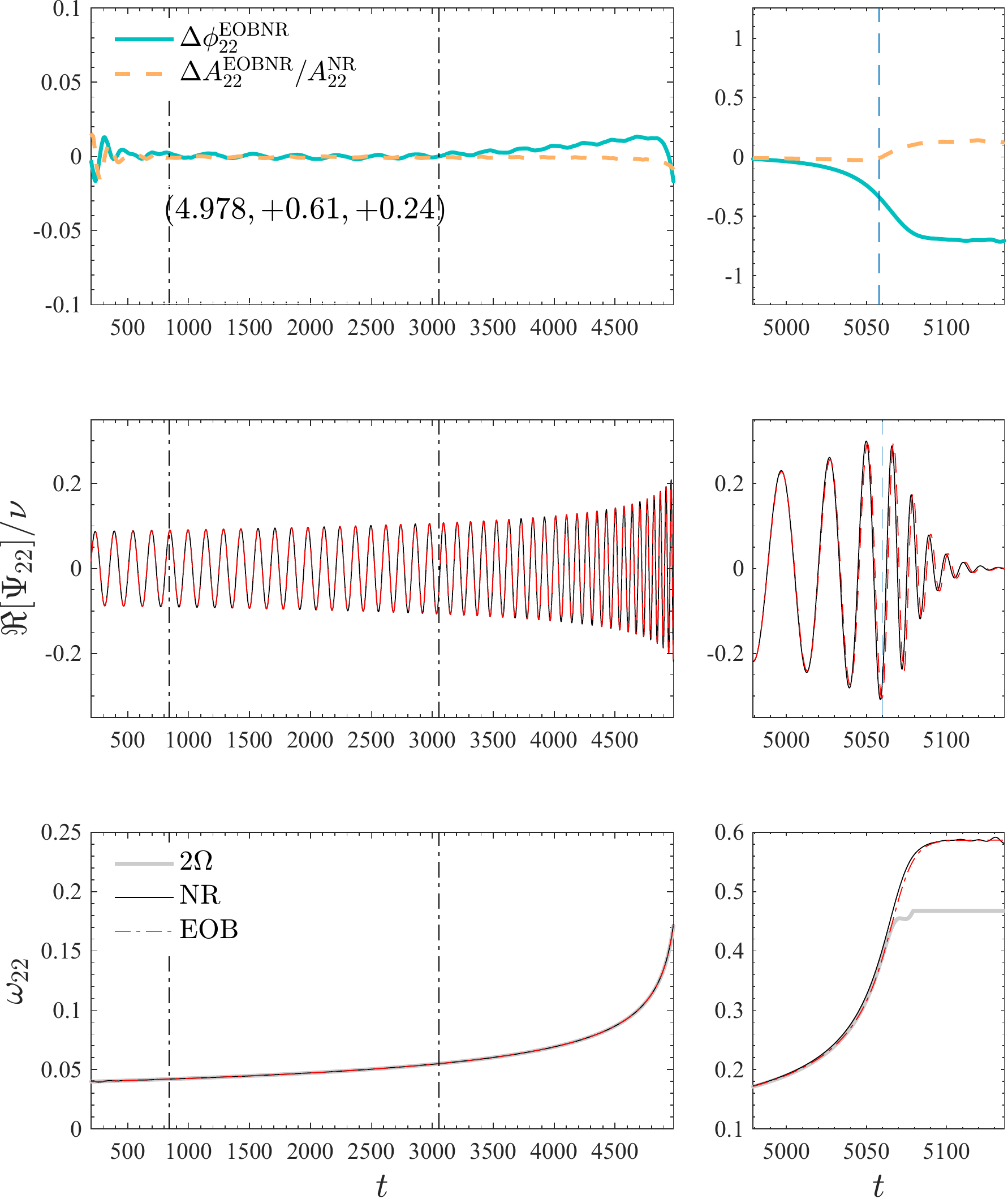}  
\includegraphics[width=0.23\textwidth]{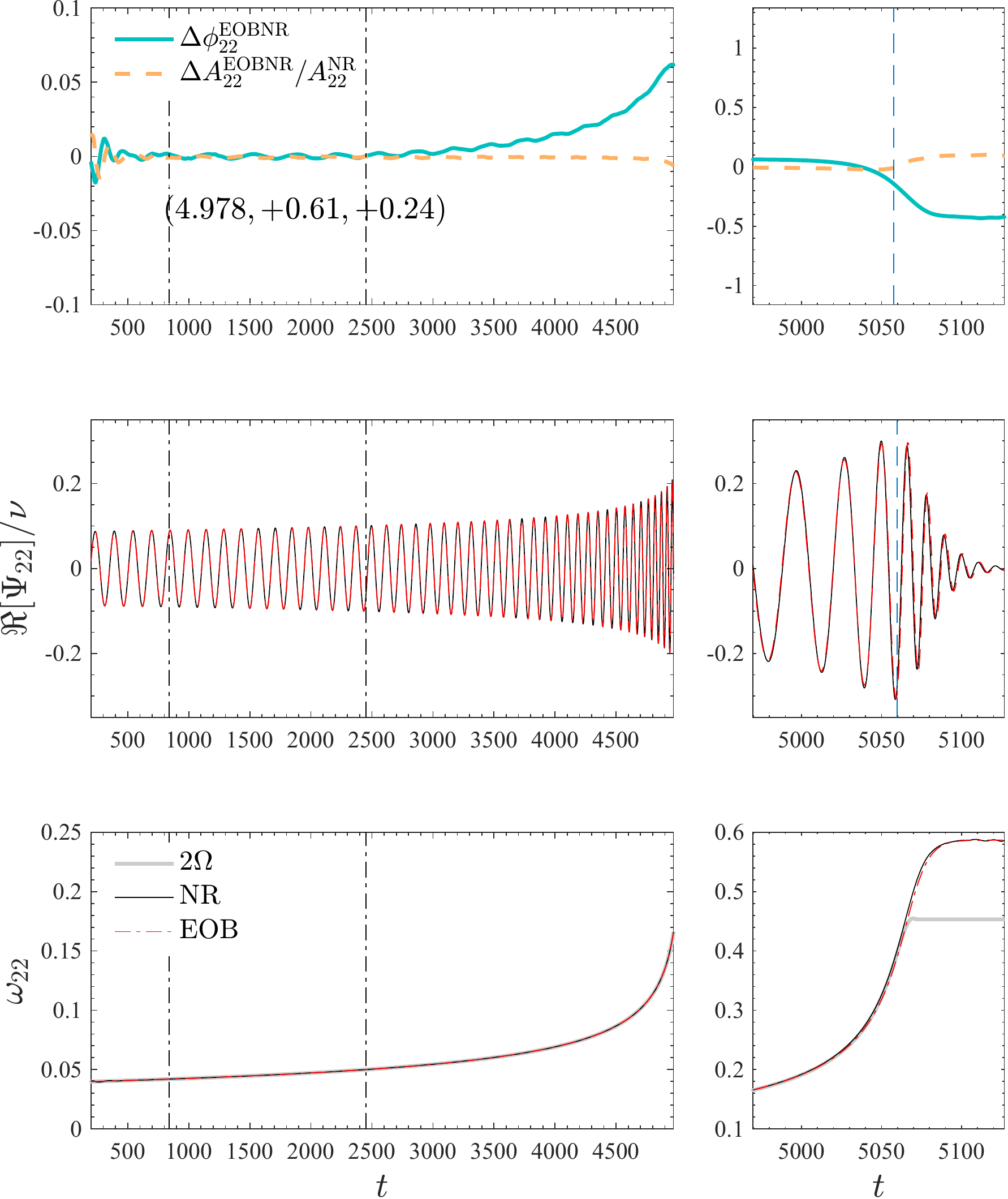} 
\includegraphics[width=0.23\textwidth]{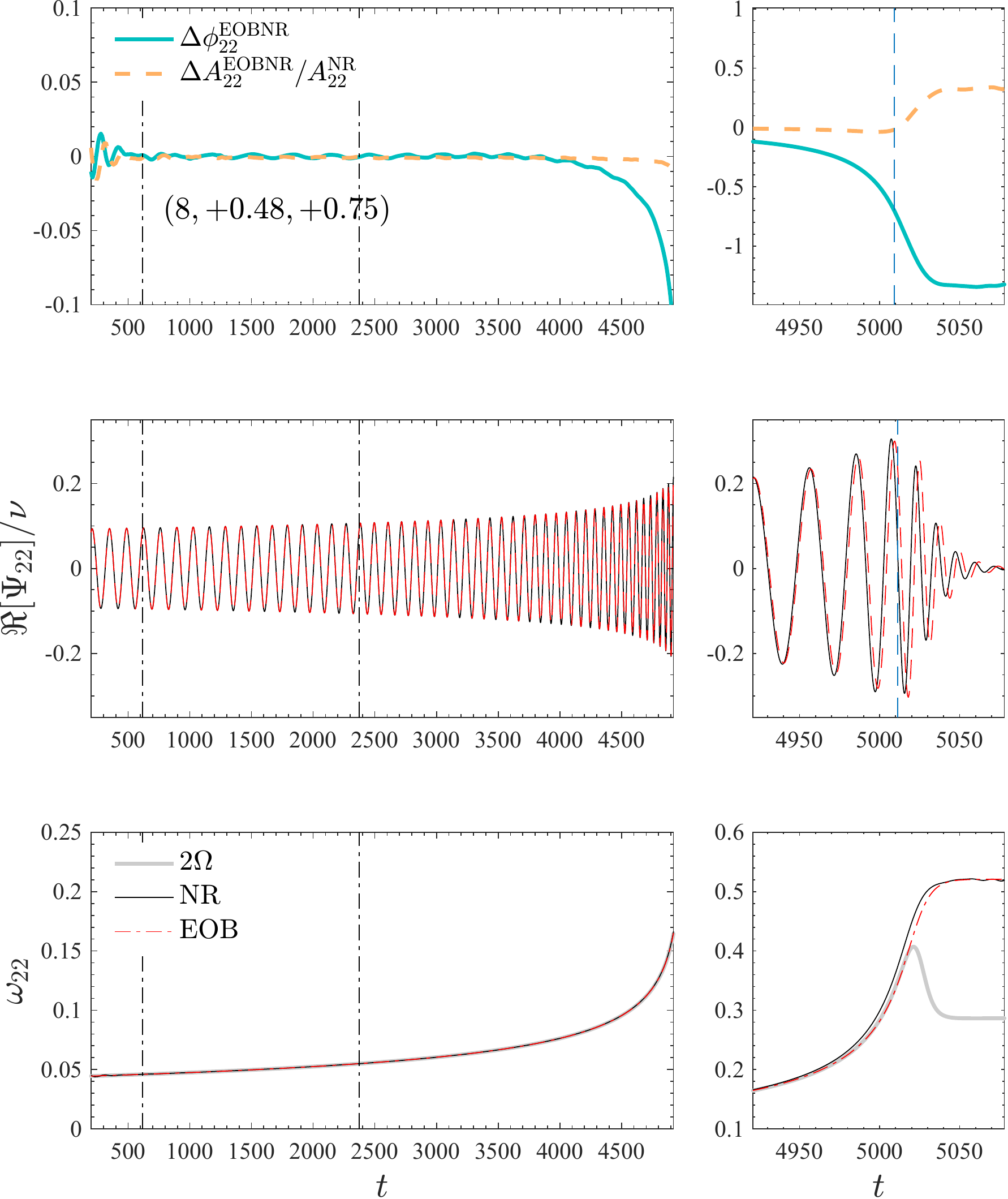}  
\includegraphics[width=0.23\textwidth]{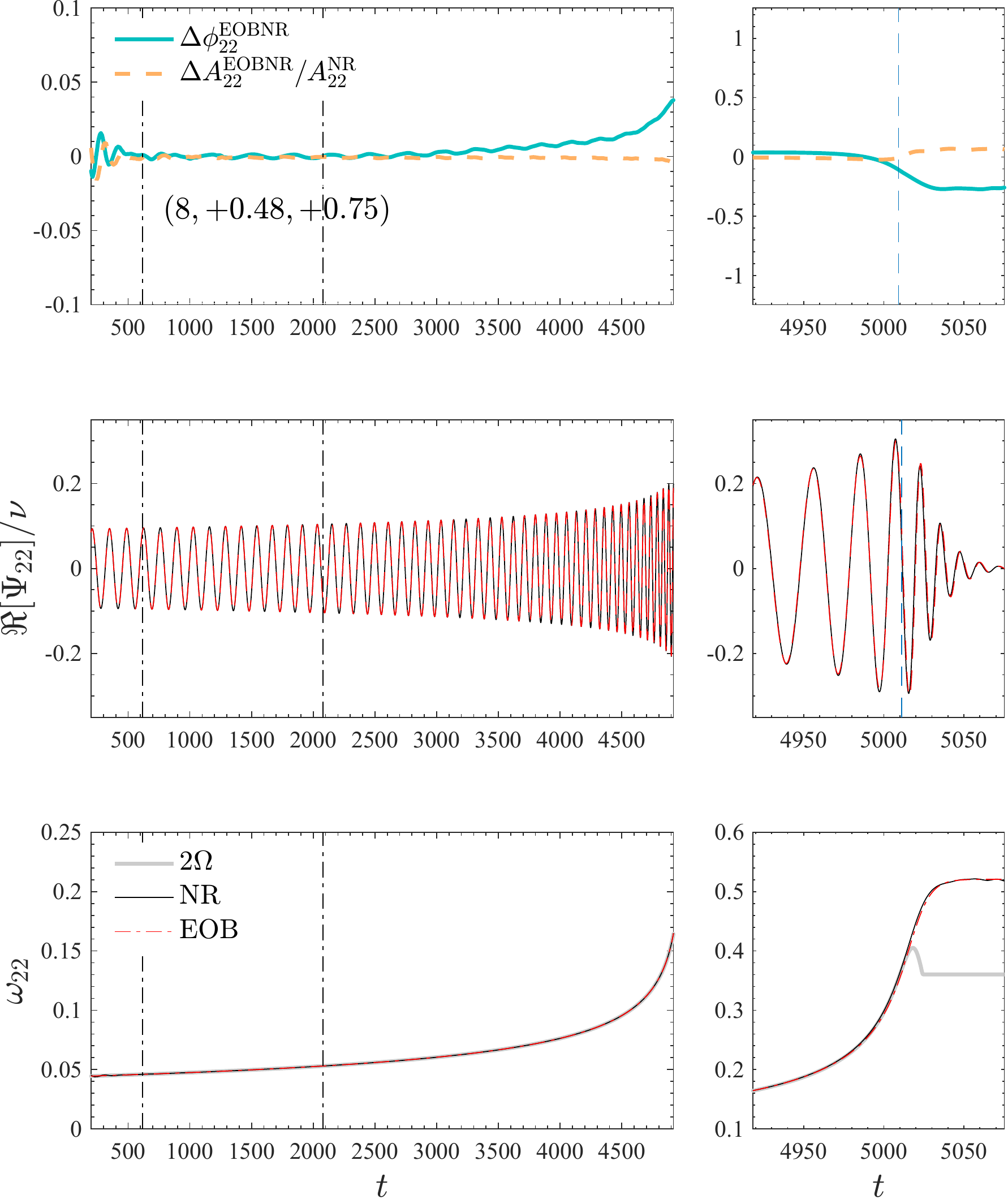} 
\caption{\label{fig:phasings} EOB/NR time-domain phasing for two illustrative datasets:
SXS:BBH:1463 with $(q, \chi_1, \chi_2) = (4.978, +0.61, +0.24)$ (top panels) and
 SXS:BBH:1426 with $(q, \chi_1, \chi_2) = (8, +0.48, +0.75)$ (bottom panels),
 using \TEOBResumS{} (left) and \TEOBResumSlm{} (right).
 Each plot shows: (i) the phase difference and the relative amplitude difference; 
 (ii) the real parts of the EOB and NR waveforms; (iii) the instantaneous GW
 frequency together with twice the orbital frequency $\Omega$. Vertical dash-dotted lines
 indicate the alignment interval. The phase differences 
 $\Delta \phi^{\rm EOBNR}_{22}$ at merger (vertical dashed blue line) are respectively $(-0.34,-0.70)$~rad 
 for \TEOBResumS{} and become $(-0.14,-0.11)$ for \TEOBResumSlm{}.
 Note that only SXS:BBH:1426 was used to inform $c_3$.}
 \end{figure}
 Another example is shown in Fig.~\ref{fig:phasings}, that focuses on time-domain phasings.
 We use here the Regge-Wheeler-Zerilli normalized waveform, defined as
$\Psi_\lm = h_\lm/\sqrt{(\ell - 1) \ell (\ell + 1) (\ell + 2)}$. The EOB waveforms
have been obtained by setting the spin values with 6 digits precision, considering the 
initial $\chi_1, \chi_2$ given in the metadata file for each simulation\footnote{We noticed a decreased phase difference 
at merger when using larger precisions.}. The figure contrasts EOB/NR waveform phasings for the
$\ell = m = 2$ multipole, considering datasets SXS:BBH:1463 (first row) and SXS:BBH:1426 (second row) 
using \TEOBResumS{} (left) and \TEOBResumSlm{} (right). As usually done, in this case we are using
$N=3$  extrapolation order for the SXS waveforms. In each figure, the top panels show the phase and amplitude 
difference, where $\Delta \phi^{\rm EOBNR}_{22} \equiv \phi^\EOB_{22}- \phi^{\rm NR}_{22}$.
The EOB/NR phasing agreement is better for \TEOBResumSlm{} than for \TEOBResumS{}, although the 
SXS:BBH:1426 dataset is among those used to inform the new expression of $c_3$.

\section{EOB/NR $\ell=m=2$ unfaithfulness}
\label{sec:barF}
\begin{figure}[t]
\center
\includegraphics[width=0.45\textwidth]{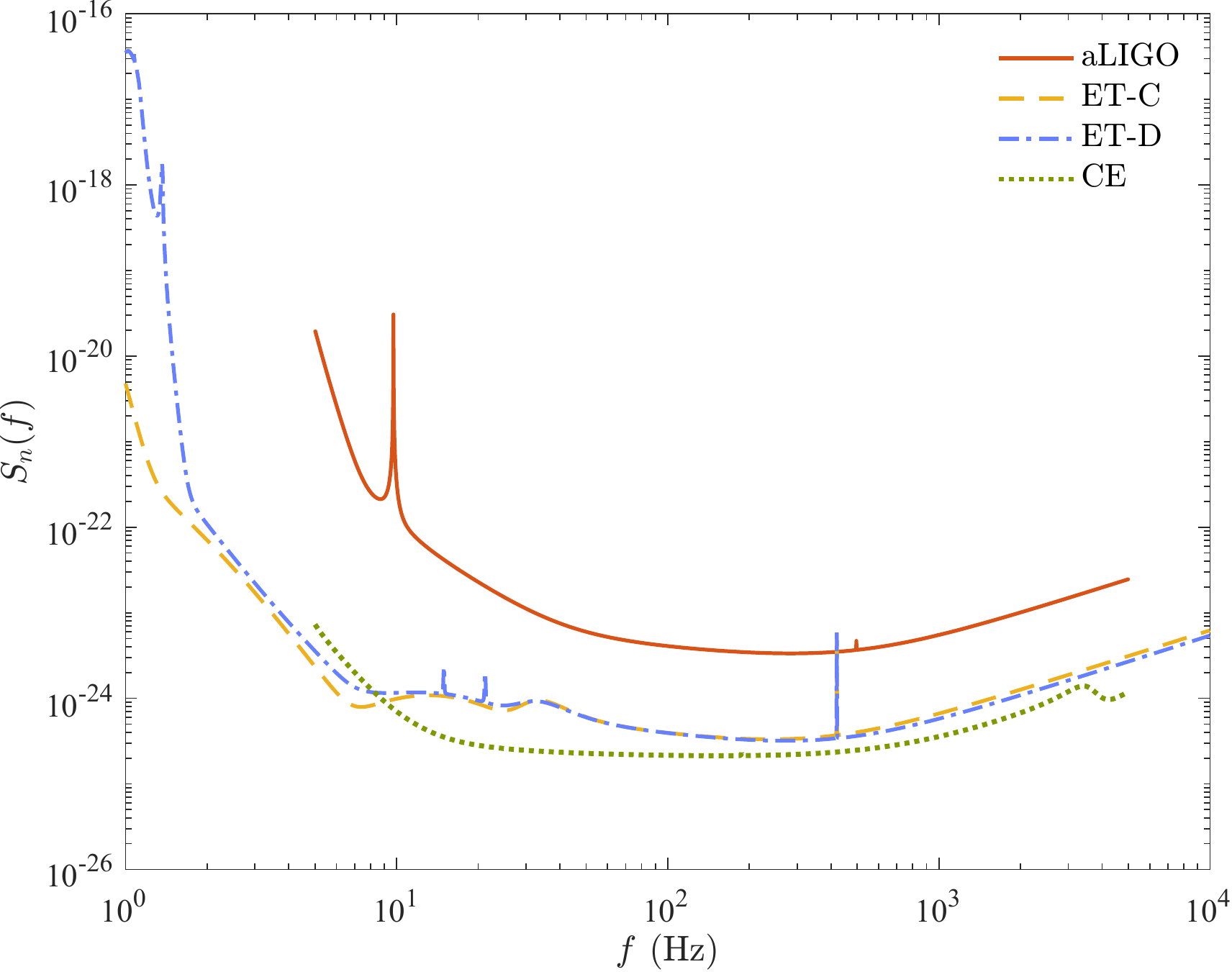} 
\caption{\label{fig:noises} Sensitivity curves for the three detectors we take into consideration
in computing the unfaithfulness for the two versions of our model: Advanced LIGO, Einstein Telescope (ET) and 
Cosmic Explorer (CE). Here ET-C is the sensitivity model described in Ref.~\cite{Hild:2009ns}, while ET-D is the latest version~\cite{Hild:2010id}.}
\end{figure}
\begin{figure*}[t]
\center
\includegraphics[width=0.32\textwidth]{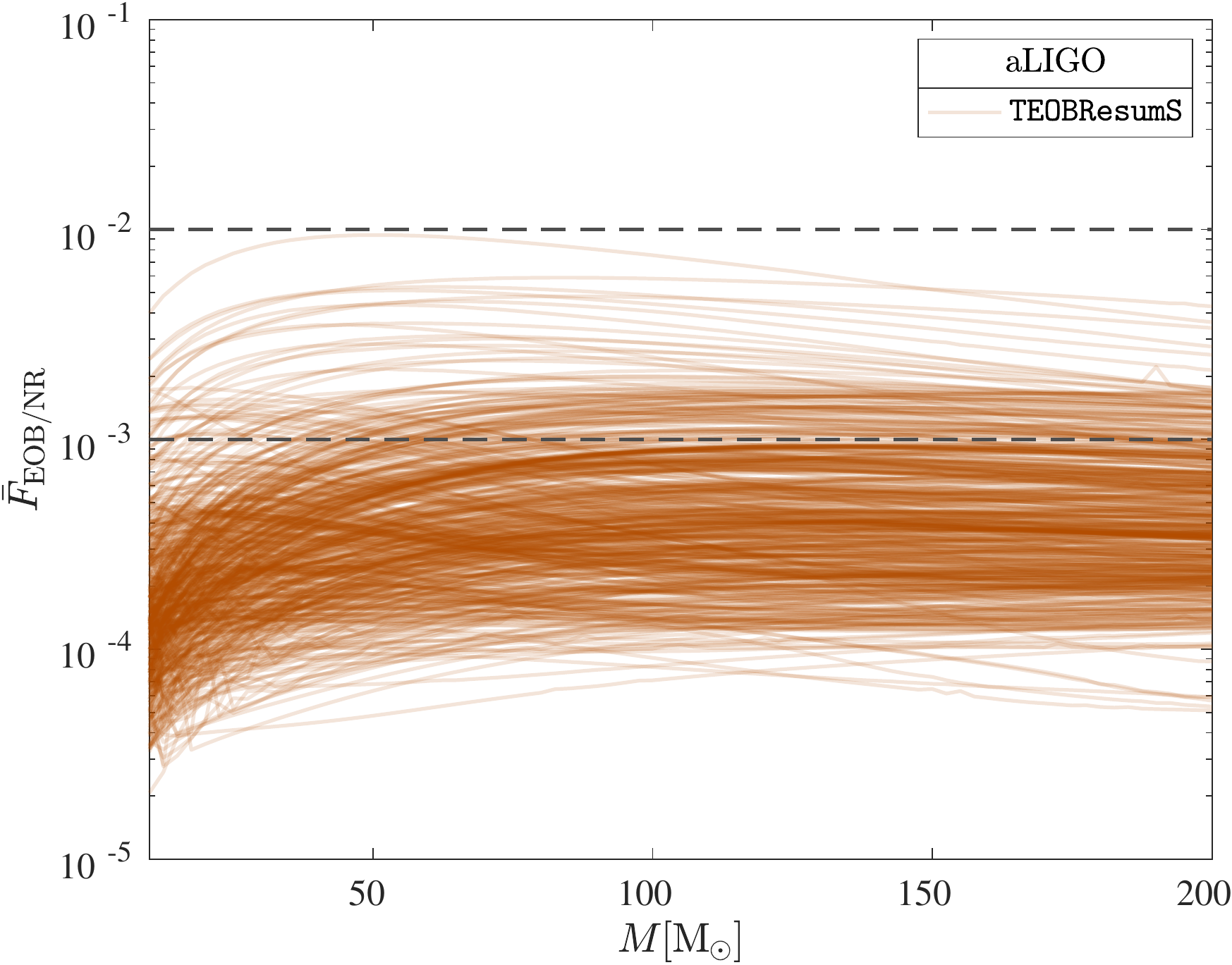} 
\includegraphics[width=0.32\textwidth]{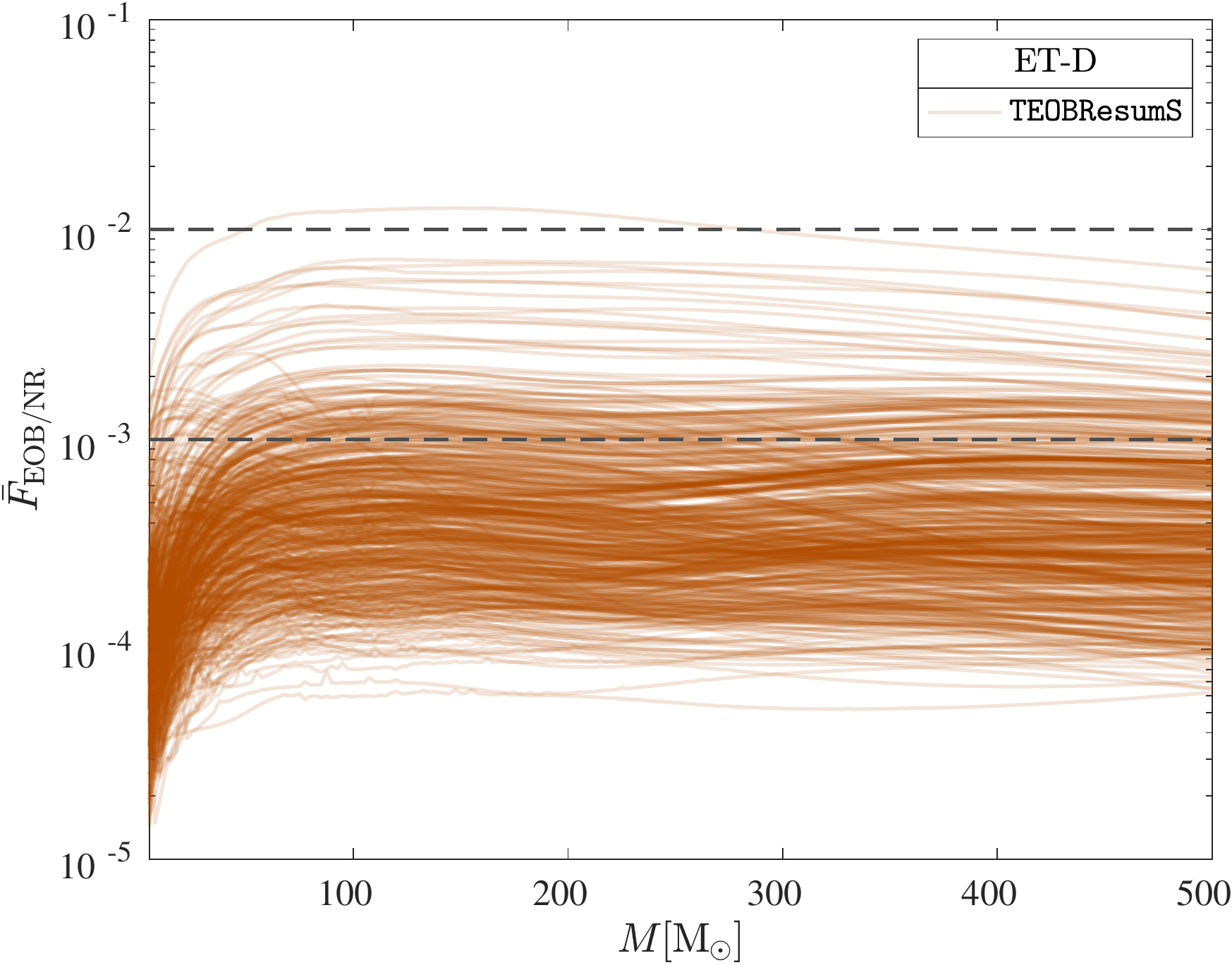}
\includegraphics[width=0.32\textwidth]{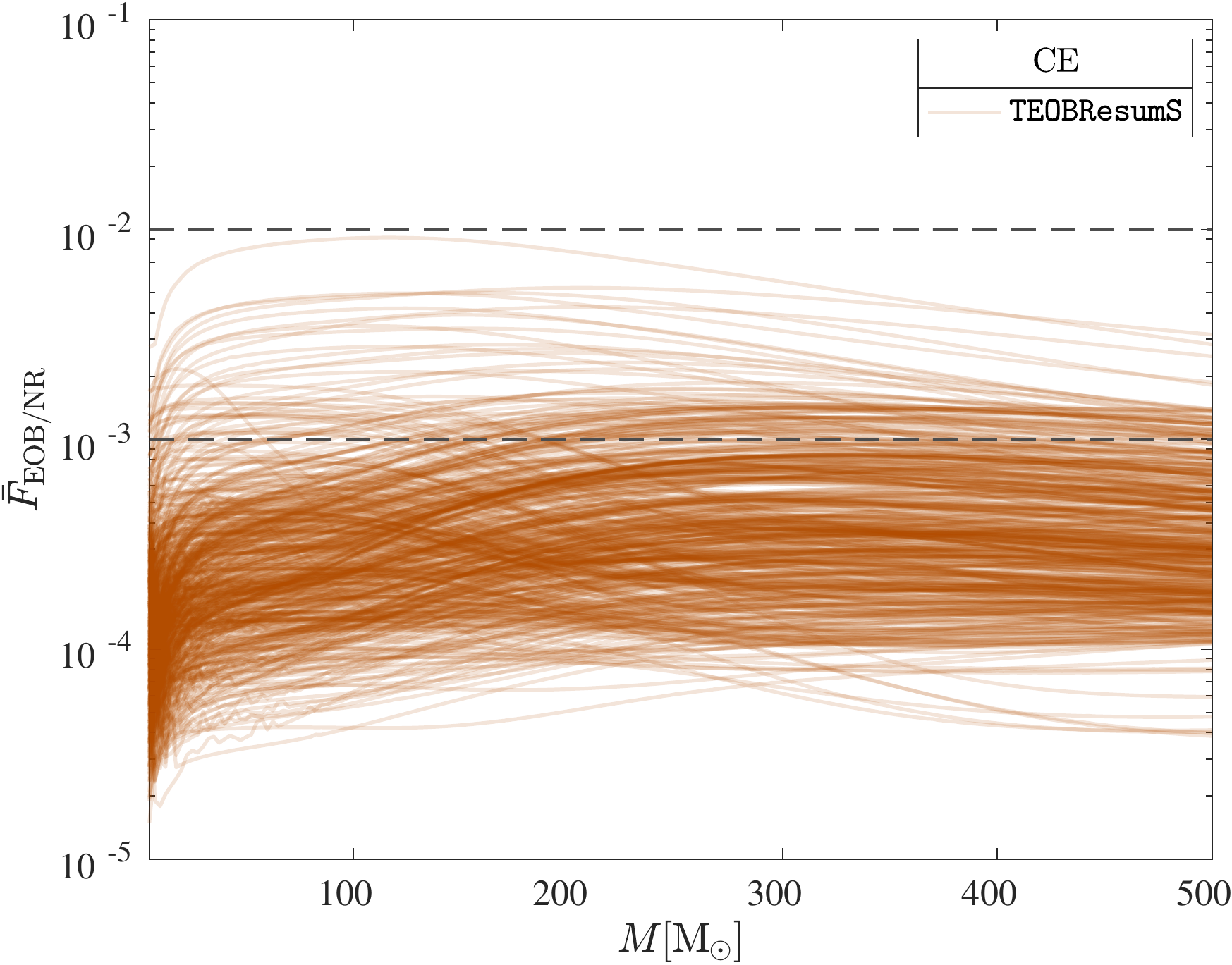} \\
\includegraphics[width=0.32\textwidth]{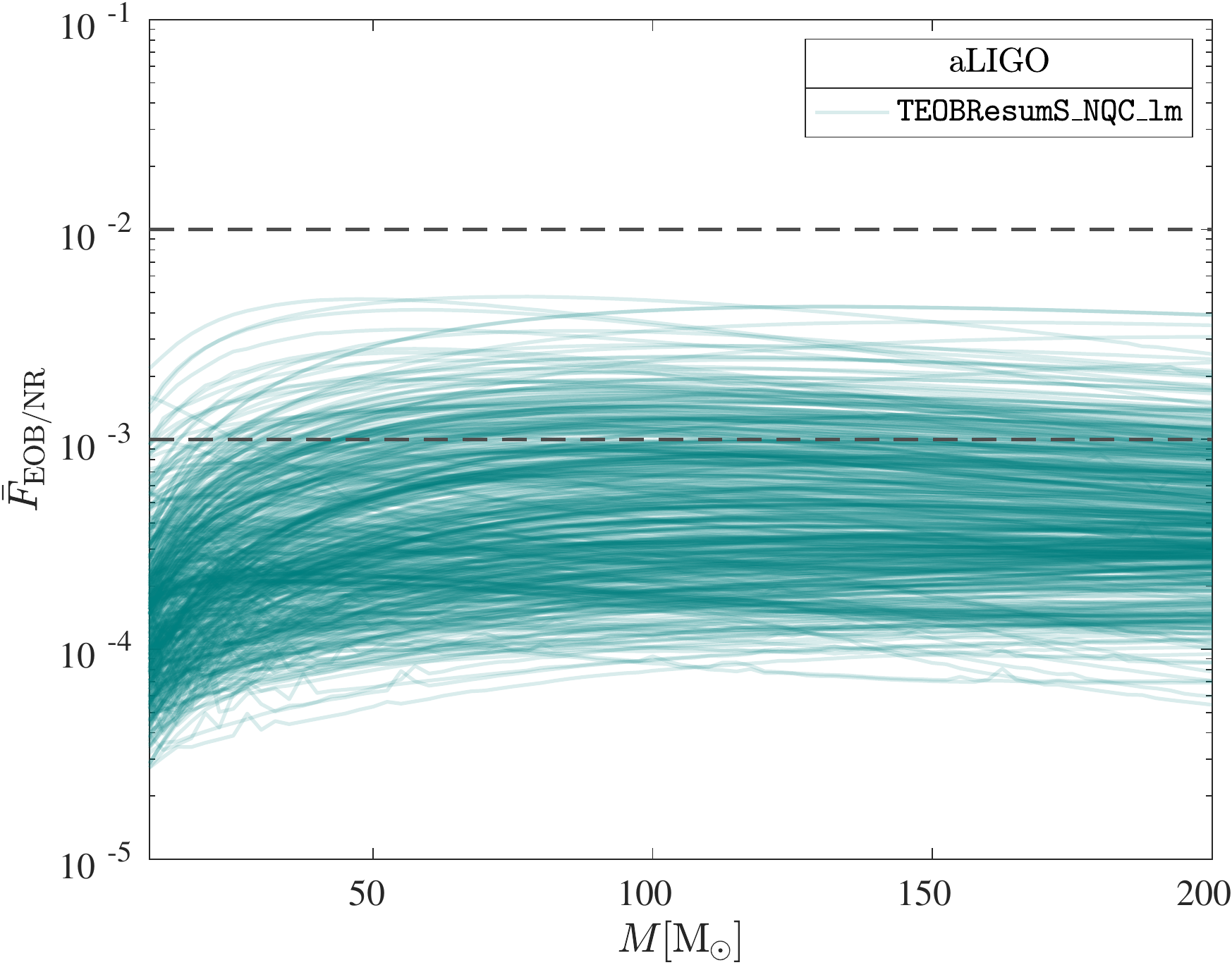} 
\includegraphics[width=0.32\textwidth]{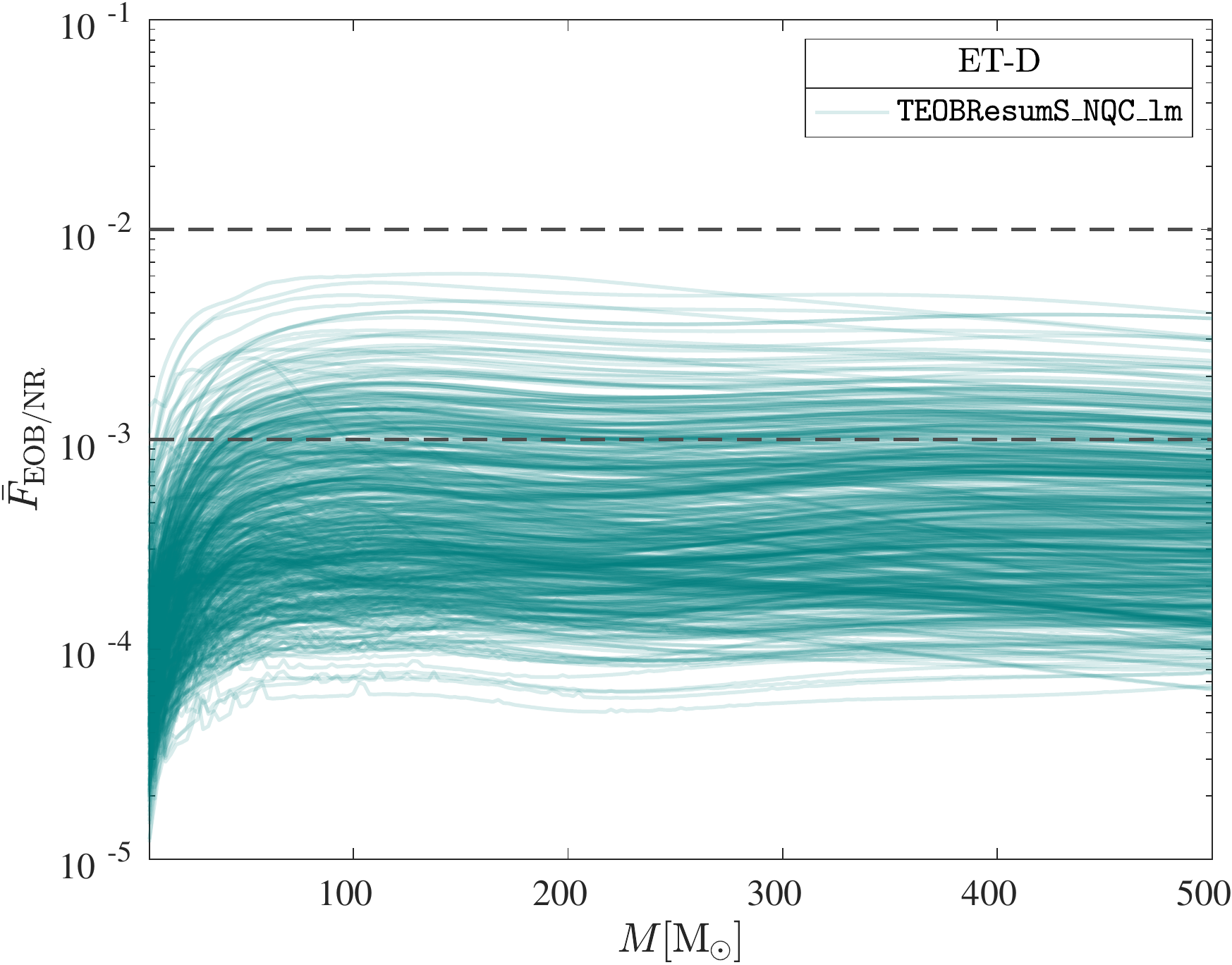} 
\includegraphics[width=0.32\textwidth]{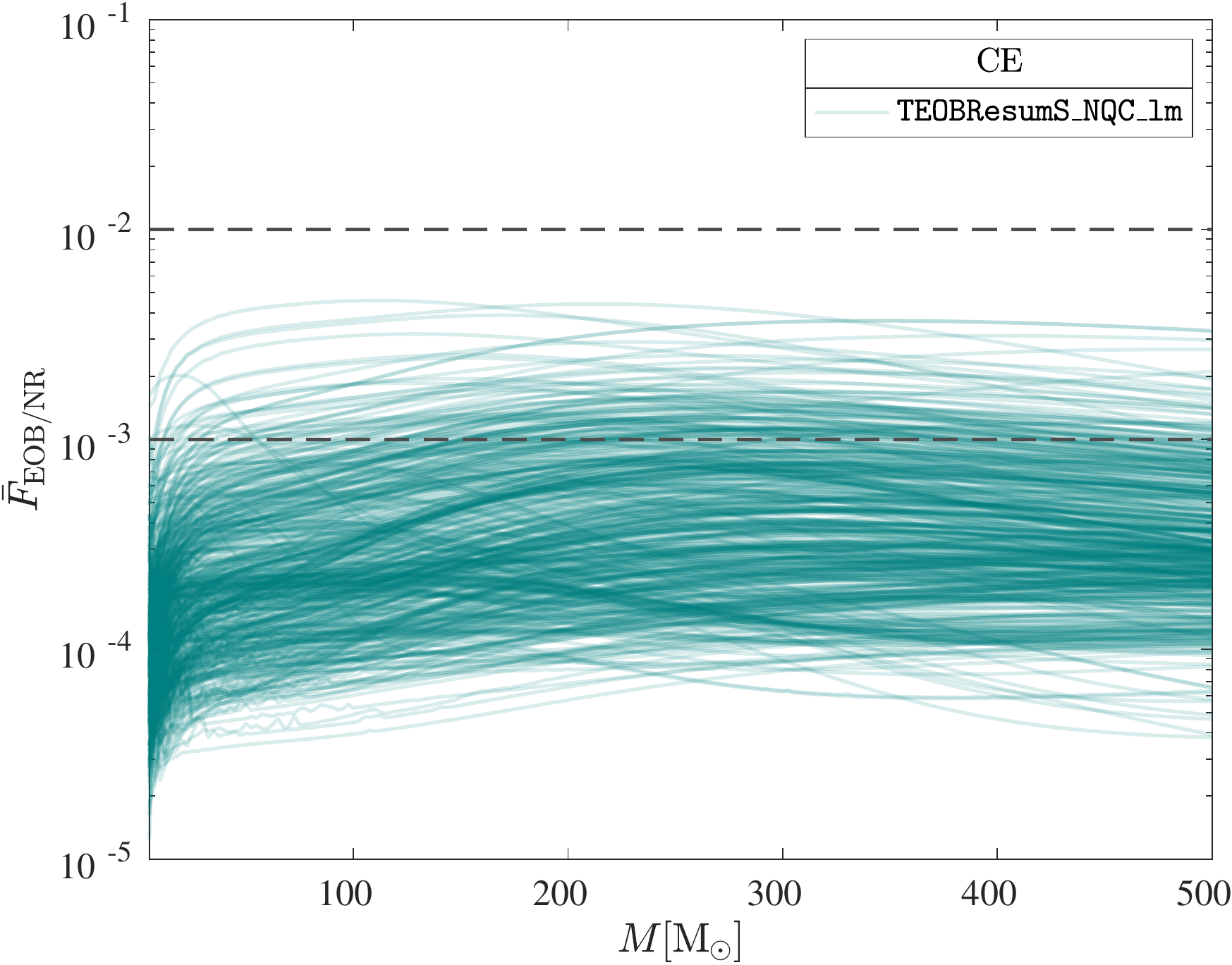} 
\caption{\label{fig:barF} EOB/NR unfaithfulness for \TEOBResumS{} (top panels) 
and \TEOBResumSlm{} (bottom panels) evaluated over the
sample of 534 nonprecessing quasicircular datasets of the SXS catalog already considered in Ref.~\cite{Riemenschneider:2021ppj}, using: 
(i) the zero-detuned, high-power noise spectral density of Advanced LIGO (first column), 
(ii) the latest version of the expected noise for the  Einstein Telescope (second column),
(iii) the expected noise for Cosmic Explorer (third column).
We observe here how the changes implemented in the new version of our model ensure a slight decrease 
in $\bar{F}_{\rm EOB/NR}$, whose average is between $10^{-3}$ and $10^{-4}$.}
\end{figure*}
\begin{figure}[t]
	\begin{center}
		\includegraphics[width=0.43\textwidth]{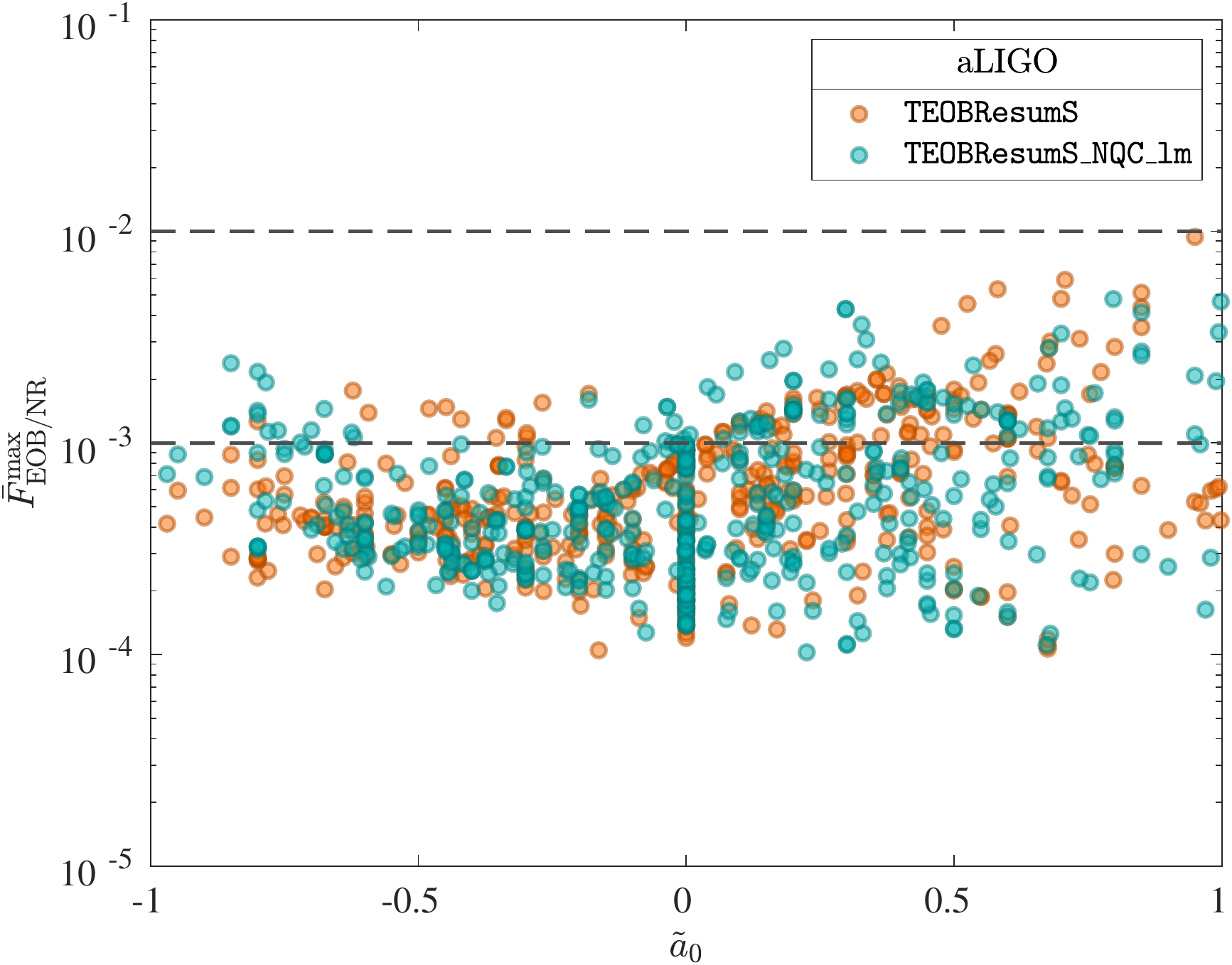} \\
		\includegraphics[width=0.43\textwidth]{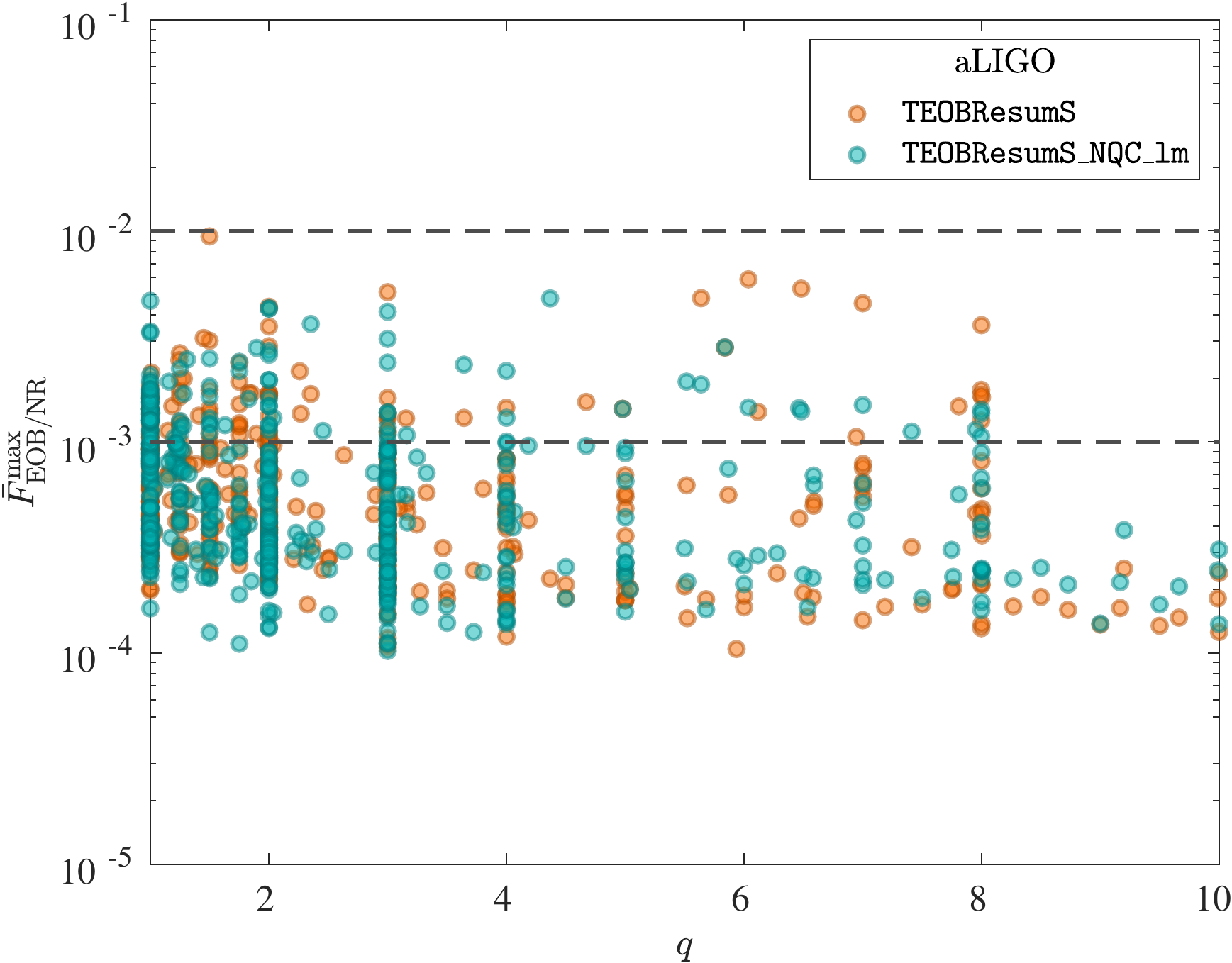}
		\caption{\label{fig:maxF}Contrasting $\bar{F}^{\rm max}_{\rm EOB/NR}$
			for \TEOBResumS{} and \TEOBResumSlm{} versus $\tilde{a}_0$ and $q$, using the 
			PSD of Advanced LIGO. This complements the top panels of Fig.~\ref{fig:barF}.}
	\end{center}
\end{figure}
\begin{figure}[t]
\begin{center}
\includegraphics[width=0.23\textwidth]{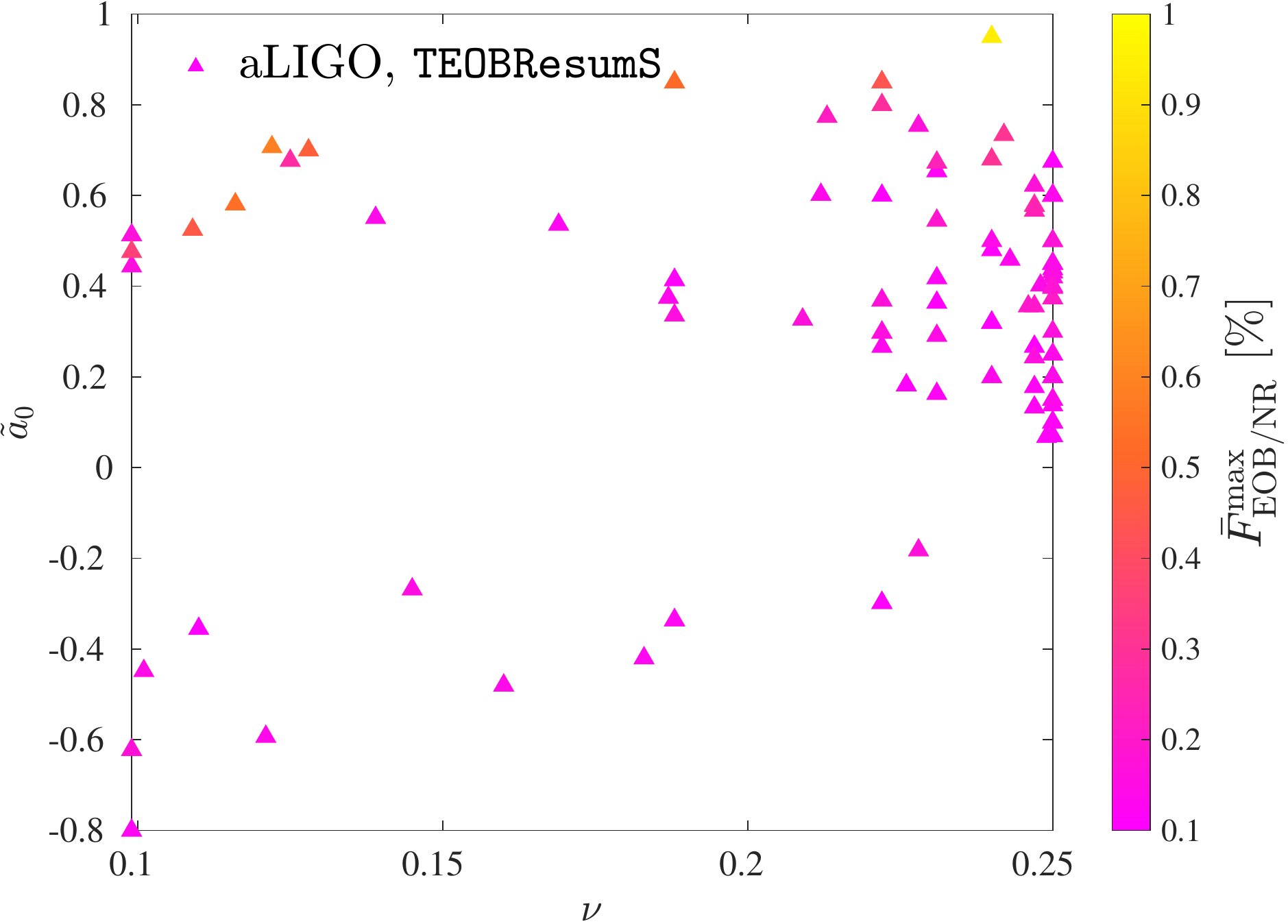}
\includegraphics[width=0.23\textwidth]{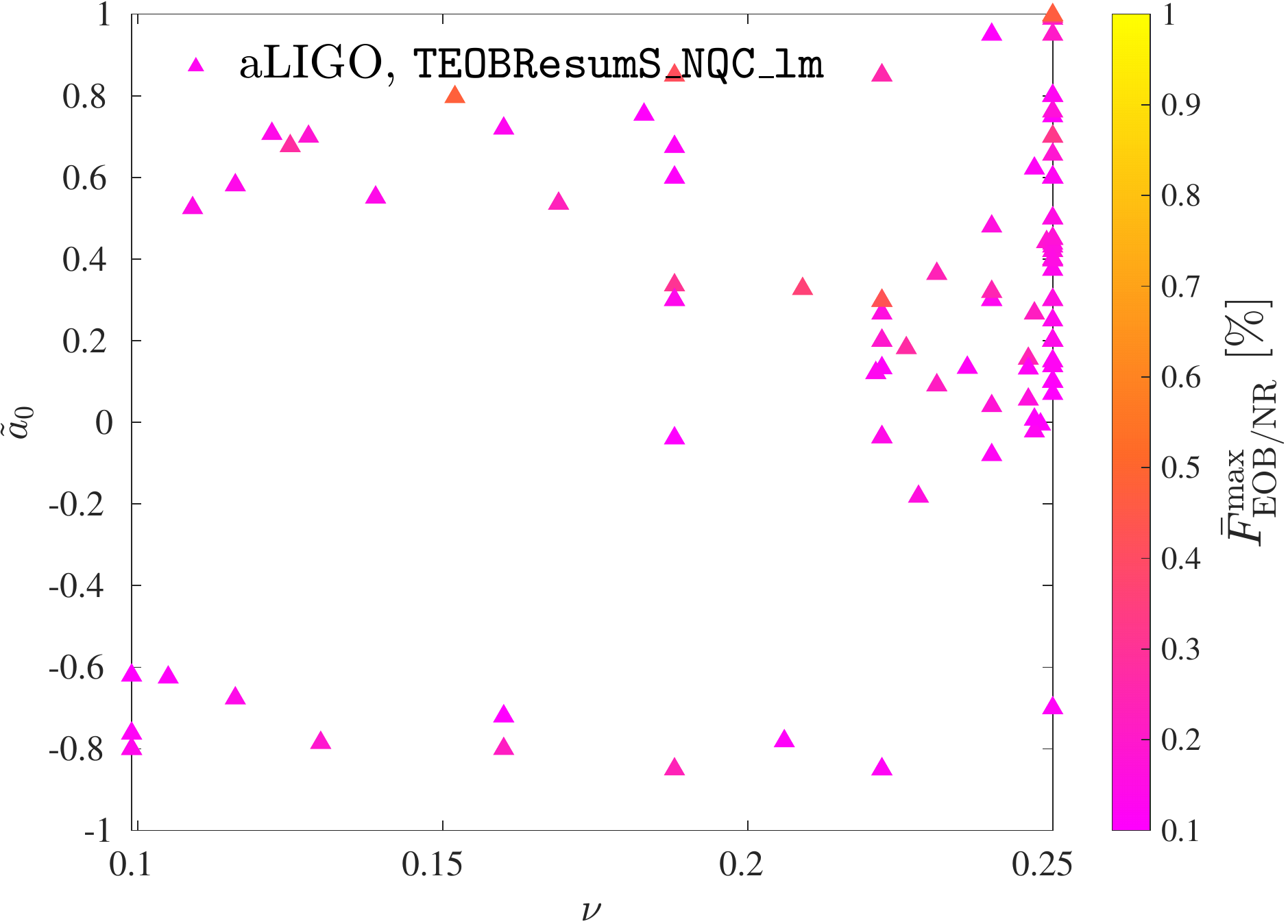} \\
\includegraphics[width=0.23\textwidth]{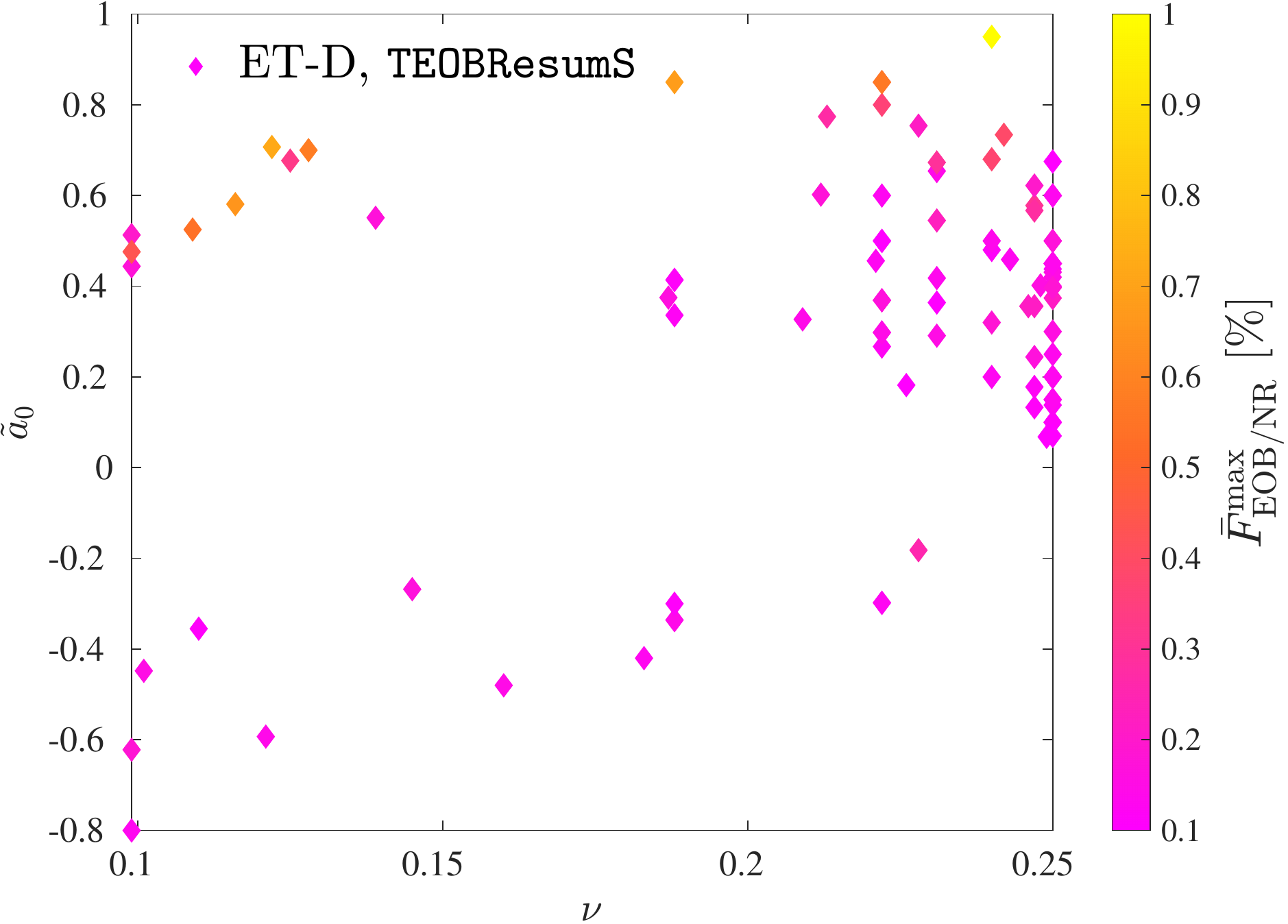}
\includegraphics[width=0.23\textwidth]{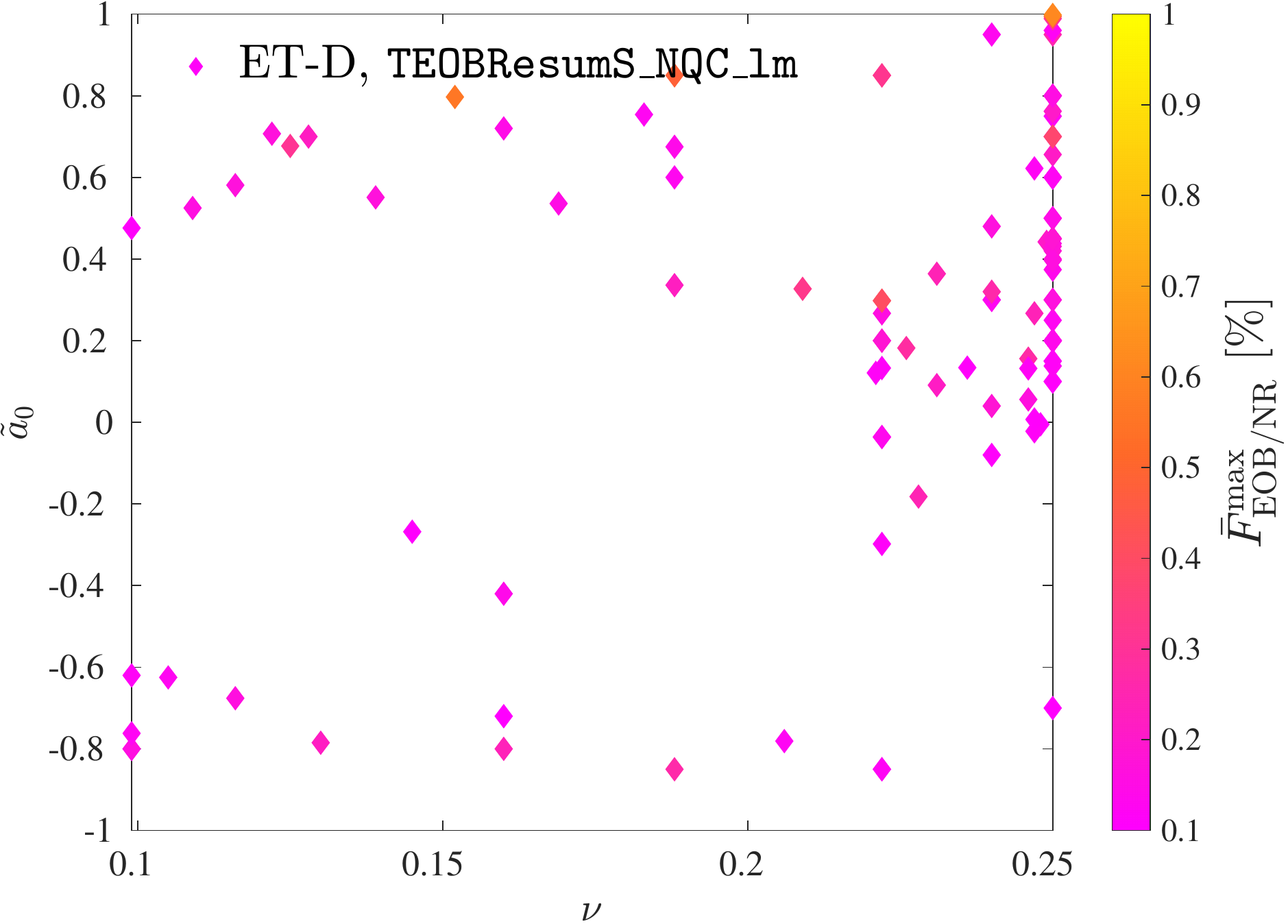} \\
\includegraphics[width=0.23\textwidth]{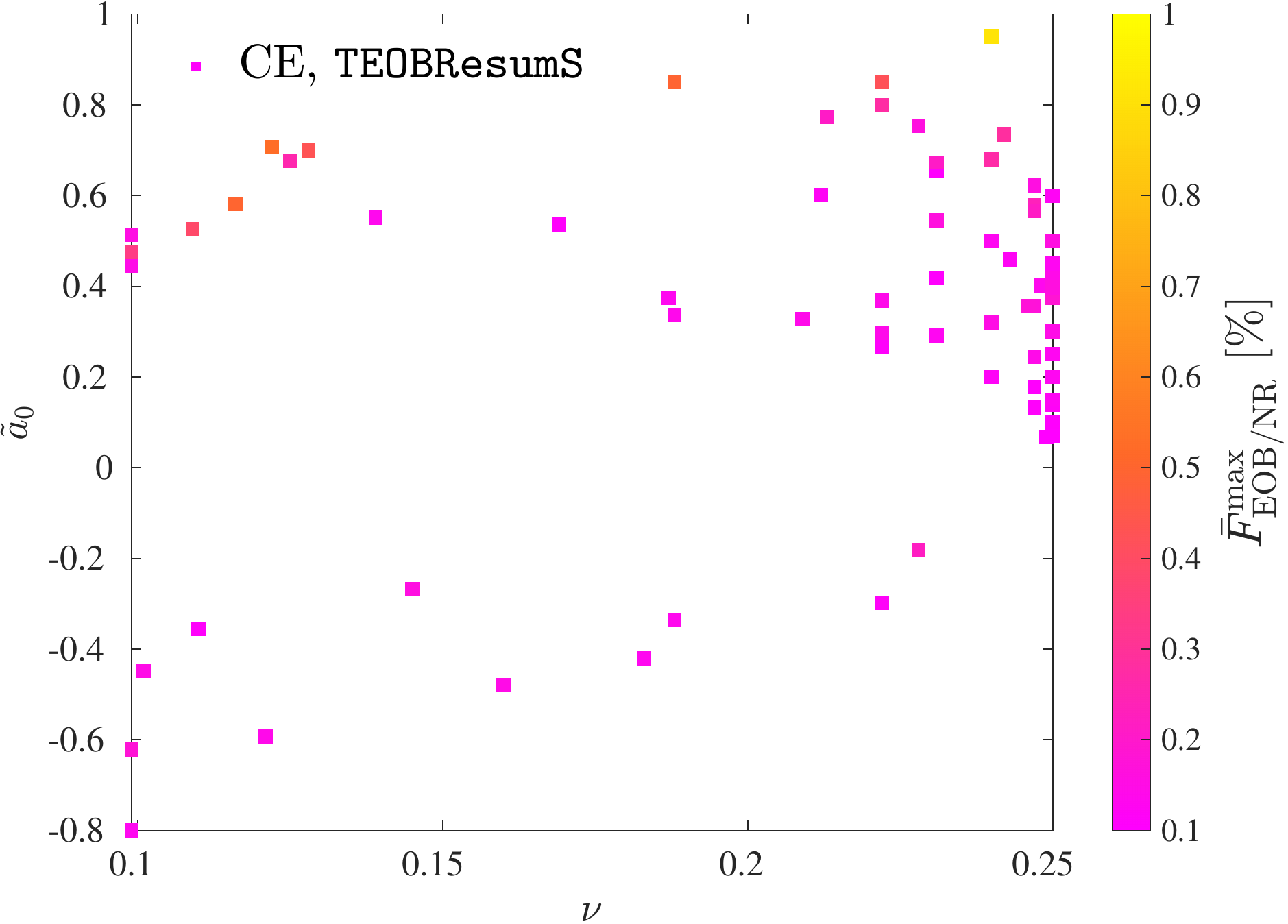}
\includegraphics[width=0.23\textwidth]{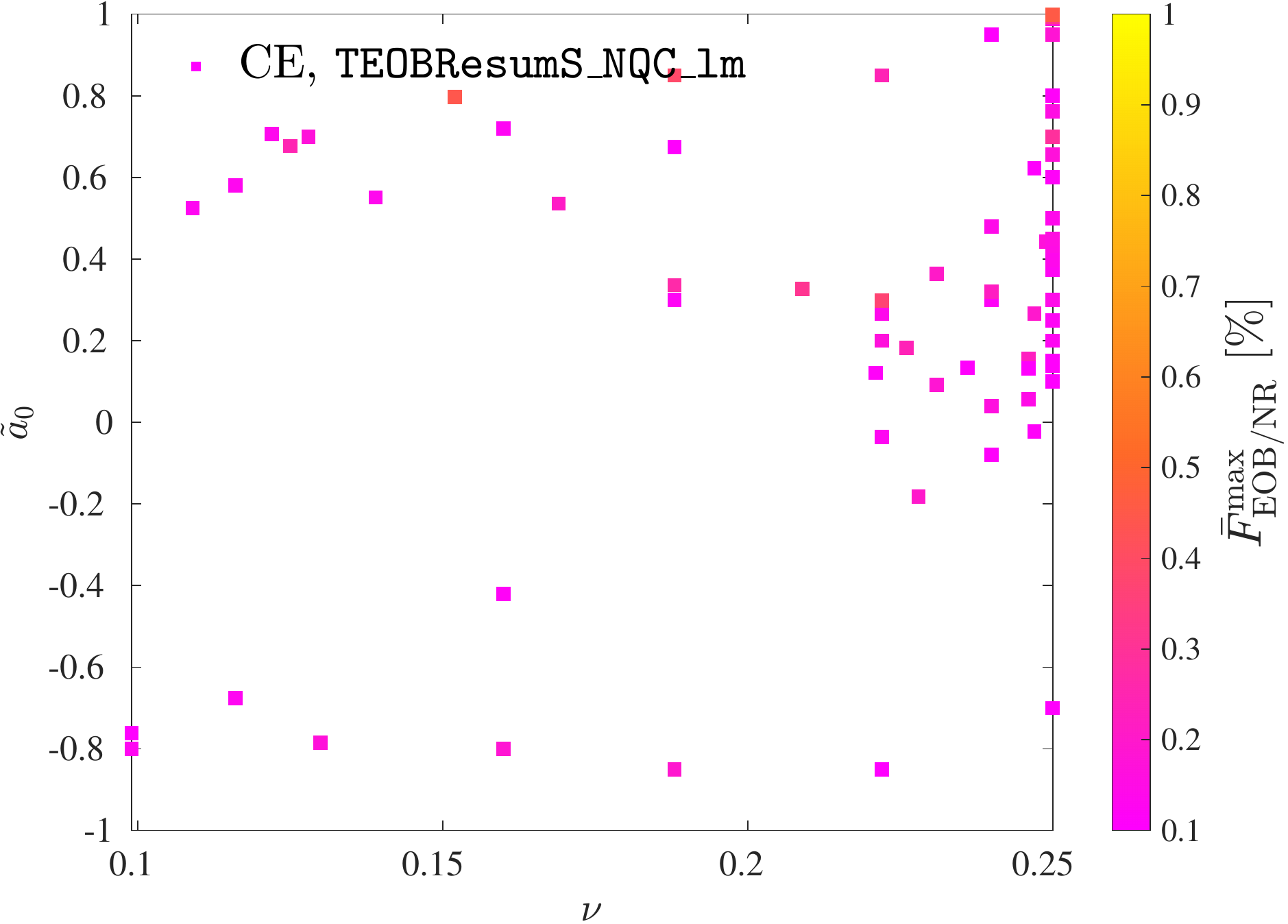} 
\caption{Distribution over the parameter space $(\nu, \tilde{a}_0)$ of those configurations whose  
$\bar{F}^{\rm max}_{\rm EOB/NR}$ exceeds $10^{-3}$, for aLIGO (first row), ET-D (second row), CE (third row),
both for \TEOBResumS{} (left column) and its updated version (right column). Notably,
the changes implemented in \TEOBResumSlm{} lower the maximum unfaithfulness,
although we see that higher values of the spin remain the most challenging ones to be modeled,
along with the comparable-mass case.}
\label{fig:outliers}
\end{center}
\end{figure}

A global view of the EOB/NR agreement is given by the computation of
the EOB/NR unfaithfulness as a function of the total mass of the system. 
As done for the time-domain phasing, for the EOB spin values we take the initial 
$(\chi_1, \chi_2)$ given in the metadata file for each simulation with 6 digits precision. 
For simplicity, here we focus only on the $\ell=m=2$ mode.  
Considering two waveforms $(h_1,h_2)$ with same fixed mass ratio and spins,  
the unfaithfulness is a function of the total mass 
$M$ of the binary and is defined as
\be
\label{eq:barF}
\bar{F}(M) \equiv 1-F=1 -\max_{t_0,\phi_0}\dfrac{\langle h_1,h_2\rangle}{||h_1||||h_2||},
\ee
where $(t_0,\phi_0)$ are the initial time and phase, $||h||\equiv \sqrt{\langle h,h\rangle}$,
and the inner product between two waveforms is defined as 
$\langle h_1,h_2\rangle\equiv 4\Re \int_{f_{\rm min}^{\rm NR}(M)}^\infty \tilde{h}_1(f)\tilde{h}_2^*(f)/S_n(f)\, df$,
where $\tilde{h}(f)$ denotes the Fourier transform of $h(t)$, $S_n(f)$
is the detector's power spectral density (PSD) and $f_{\rm min}^{\rm NR}(M)=\hat{f}^{\rm NR}_{\rm min}/M$ 
is the initial frequency of the NR waveform. In practice, the integral is done up to a maximal NR frequency
$f_{\rm max}^{\rm NR}$ that is chosen as the frequency where the amplitude of $\tilde{h}_{\rm NR}$ is $10^{-3}$.
Waveforms are tapered in the time-domain at the beginning of the inspiral 
so as to reduce the presence of high-frequency oscillations in the corresponding Fourier transforms.
As a step forward to previous work, we here consider for this calculation 
not only the standard zero-detuned, high-power noise spectral density of 
Advanced LIGO~\cite{aLIGODesign_PSD}, but also the anticipated PSD
of Einstein Telescope, considering its latest sensitivity model ET-D~\cite{Hild:2010id}, 
and of Cosmic Explorer~\cite{Evans:2021gyd}. The corresponding PSDs are shown in Fig.~\ref{fig:noises},
together with the less recent ET-C version of the PSD of Einstein Telescope~\cite{Hild:2009ns}. As a complementary analysis
we perform the unfaithfulness computation for this PSD in Appendix~\ref{sec:ET-C}.

The outcome of the $\bar{F}(M)$ computation is shown in Fig.~\ref{fig:barF}, where 
we used Eq.~\eqref{eq:barF}  with $h_1=h_{\rm EOB}$ and $h_2=h_{\rm NR}$.
For each detector choice, the top panels of the figure displays the results 
obtained with \TEOBResumS{}, while the bottom ones those pertaining to \TEOBResumSlm{}. 
For what concerns the aLIGO PSD, the first column of Fig.~\ref{fig:barF} highlights
that $\bar{F}^{\rm max}_{\rm EOB/NR}$ comfortably stays well below the $10^{-2}$ threshold,
all over the parameter space. More precisely, one finds that for \TEOBResumSlm{} 
the datasets in the range $10^{-3} < \bar{F}^{\rm max}_{\rm EOB/NR} < 10^{-2}$ are 
18.4\% (see Table~\ref{tab:maxFbar}), out of which 1.7\% have a maximum $ \bar{F}_{\rm EOB/NR}$ 
value above $3\times 10^{-3}$, where the latter percentage value is lower than the one related to \TEOBResumS{}.
The largest unfaithfulness values obtained with \TEOBResumSlm{}, $\bar{F}^{\rm max}_{\rm EOB/NR}=(0.47, 0.49)\%$,  
correspond respectively to the extremely spinning configuration SXS:BBH:1124 with $(1,+0.998,+0.998)$
and to the configuration SXS:BBH:1434 with $(4.367, +0.798, +0.795)$. 
In general, as deducible from Fig.~\ref{fig:outliers}, the largest values of $\bar{F}^{\rm max}_{\rm EOB/NR}$ 
are obtained for the datasets with individual spins large and positive, i.e. in a regime where we a priori 
expect the largest uncertainties in both the NR waveforms and in the model.
Our result already represents nonnegligible quantitative progress with respect to 
Refs.~\cite{Nagar:2020pcj,Riemenschneider:2021ppj}. Still, there exists room 
for improvement, since the NR error is estimated between $10^{-6}$ and $10^{-4}$, 
as shown in the right panel of Fig.~2 of Ref.~\cite{Nagar:2020pcj}.

For what concerns ET-D, the second column of Fig.~\ref{fig:barF} and Table~\ref{tab:maxFbar}
highlight that $\bar{F}^{\rm max}_{\rm EOB/NR}$ mostly stays below $10^{-3}$.
For \TEOBResumSlm{}, there are only 11 configurations with $\bar{F}^{\rm max}_{\rm EOB/NR} > 3 \times 10^{-3}$,
and again the highest values correspond to SXS:BBH:1124 and SXS:BBH:1434.
Moreover, 79.9\% of the total number of mismatches for \TEOBResumSlm{}
are in the range  $10^{-4} < \bar{F}_{\rm EOB/NR} < 10^{-3}$
and 3.9\% of the total mismatches are below $10^{-4}$ (see Table~\ref{tab:maxFbar}).

Finally, regarding CE, only 1.3\% of the configurations have $\bar{F}^{\rm max}_{\rm EOB/NR} > 3 \cdot 10^{-3}$,
and the percentage of those below $10^{-3}$ reaches 84.1\%. 
It is quite remarkable that for this detector $6.4\%$ of the total mismatches using \TEOBResumSlm{} are below $10^{-4}$.

 \begin{table*}[t]
   \caption{\label{tab:maxFbar} Quantifying the EOB/NR agreement. The central columns of the table contain the fraction of datasets 
   whose maximum unfaithfulness $\bar{F}^{\rm max}_{\rm EOB/NR}$ is within the indicated limits for either  \TEOBResumS{}
   or \TEOBResumSlm{}. The last two columns display percentage numbers out of \textit{all} the mismatch values. These are found 
   independently of the single simulations, by considering how many points pertaining to the curves of Fig.~\ref{fig:barF} 
   fall into a certain range of $\bar{F}$. The range in $M$ is $2.5M_{\odot}$. 
   }
   \begin{center}
 \begin{ruledtabular}
   \begin{tabular}{l l | c c c | c c}
      & & $\bar{F}^{\rm max} < 10^{-3}$ & $10^{-3} < \bar{F}^{\rm max} < 10^{-2}$ & $\bar{F}^{\rm max}> 3\times 10^{-3}$ 
      & $10^{-4} < \bar{F}< 10^{-3}$ & $\bar{F} < 10^{-4}$\\
     \hline
     aLIGO &\TEOBResumS{} & 83.1\% & 16.9\% & 2.1\% & 83.9\% & 3.1\% \\
      &\TEOBResumSlm{} & 82.0\% & 18.4\% & 1.7\% & 81.5\% & 3.8\% \\
     \hline
     ET-D &\TEOBResumS{} & 83.5\% & 15.9\% & 2.6\% & 82.9\% & 3.2\% \\
      &\TEOBResumSlm{} & 81.5\% & 18.5\% & 2.1\% & 79.9\% & 3.9\% \\
     \hline
     CE &\TEOBResumS{} & 85.6\% & 14.8\% & 1.7\% & 84.7\% & 5.2\%\\
     &\TEOBResumSlm{} & 84.1\% & 16.7\% & 1.3\% & 82.8\% & 6.4\%\\
 \end{tabular}
 \end{ruledtabular}
 \end{center}
 \end{table*}  

Concerning the two configurations displayed in Fig.~\ref{fig:phasings}, the lowered phase difference at merger for \TEOBResumSlm{}
 reflects in a slightly lower value of $\bar{F}^{\rm max}_{\rm EOB/NR}$. Namely, for the dataset SXS:BBH:1463, the [\%] unfaithfulness
 switches from $(0.1437, 0.1736, 0.1323) $ respectively for aLIGO, ET-D and CE to $(0.1434, 0.1703, 0.1323)$, while
 for the dataset SXS:BBH:1426, the values lower from $(0.1671, 0.1985, 0.1546)$ to $(0.0675, 0.0731, 0.0613)$. 
 

Figures~\ref{fig:barF},~\ref{fig:maxF} and~\ref{fig:outliers} represent, to our knowledge, 
the first systematic assessment of the quality of a state-of-the-art waveform model in view 
of the 3G detector effort~\cite{Reitze:2021gzo, Couvares:2021ajn, Punturo:2021ryo, Katsanevas:2021fzj, 
 Kalogera:2021bya, McClelland:2021wqy}.
Our plots look a bit more optimistic than the conclusions of Ref.~\cite{Purrer:2019jcp}, 
that assessed the quality of the phenomenological waveform model 
{\tt IMRPhenomPv2} for specific configurations, and concluded that the accuracy
of current waveform models needs to be improved by at least three orders of magnitude. 
If this is certainly true of {\tt IMRPhenomPv2}, it doesn't seem to be the case for the spin-aligned model
that we are discussing here, as it already grazes the expected detector calibration
uncertainty, $\sim 10^{-5}$, for masses up to $20M_\odot$. For larger values of $M$, 
where the detector is mostly sensitive to the ringdown,
$\bar{F}_{\rm EOB/NR}$ goes up to $10^{-3}$ for
several configurations. This however should be carefully interpreted,
since it is related to two physical facts: (i) on the one hand,
the quality of the late part of the NR ringdown might be more or less noisy depending on 
the configuration, thus affecting the unfaithfulness calculation; (ii) on the other hand, even 
if there was no relevant numerical noise, there are differences between the EOB modeled 
ringdown and the actual one. In particular, the absence of mode mixing between positive 
and negative frequency QNMs (a phenomenon that is present especially for spins 
anti-aligned with the angular momentum) can play a role in this context.
In addition, one should also be aware of the fact that the NR-informed postmerger was constructed
using SXS data extrapolated with $N=2$~\cite{Nagar:2020pcj}, since this reduces the amount
of NR noise during this specific part of the waveform. However, the EOB/NR comparison is done
using $(N=3)$-extrapolated waveform data, that gives a good compromise between the inspiral
and the merger-ringdown part of the signal. This means that the differences that we 
see in Fig.~\ref{fig:barF} for large masses are {\it partly} coming from the NR simulations and not
from the model. We thus expect that our EOB/NR comparisons will benefit of improved NR
simulations that use Cauchy Characteristic Extraction~\cite{Moxon:2021gbv, Fischer:2021qbh, Zertuche:2021xkb}.

On a more general ground, a precise assessment of the accuracy of the current version(s) 
of \TEOBResumS{} for ET will require dedicated injection/recovery campaigns.  Nonetheless 
our analysis seems  to indicate that both versions of \TEOBResumS{}, either the standard or
the NQC-improved one, already offer a reliable starting point to investigate PE having in 
mind 3G detectors. To obtain such result it was crucial to improve the self-consistency of the
model and to provide a new analytical representation of the $c_3$ function carefully selecting
a new sample of useful NR datasets.

\section{Contrasting \TEOBResumS{} and \SEOB{} waveform models}
\label{sec:seob}
Now that we have explored the performance of \TEOBResumS{} under a different point of 
view and shown how to improve it further, let us shift to compare it with 
\SEOB{}~\cite{Bohe:2016gbl,Cotesta:2020qhw, Ossokine:2020kjp}. This model is another state-of-the-art
EOB model informed by NR simulations and differs from  \TEOBResumS{} for several
structural choices, that involve the structure of the Hamiltonian, the gauge, the analytic 
content and the resummation strategies. A comprehensive analysis of what distinguishes the 
Hamiltonians of \TEOBResumS{} and of \SEOB{} is presented in Ref.~\cite{Rettegno:2019tzh}. 
The {\tt SEOBNRv4}
model was presented in 2016 and never structurally updated since, except for the addition
of higher modes~\cite{Cotesta:2020qhw}, without any change to the dynamics, and 
precession~\cite{Ossokine:2020kjp}.
The purpose of this section is to discuss more specific comparisons between the two models, 
especially focusing on frequencies and angular momentum fluxes.
Moreover, even if \TEOBResumS{} has been publicly available for many 
years~\cite{Nagar:2018zoe}, direct comparisons involving both EOB models and the full 
NR catalog do not seem to exist in the literature. Note however that \SEOB{} was compared 
to the most recent generation of phenomenological models (see in particular Fig.17 of Ref.~\cite{Pratten:2020fqn}).
We aim at filling this gap by providing one-to-one comparisons between
\SEOB{} and \TEOBResumS{} that involve the important observables discussed so 
far: (i) angular momentum fluxes; (ii) waveform amplitude and frequency and the consistency
of this latter with the dynamics; (iii) EOB/NR unfaithfulness computations taking into account also 3G detectors.
In this section we will use {\it only} the standard version of \TEOBResumS{}.
In addition, for the unfaithfulness calculation we will use the publicly available $C$ 
implementation\footnote{The same code is going to be released also via {\tt LALSimulation}.}, 
that employs  fits for the $\ell=m=2$ NQC parameters entering the flux as well 
as the (iterated) post-adiabatic approximation~\cite{Nagar:2018gnk} to efficiently describe the inspiral, 
as detailed in Ref.~\cite{Riemenschneider:2021ppj}.

\subsection{Angular momentum fluxes}
\begin{figure}[t]
\center
\includegraphics[width=0.23\textwidth]{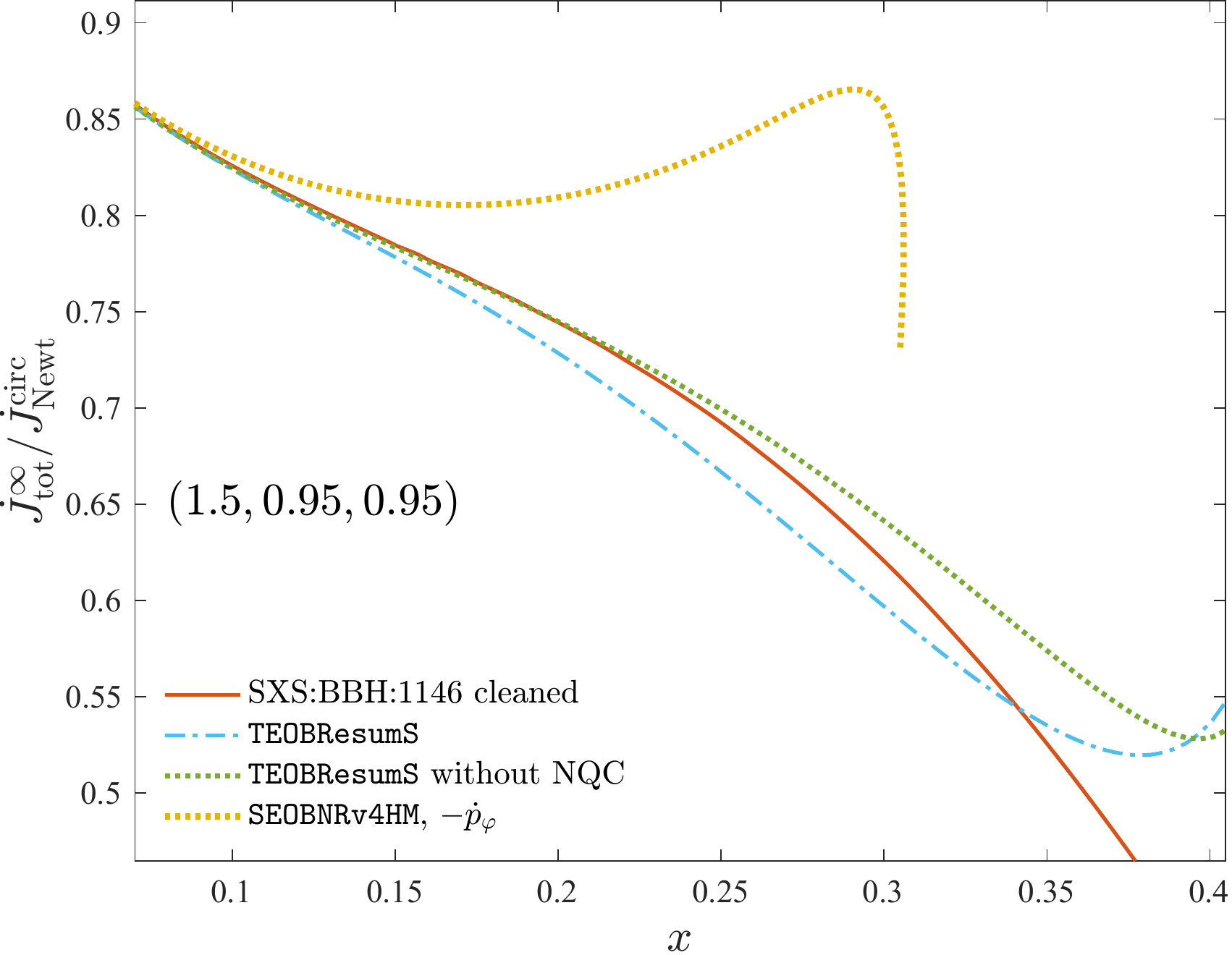} 
\includegraphics[width=0.23\textwidth]{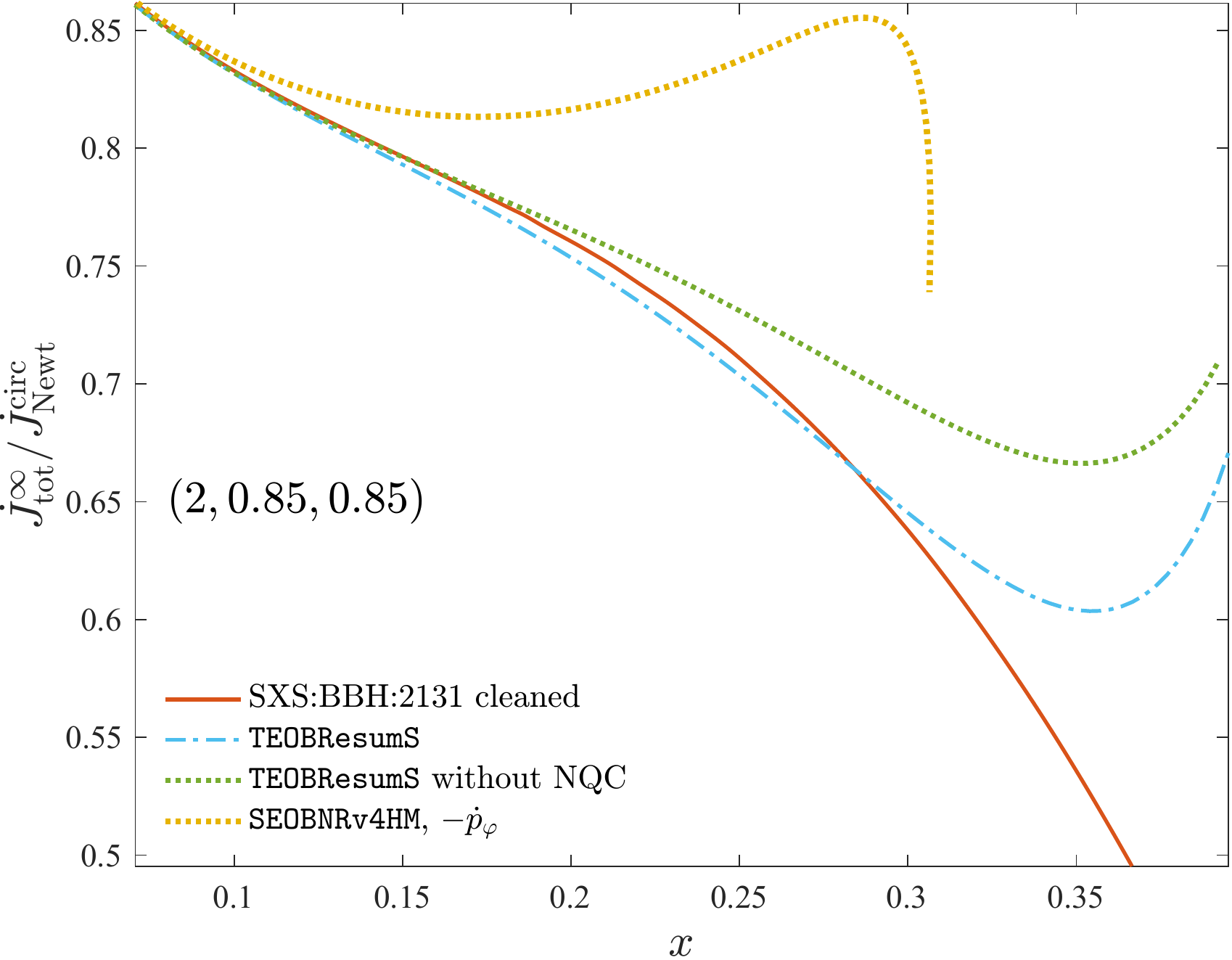} \\
\vspace{1mm}
\includegraphics[width=0.23\textwidth]{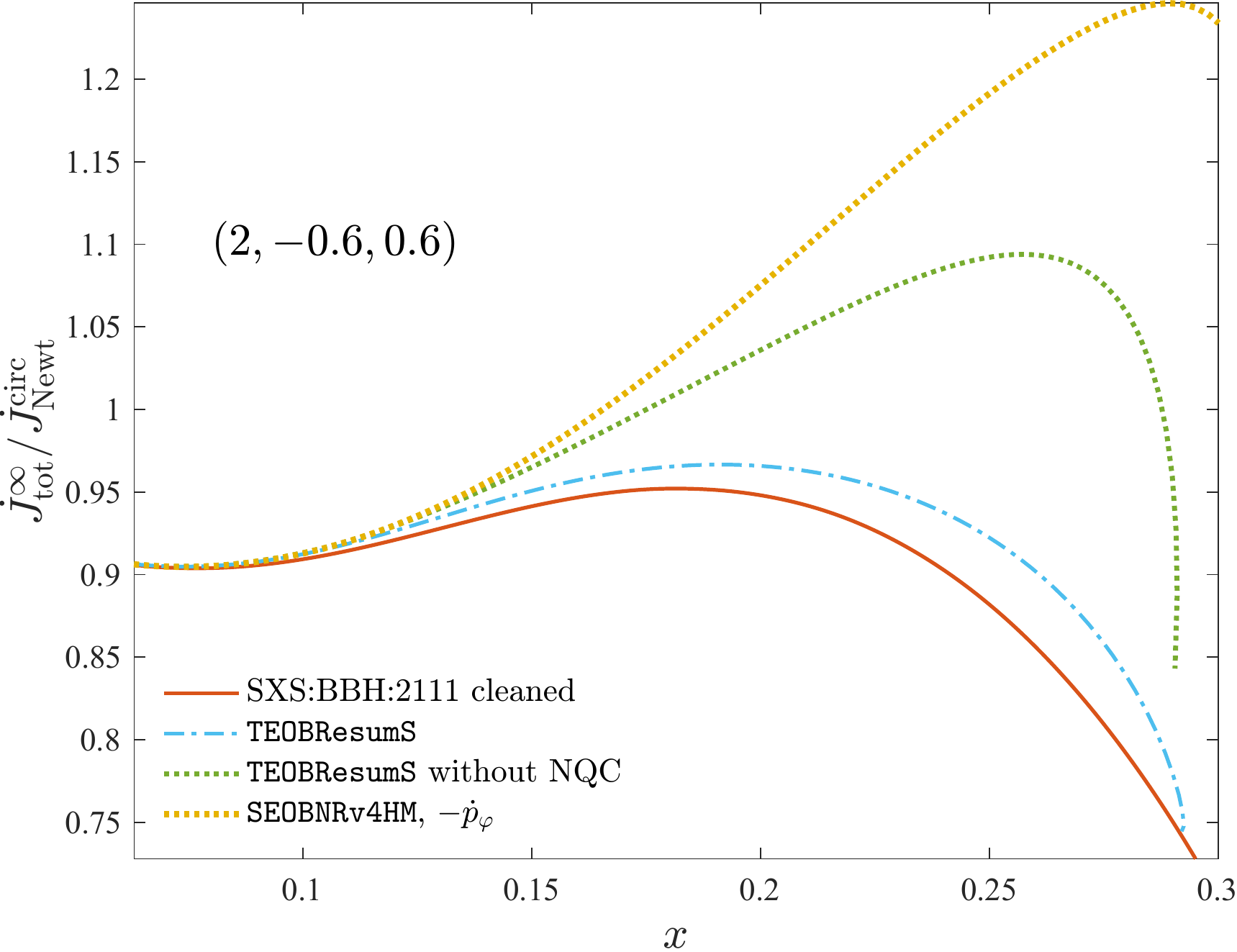} 
\includegraphics[width=0.23\textwidth]{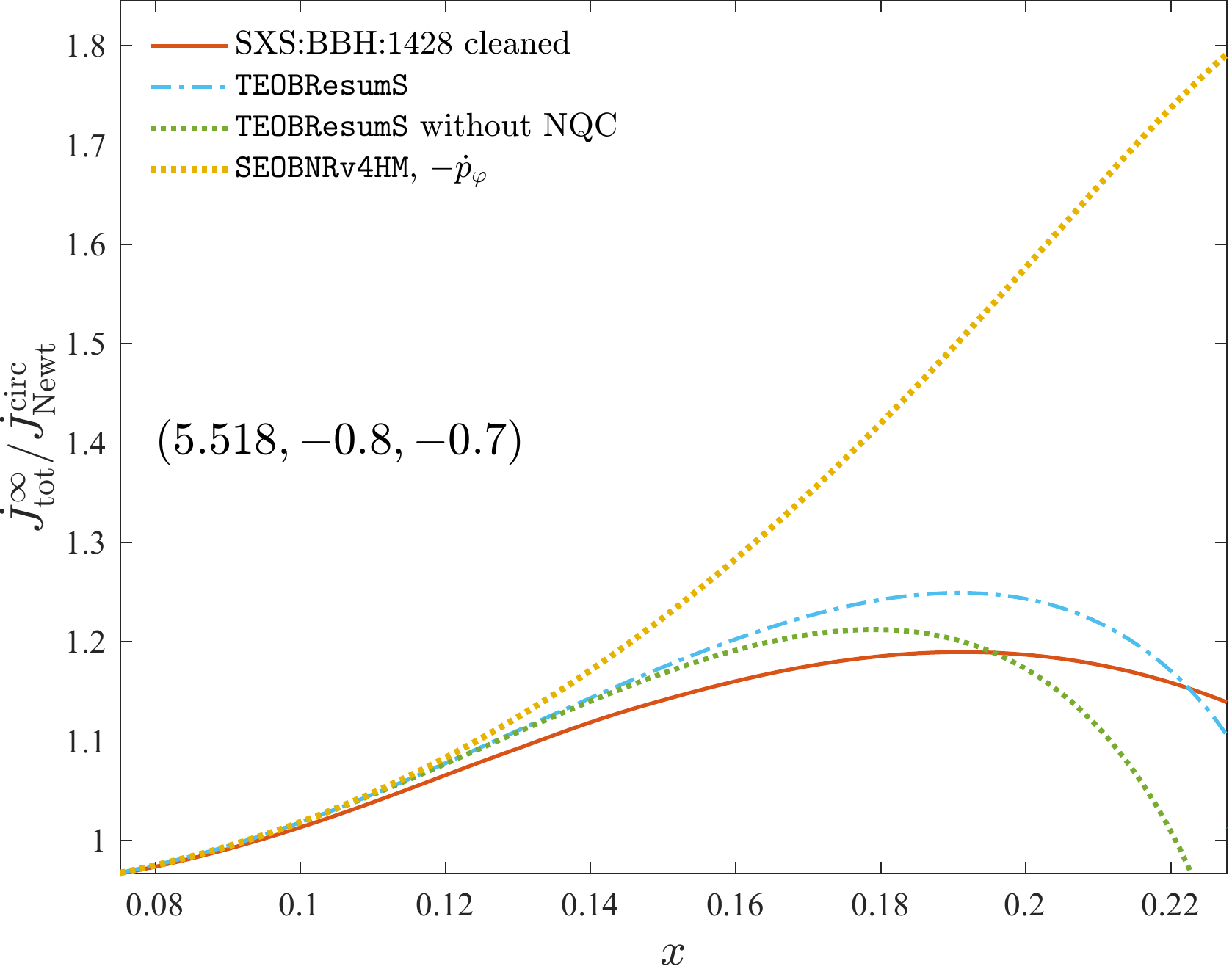} 
\caption{\label{fig:fluxes_seob_comparison} For the configurations corresponding to simulations SXS:BBH:1146,
SXS:BBH:2131, SXS:BBH:2111, SXS:BBH:1428 we show several angular momentum fluxes: 
(i) the NR one (orange),
(ii) the \TEOBResumS{} one (dash-dotted light blue), (iii) the  \TEOBResumS{} one 
without the NQC correction in the $\ell=m=2$ mode (dash-dotted green),
(iii) the corresponding flux from \SEOB{} computed as $-\dot{p}_\ph$ (dotted yellow). 
}
\end{figure}

Let us firstly discuss the flux of angular momentum. To begin with, one has to be aware
that -- to the best of our knowledge -- the dynamical phase-space variables are not among the standard outputs 
of the \SEOB{} implementation within {\tt LALSimulation}, so that some modifications of the code are needed\footnote{By contrast, 
let us remind that the standalone \TEOBResumS{} $C$ code can optionally output several dynamical quantities.}.
This was done and explicitly described already in Ref.~\cite{Nagar:2019wds}.
The simplest way to compute the angular momentum flux for \SEOB{} is by 
taking the time derivative of the angular momentum $p_\varphi$, 
i.e. using the relation
\be
\dot{J}_{\tt SEOB}= -\dot{p}_\varphi^{\tt SEOB} = -\hat{\cal F}_{\varphi}^{\tt SEOB}.
\ee
Figure \ref{fig:fluxes_seob_comparison} displays the related fluxes for the configurations
$(1.5, 0.95, 0.95)$, $(2, 0.85, 0.85)$, $(2, -0.6, 0.6)$ and $(5.52, -0.8, -0.7)$,
corresponding to SXS datasets SXS:BBH:1146, SXS:BBH:2131, SXS:BBH:2111 and SXS:BBH:1428.
Each panel compares five curves: (i) the NR flux (red); (ii) the standard \TEOBResumS{} flux; 
(iii) the flux from \TEOBResumS{} without the $\ell=m=2$ NQC corrections; (iv) the \SEOB{} flux.
Let us firstly focus on the two cases with the largest spins, top 
row of Fig.~\ref{fig:fluxes_seob_comparison}: the figure highlights the differences
between the \SEOB{} and NR fluxes. 
We believe this is related to the \SEOB{} dynamics for these
two configurations, as we will further point out in Sec.~\ref{sec:wave_amp_freq} below.
By contrast, the \TEOBResumS{} fluxes look consistent with
the NR one. In particular, the agreement that can
be reached between \TEOBResumS{} and NR {\it without} the NQC correction 
factor  is remarkable. However, this also shows that the NQC implementation should
 be revised for large spins, since it introduces nonnegligible differences already during 
 the inspiral\footnote{As already suggested
 in Ref.~\cite{Chiaramello:2020ehz} it would be better to see the NQC corrections as an
 effective way of improving the EOB analytical waveform only very close to merger,
 and as such they should be progressively switched on only during the plunge.} (see also Appendix~\ref{sec:NQCissues}).

The differences between the \SEOB{} and NR fluxes remain large also in the other
two cases (bottom row of Fig.~\ref{fig:fluxes_seob_comparison}).
Given the many structural differences between the \SEOB{} and \TEOBResumS{} models, 
it is difficult to precisely track what are the elements within \SEOB{} that are responsible of the flux behavior.
The lack of the NQC factor in the \SEOB{} flux is seemingly not enough to explain the
differences that appear in the bottom panels of Fig.~\ref{fig:fluxes_seob_comparison}, 
since the \SEOB{} curve differs even from the NQC-free flux of \TEOBResumS{}. Let us
mention at least two other differences that may be relevant in strong field. First of all,
although the \SEOB{} flux shares the same formal functional form of the \TEOBResumS{}
one, the definition of $r_\omega$ is different (see e.g.~\cite{Cotesta:2018fcv}). In addition,
the PN truncation and the resummation of each waveform multipole, including the quadrupole
one, differs between one model and the other.

\subsection{Waveform amplitude and frequency}
\label{sec:wave_amp_freq}
\begin{figure}[t]
\center
\includegraphics[width=0.23\textwidth]{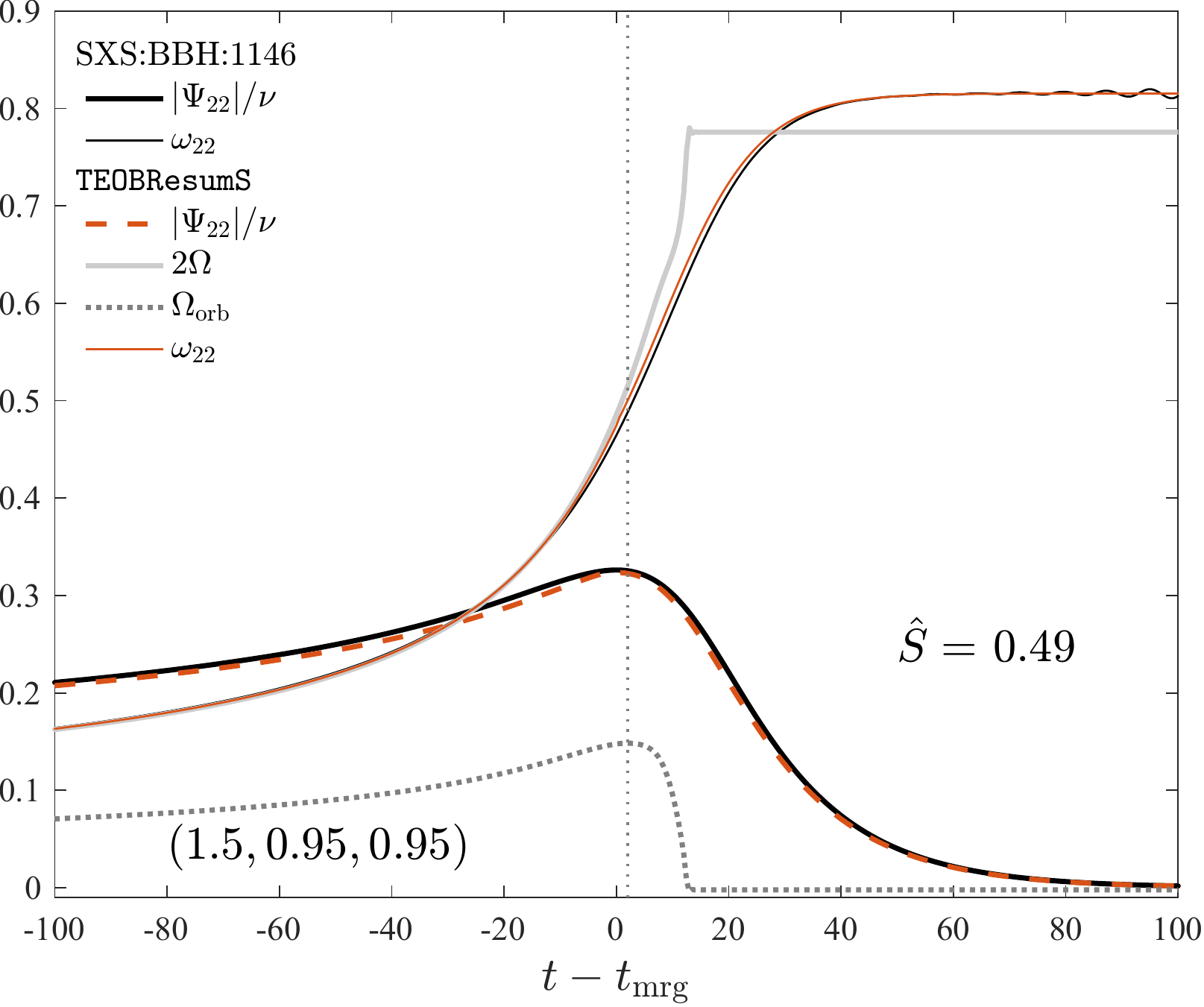} 
\includegraphics[width=0.23\textwidth]{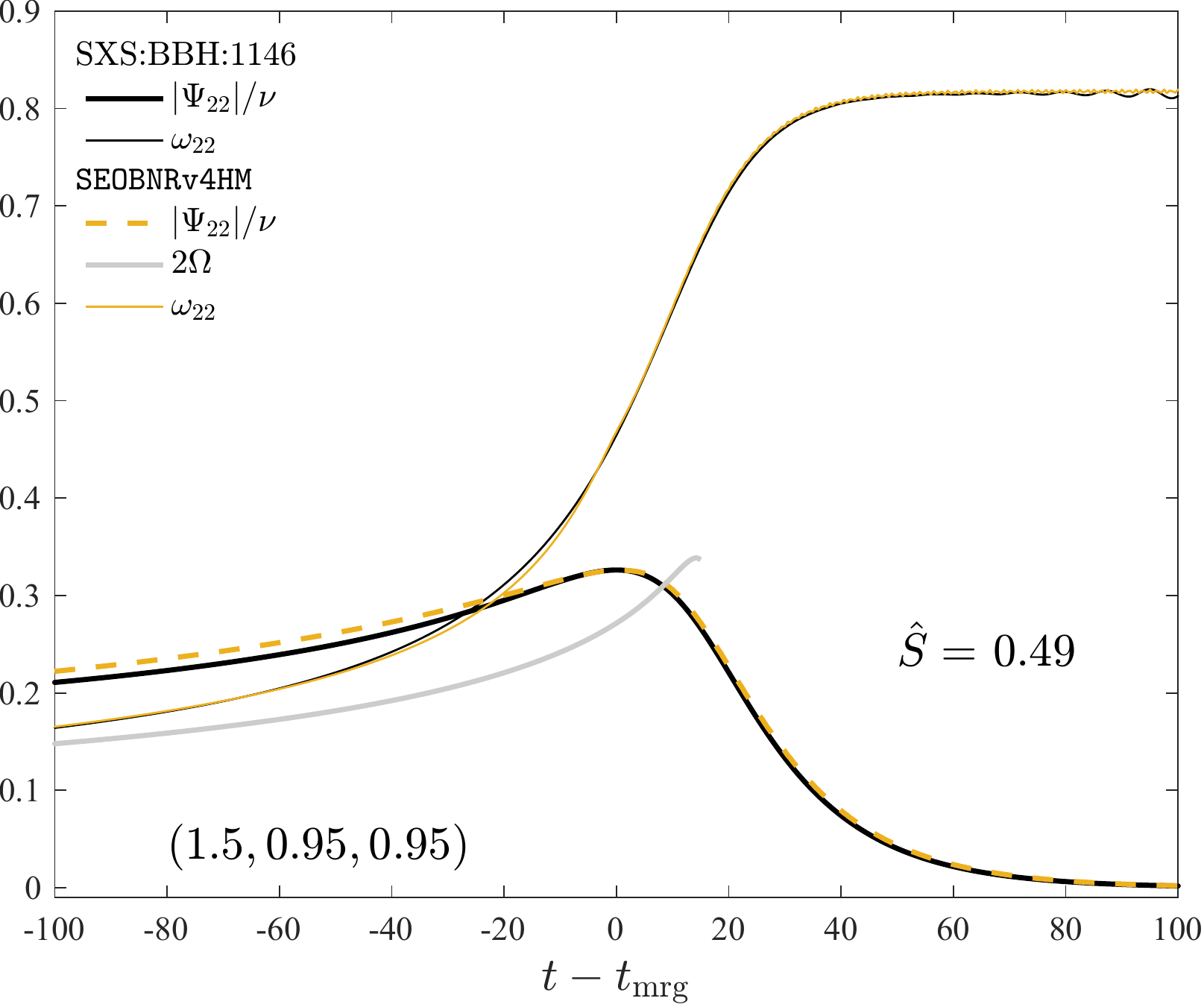} \\
\includegraphics[width=0.23\textwidth]{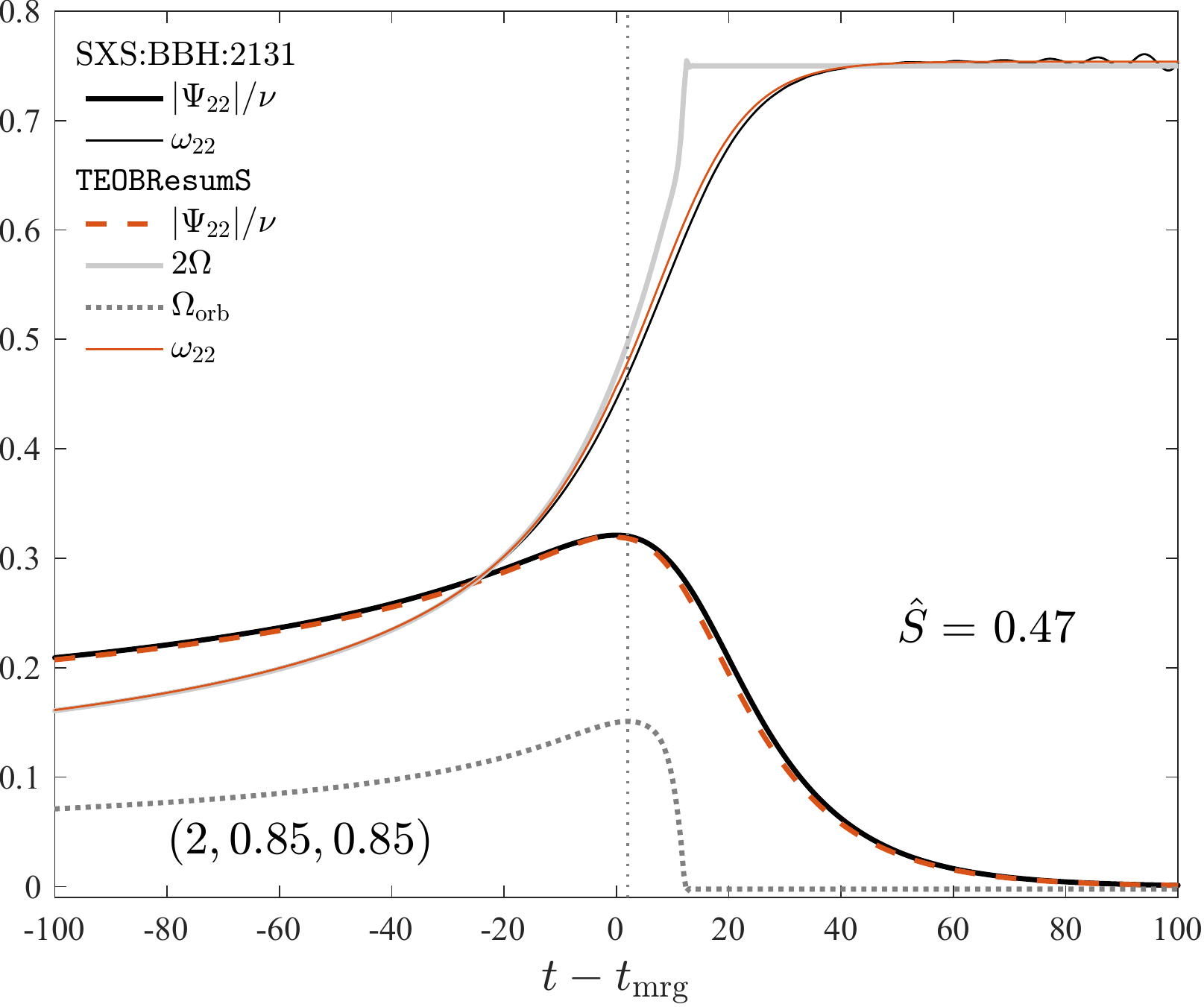} 
\includegraphics[width=0.23\textwidth]{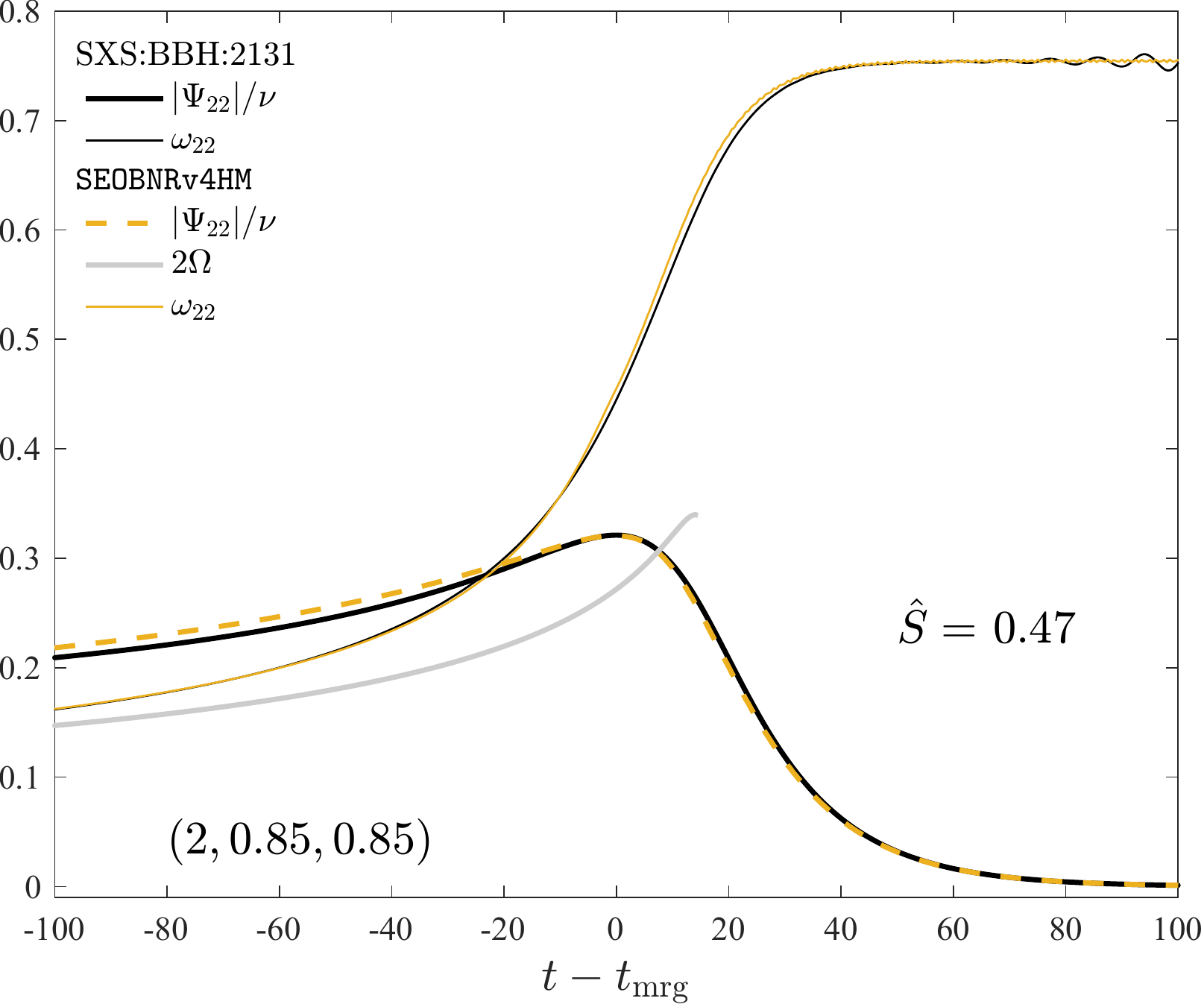} \\
\includegraphics[width=0.23\textwidth]{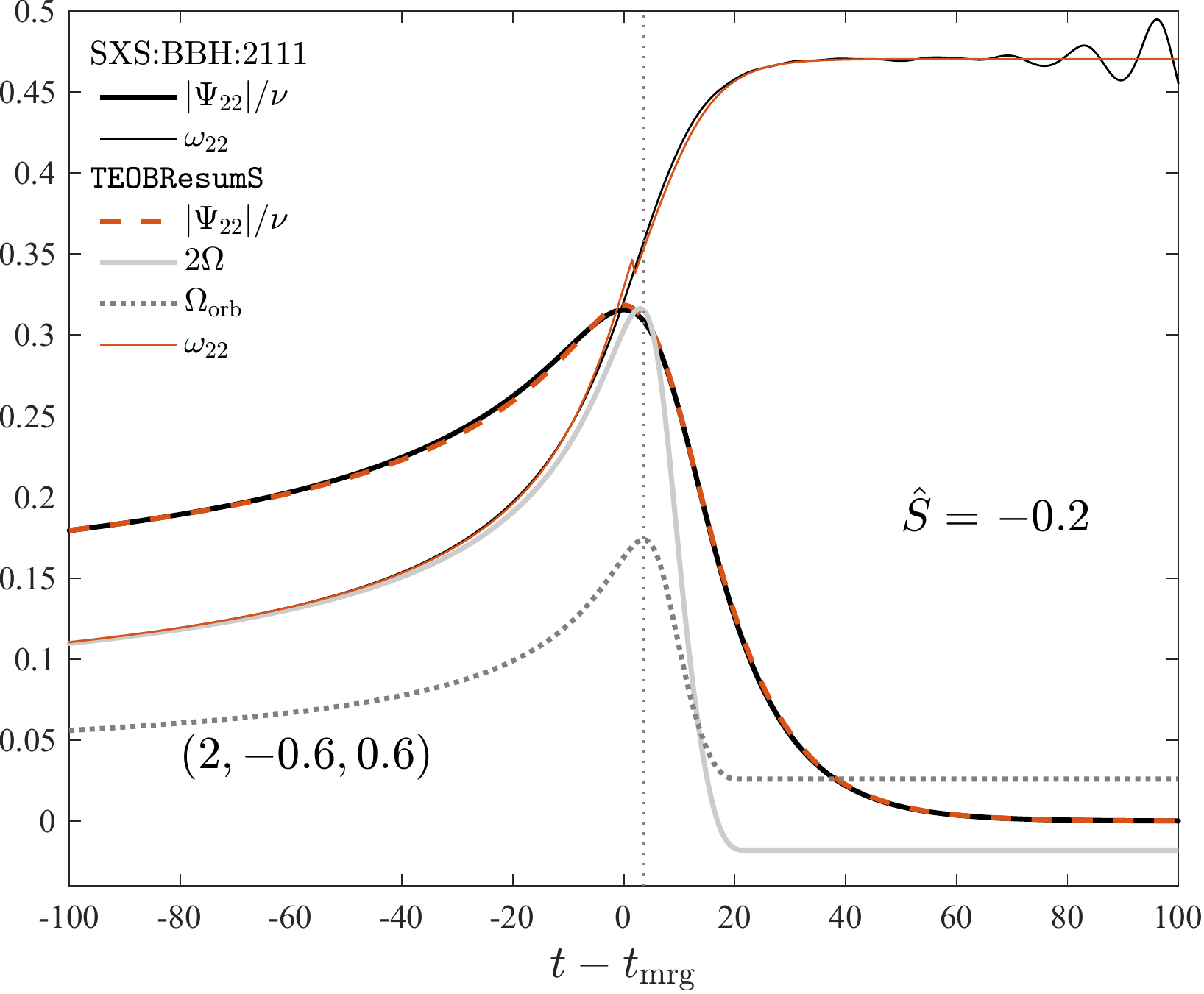} 
\includegraphics[width=0.23\textwidth]{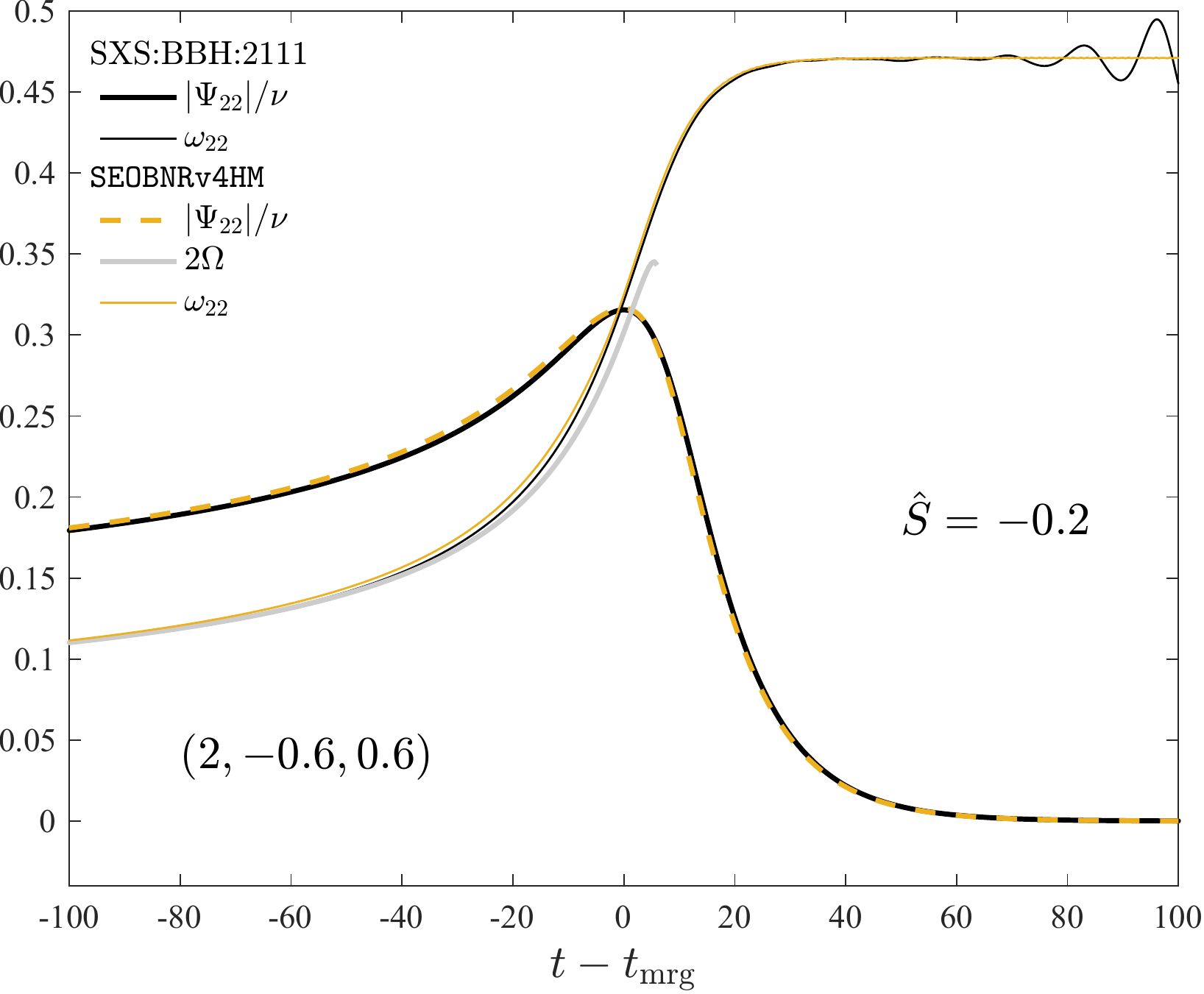} \\
\includegraphics[width=0.23\textwidth]{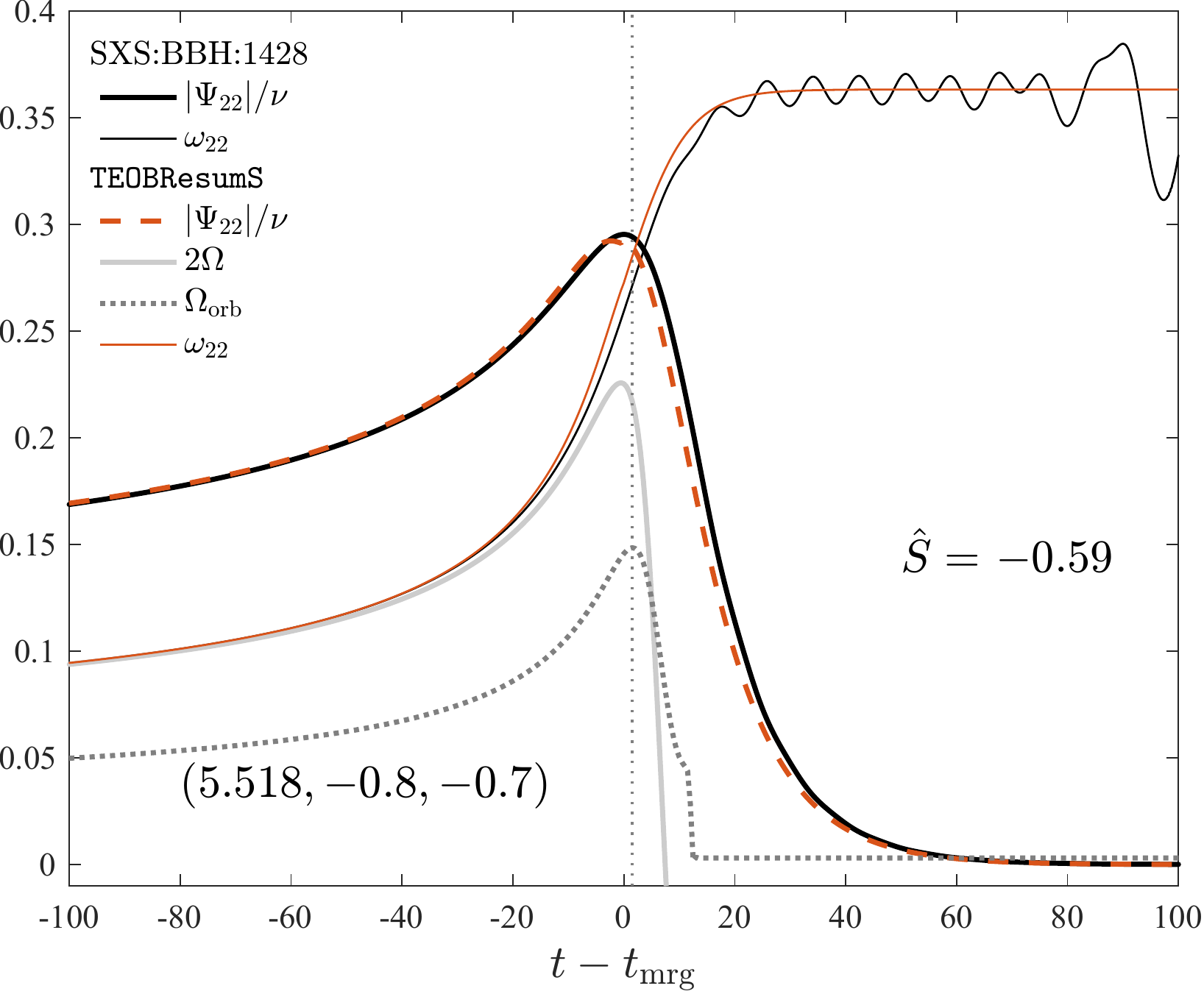}
\includegraphics[width=0.23\textwidth]{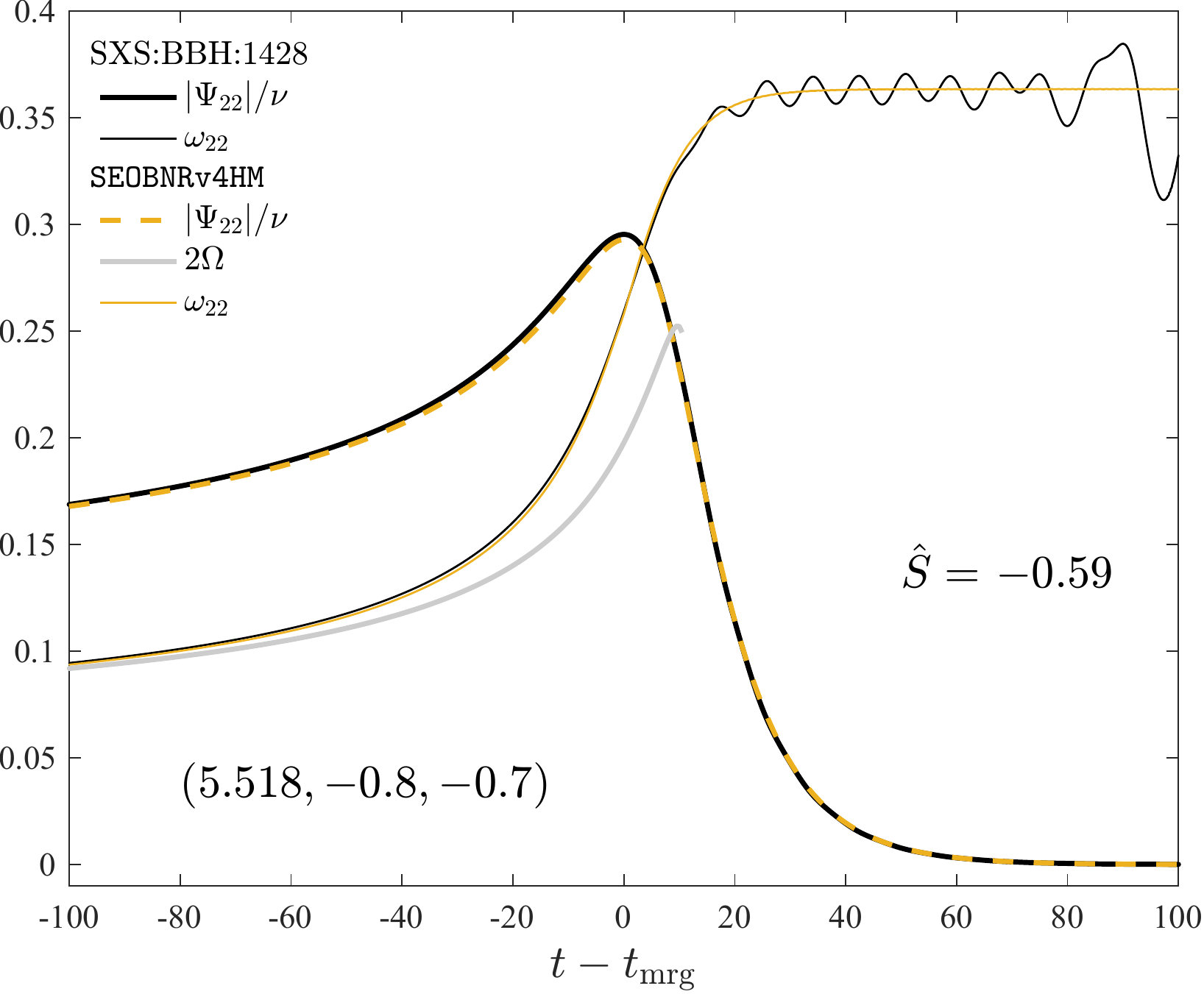} 
\caption{\label{fig:freq_comparison_seob}Contrasting \TEOBResumS{} (left)
 and \SEOB{} (right). For each configuration we show: (i) the waveform amplitude, (ii) the instantaneous gravitational
 wave frequency, (iii) twice the orbital frequency $\Omega$ and (iv) the pure orbital frequency $\Omega_{\rm orb}$ 
 (i.e., without the spin-orbit contribution). Each binary is also labeled by its effective spin $\hat{S}\equiv (S_1+S_2)/M^2$.
 For any configuration \TEOBResumS{} maintains an excellent consistency 
 between (twice) the orbital frequency and the gravitational wave frequency. This is especially true, as a priori expected, 
 in the highly adiabatic cases with large positive spins where NQC corrections have a very limited effect. 
 By contrast, for \SEOB{} this holds only for the configuration $(2, -0.6, 0.6)$. In the other cases, $\omega_{22}\neq 2\Omega$ and the correct 
 behavior of the waveform frequency is guaranteed only by the action of NQC corrections. }
\end{figure}
Let us now provide a direct comparison between \TEOBResumS{}, \SEOB{} and NR waveforms
for the configurations considered above. We focus on the $\ell=m=2$ waveform amplitude and frequency.
Figure~\ref{fig:fluxes_seob_comparison} contrasts the EOB/NR performance 
for \TEOBResumS{} (left panels) and \SEOB{} (right panels). The figure focuses around 
merger time and the waveforms are aligned in the late inspiral, just before merger.
We recall that among the configurations presented in the figure, only the $(2,+0.85,+0.85)$
was used to inform $c_3$ for \TEOBResumS{} and similarly only this was used to calibrate 
the spin sector of \SEOB{}~\cite{Bohe:2016gbl}. Both models deliver an excellent agreement 
with the NR waveform amplitude and frequency.
However, there are relevant differences in the underlying dynamics, as suggested by the
behavior of twice the orbital frequency, $2\Omega$, that is also displayed on the figure.
In particular one sees that while for \TEOBResumS{} $\omega_{22^{\rm EOB}}\simeq 2\Omega$ is always true
up to the merger point, for \SEOB{} this is approximately true only for the $(2,-0.6,+0.6)$ configuration.
For the other cases, the dynamics seems to point to a delayed plunge, but
the NR calibration of the \SEOB{} model manages
to have the analytical waveform on top of the NR one. Let us remember in fact that 
Ref.~\cite{Bohe:2016gbl} also calibrates the time shift between the EOB orbital frequency
and the peak of the EOB waveform where NQC corrections are determined and the ringdown attached. 
This feature is not needed in the \TEOBResumS{} model, that uses as natural anchor point 
to determine NQC corrections the peak of the {\it pure} orbital 
frequency\footnote{In fact, we use $t_{\rm NQC}=t_{\Omega_{\rm orb}}^{\rm peak}-1$, see 
in particular Eqs.~(3.46)-(3.47) of Ref.~\cite{Nagar:2017jdw} 
and Eqs.~(102)-(105) and (108)  of Ref.~\cite{Damour:2014sva}.} $\Omega_{\rm orb}$
(also shown in the figure), a quantity that is obtained subtracting the spin-orbit contribution from the total frequency. 
This structure  is the effective generalization to the comparable-mass case of what is found in the 
test-mass limit~\cite{Damour:2014sva}, where the maximum of $\Omega_{\rm orb}$ is always
very close to the peak of the $\ell=m=2$ waveform amplitude, as we also remind in Fig.~\ref{fig:testmass_freq} below.

\subsubsection{The large-mass-ratio limit}

\begin{figure}[t]
\includegraphics[width=0.23\textwidth]{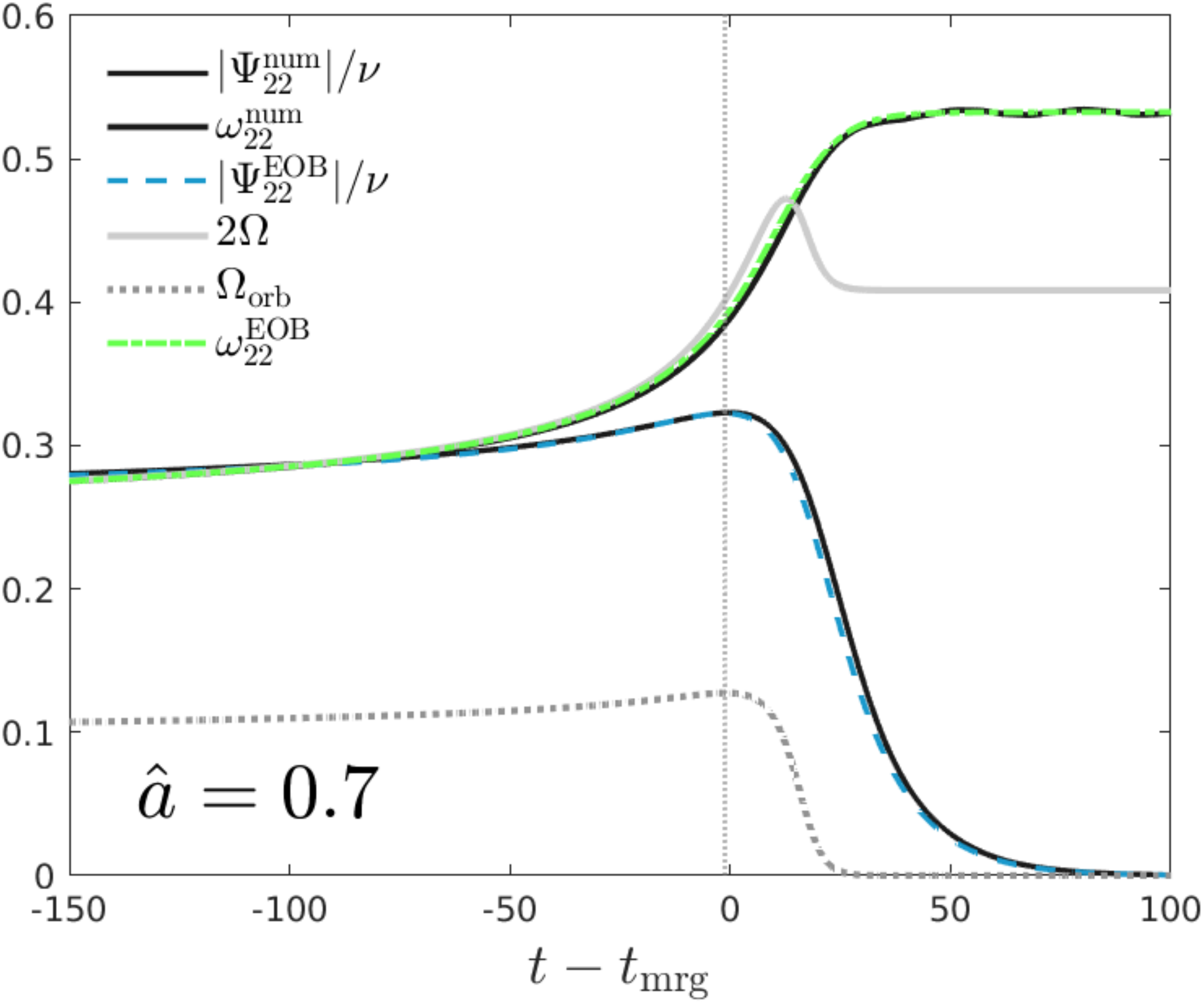}  
\includegraphics[width=0.23\textwidth]{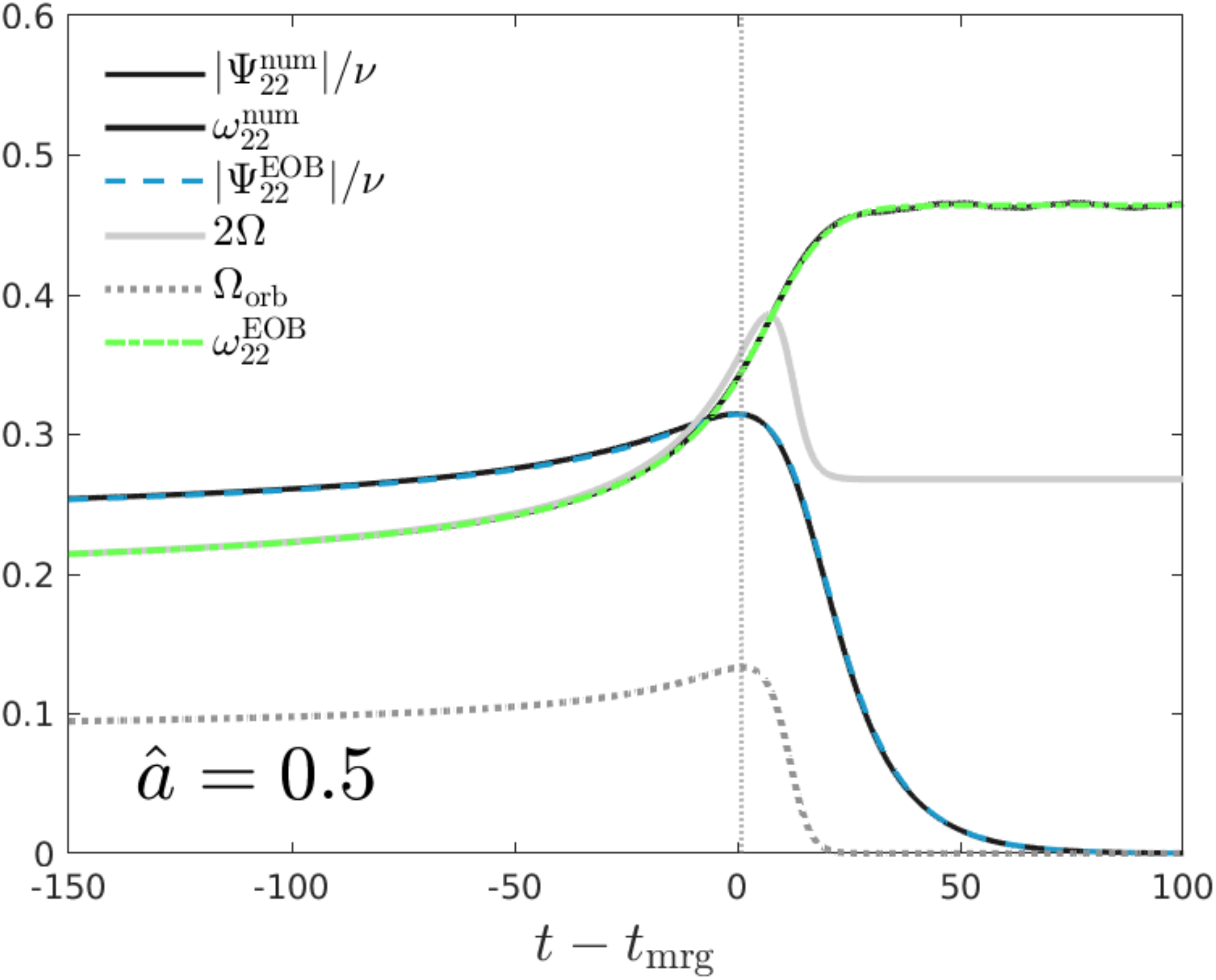} \\
\includegraphics[width=0.23\textwidth]{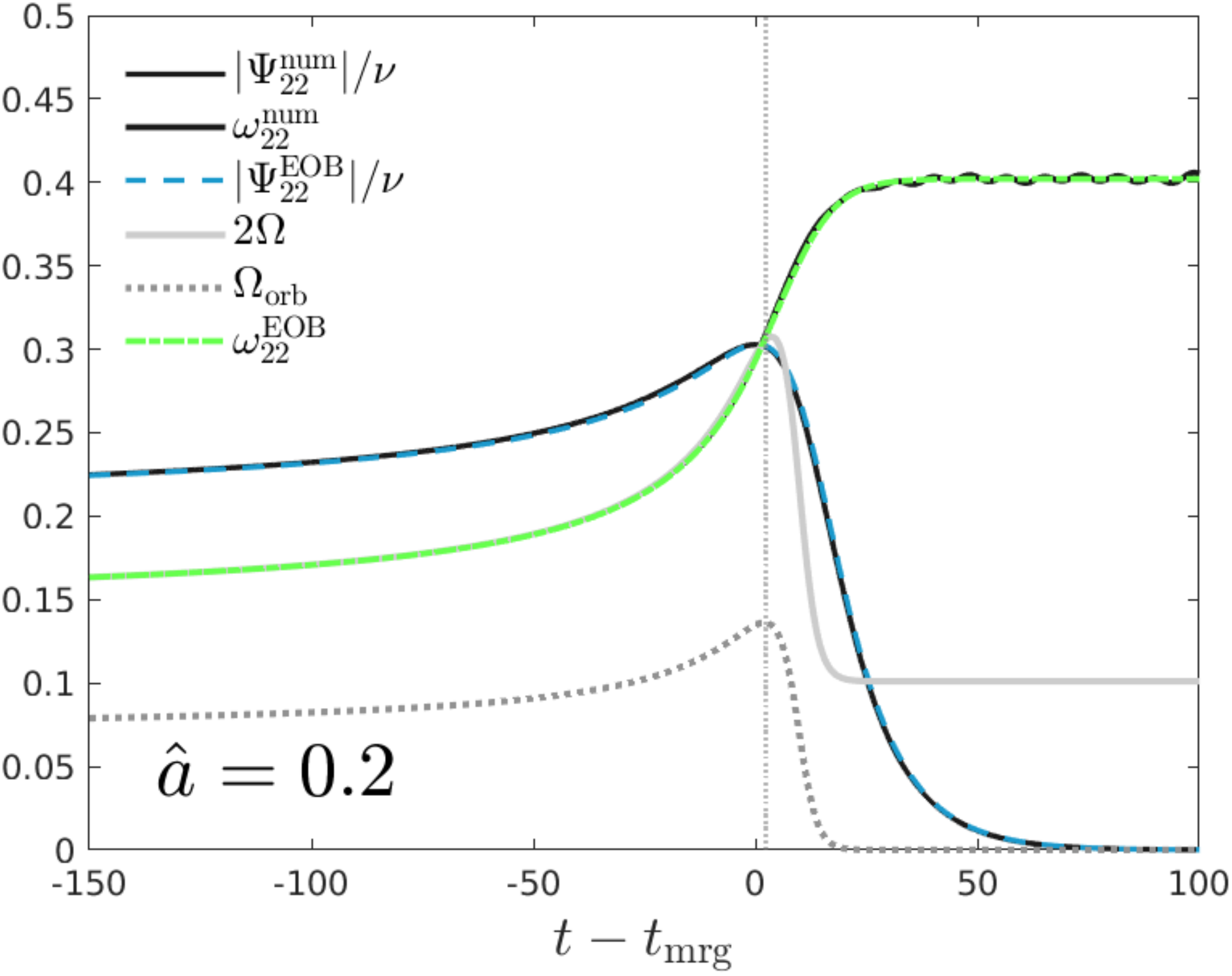} 
\includegraphics[width=0.23\textwidth]{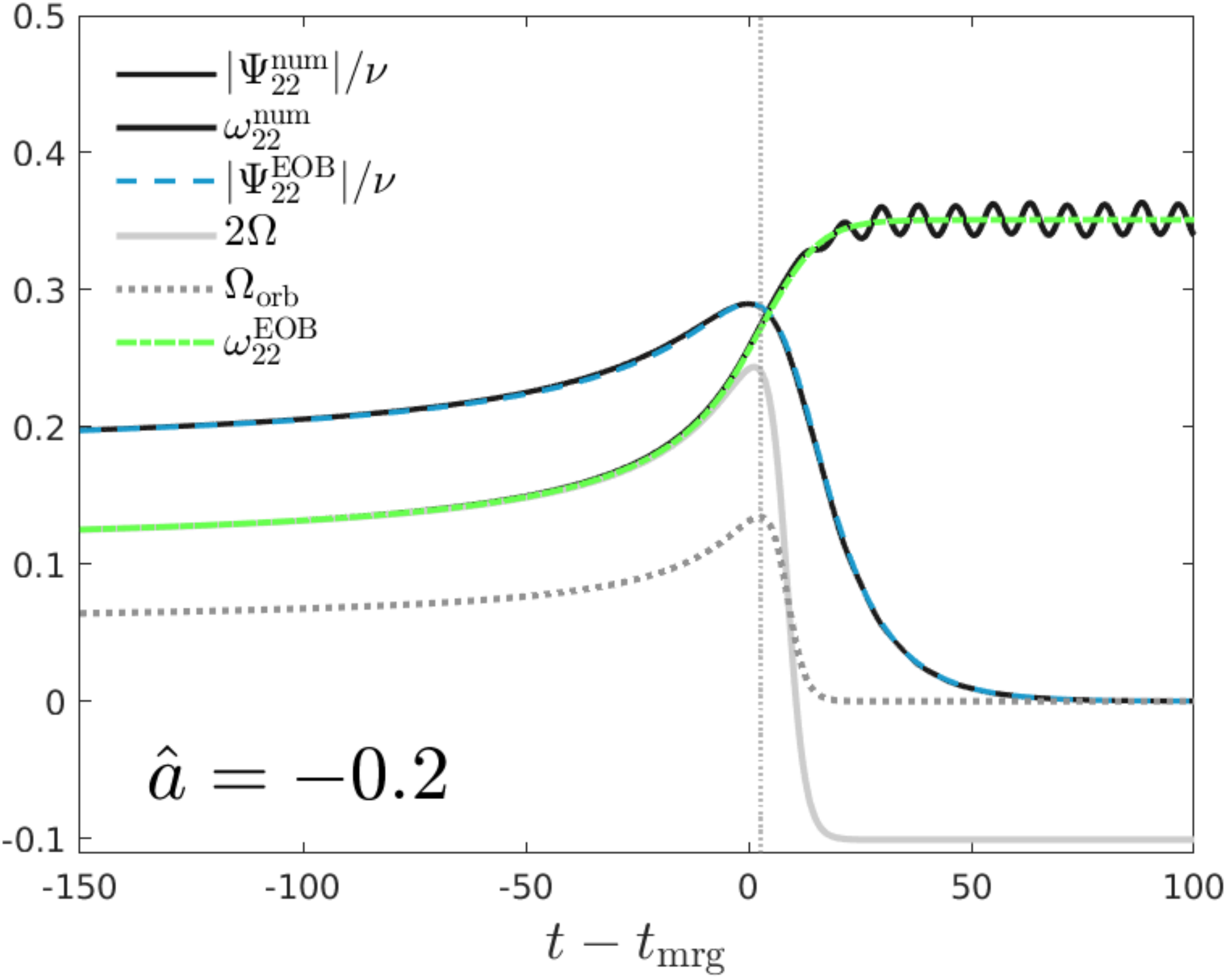}\\ 
\includegraphics[width=0.23\textwidth]{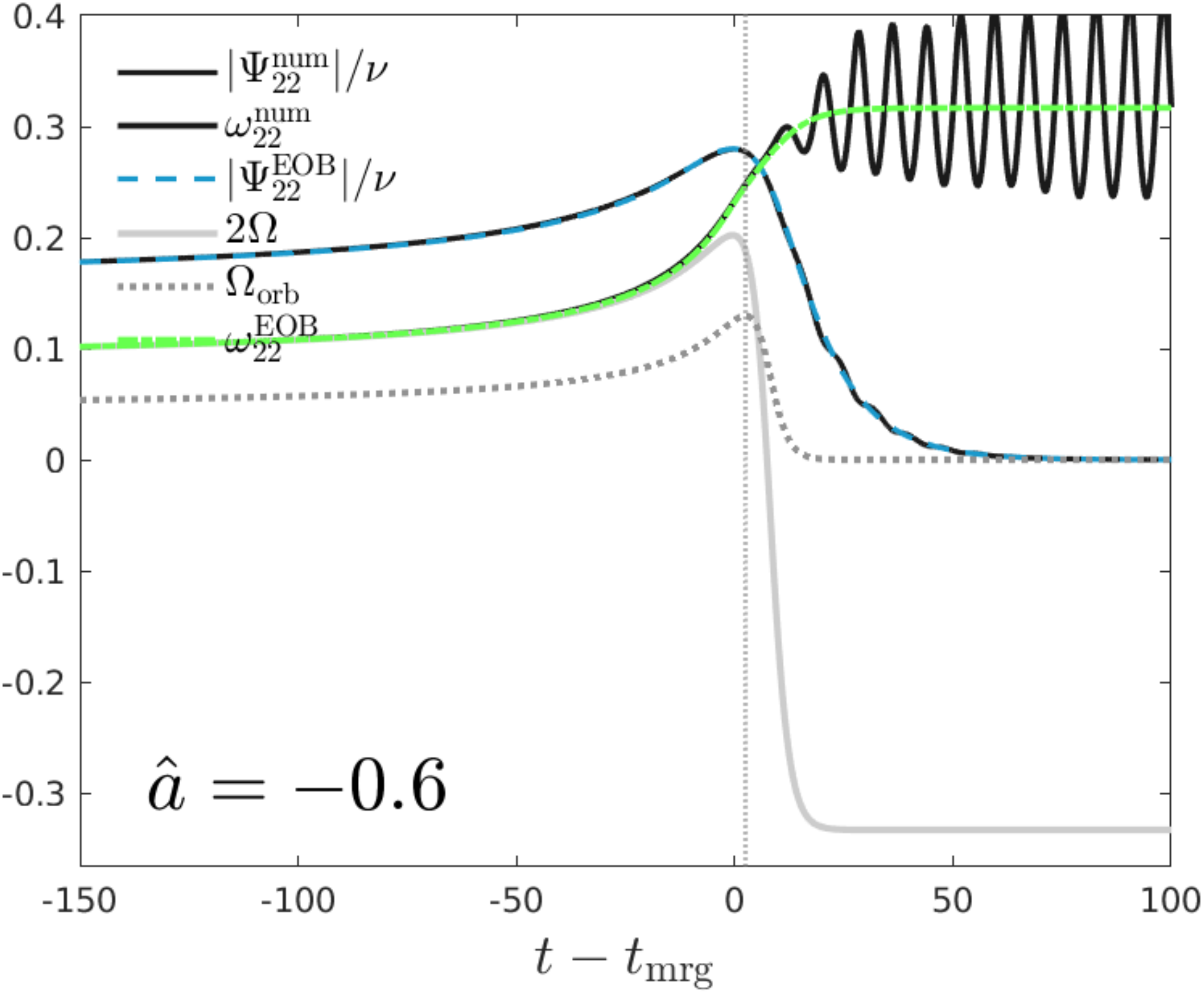} 
\includegraphics[width=0.23\textwidth]{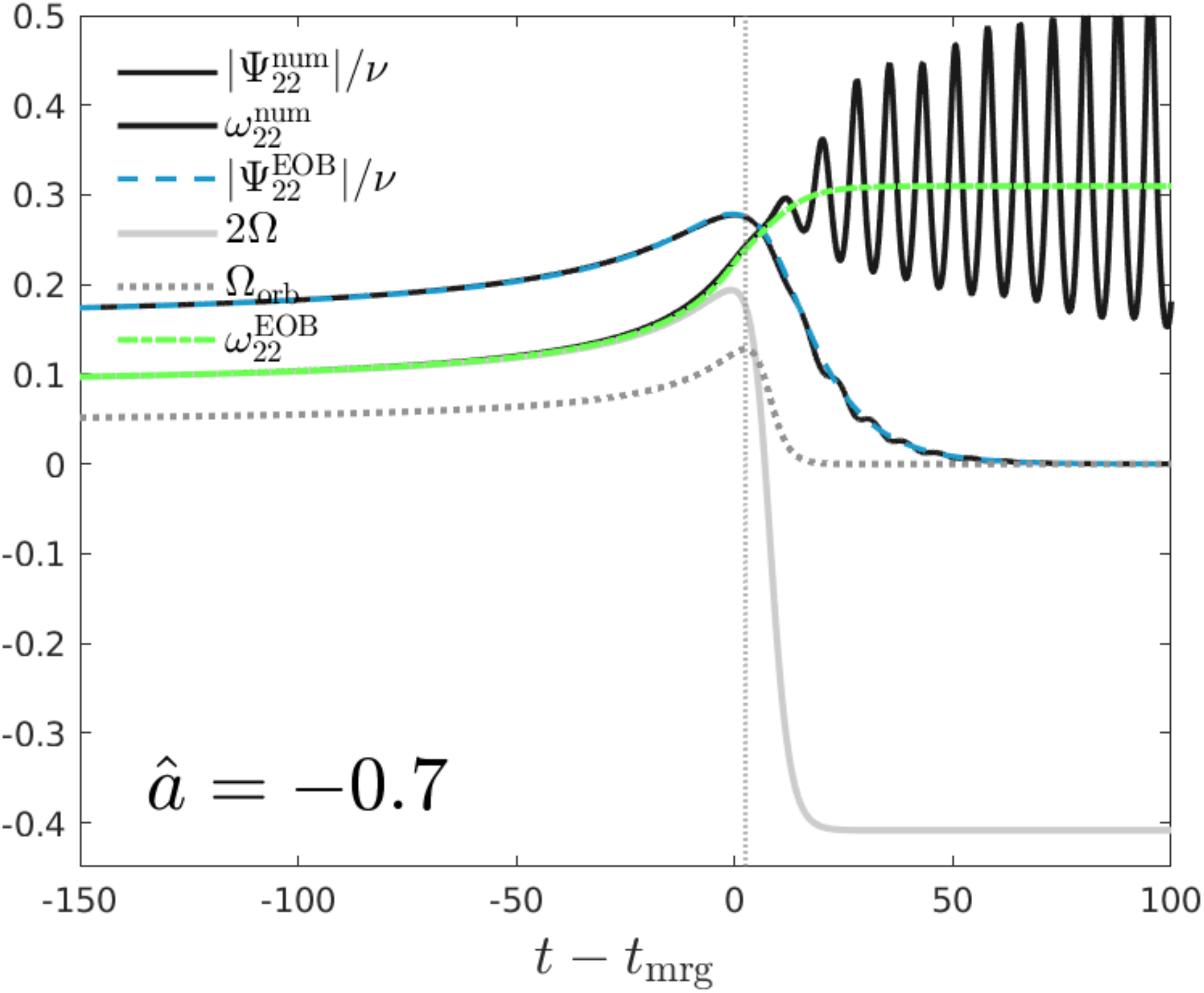} 
\caption{\label{fig:testmass_freq} Comparing EOB and numerical amplitude
and frequencies in the large-mass-ratio limit ($\nu=10^{-3}$) for different values of the 
Kerr dimensionless spin parameter $\hat{a}$. As can be seen,
$\omega_{22}\simeq 2 \Omega$ throughout the whole evolution up to merger.
This behavior is also qualitatively shared by \TEOBResumS{}, 
as shown in Fig.~\ref{fig:freq_comparison_seob}. 
We also show the reliability of the analytical
prescription overlapping the EOB amplitudes and frequencies to numerical results.}
\end{figure}

Let us now consider the case of binary black hole coalescences in the large mass ratio limit
and highlight the qualitative and quantitative features that are shared by \TEOBResumS{}.
Figure~\ref{fig:testmass_freq} shows amplitude and frequencies for a nonspinning test-particle 
(used to model the smaller black hole) inspiralling and plunging in the equatorial plane of a Kerr black hole.  
The analytical waveforms are generated using the test-mass limit version of {\TEOBResumS} 
presented in Ref.~\cite{Albanesi:2021rby}, while the numerical waveforms are computed using 
the 2+1 time-domain code \texttt{Teukode}~\cite{Harms:2014dqa} that solves the Teukolsky equation
(see also Ref.~\cite{Barausse:2011kb,Taracchini:2013wfa} for an earlier EOB model in the test-particle limit). 
Note that, as usual, the dynamics generating the EOB and Teukolsky waveforms is the same.
The analytical/numerical comparisons show that the condition $\omega_{22}\simeq 2\Omega$ 
is satisfied throughout the full evolution of the binary up to merger
\footnote{For $\hat{a}<0$ we have mode-mixing in the ringdown waveform but this is not
relevant for the discussion of this paper.}.
Figure~\ref{fig:testmass_freq} collects a few, non extremal, values of the dimensionless Kerr parameter
$\hat{a}$ so to have a global view of the waveform phenomenology. It is useful to drive a qualitative
and semi-quantitative comparison with Fig.~\ref{fig:freq_comparison_seob}. First, one notices the qualitative
similarities between \TEOBResumS{} and Teukolsky waveforms and dynamics, in particular the location
of the peak of $\Omega_{\rm orb}$. This is a feature that was included within \TEOBResumS{} by construction
and seems to be one of the key points that allows one to have robust and consistent waveforms all over
the parameter space. It is suggestive that the agreement is also semi quantitative for those cases
that have $\hat{a}\simeq \hat{S}$. For example, the configuration with $\hat{S}=-0.2$ in Fig.~\ref{fig:freq_comparison_seob}
shows a behavior of $\Omega$ and
$\Omega_{\rm orb}$ that is qualitatively and quantitatively consistent with the $\hat{a}=-0.2$ case.
Similarly, the $\hat{a}=0.5$ configuration shows a behavior close to the ones with $\hat{S}=0.47$ and $\hat{S}=0.49$
(although the EOB frequency $\Omega$ does not deliver a local maximum), while the $\hat{a}=-0.6$ configuration is
consistent with the $\hat{S}=-0.59$ one, with $\Omega$ becoming negative after merger.
This similarity between test-mass and comparable-mass frequencies can be traced back to the 
quasi-universal behavior of $\omega_{22}$ at merger when plotted versus $\hat{S}$, already 
shown for NR data in Fig.~33 of Ref.~\cite{Nagar:2018zoe}. Although at the moment this is nothing 
more than a suggestive semi-quantitative analogy, if taken seriously it could be helpful to further 
improve the dynamics of \TEOBResumS{} and increase its consistency with the test-mass one,
especially for high spins.
The most obvious thing that needs to be improved is the frequency behavior for the shown 
high-spin configurations, $(1.5,0.95,0.95)$ and $(2,+0.85,+0.85)$, where $\Omega$ keeps 
growing (until the evolution is stopped well after merger), which is in contrast with the local
maximum present in the test-mass case for $\hat{a}=0.5$ (and $\hat{a}=0.7$ as well). 
This is related to the well known problem of the absence of a LSO in \TEOBResumS{} 
for large, positive, spins and it might be solved using a different factorization and gauge for 
the spin-orbit sector~\cite{Rettegno:2019tzh}. Still, the current coherence between 
frequencies that is proper of \TEOBResumS{} looks like an encouraging 
starting point for any future development. 

\begin{figure*}[t]
\begin{center}
\includegraphics[width=0.32\textwidth]{fig12a.pdf} 
\includegraphics[width=0.32\textwidth]{fig12b.pdf} 
\includegraphics[width=0.32\textwidth]{fig12c.pdf} \\
\includegraphics[width=0.32\textwidth]{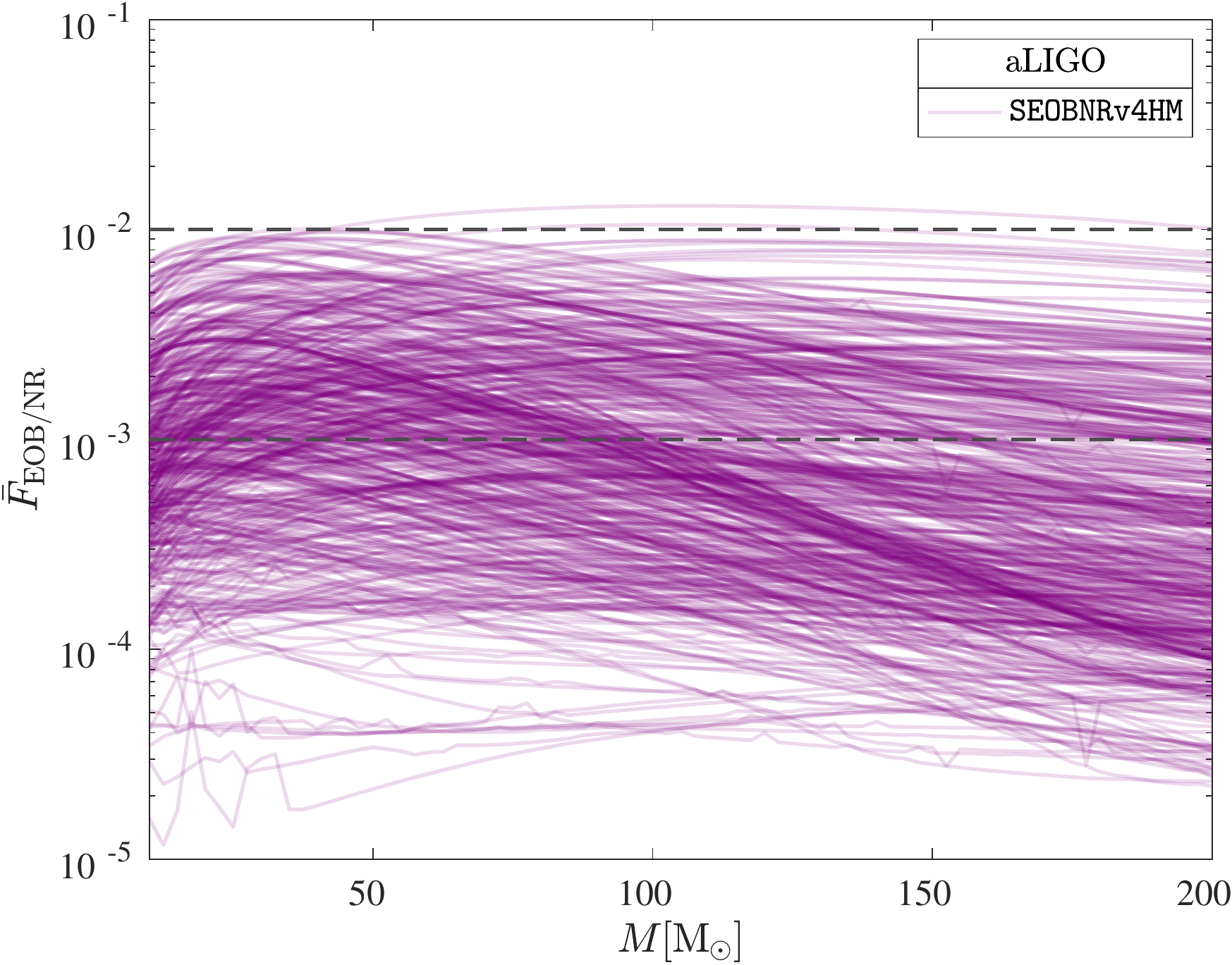} 
\includegraphics[width=0.32\textwidth]{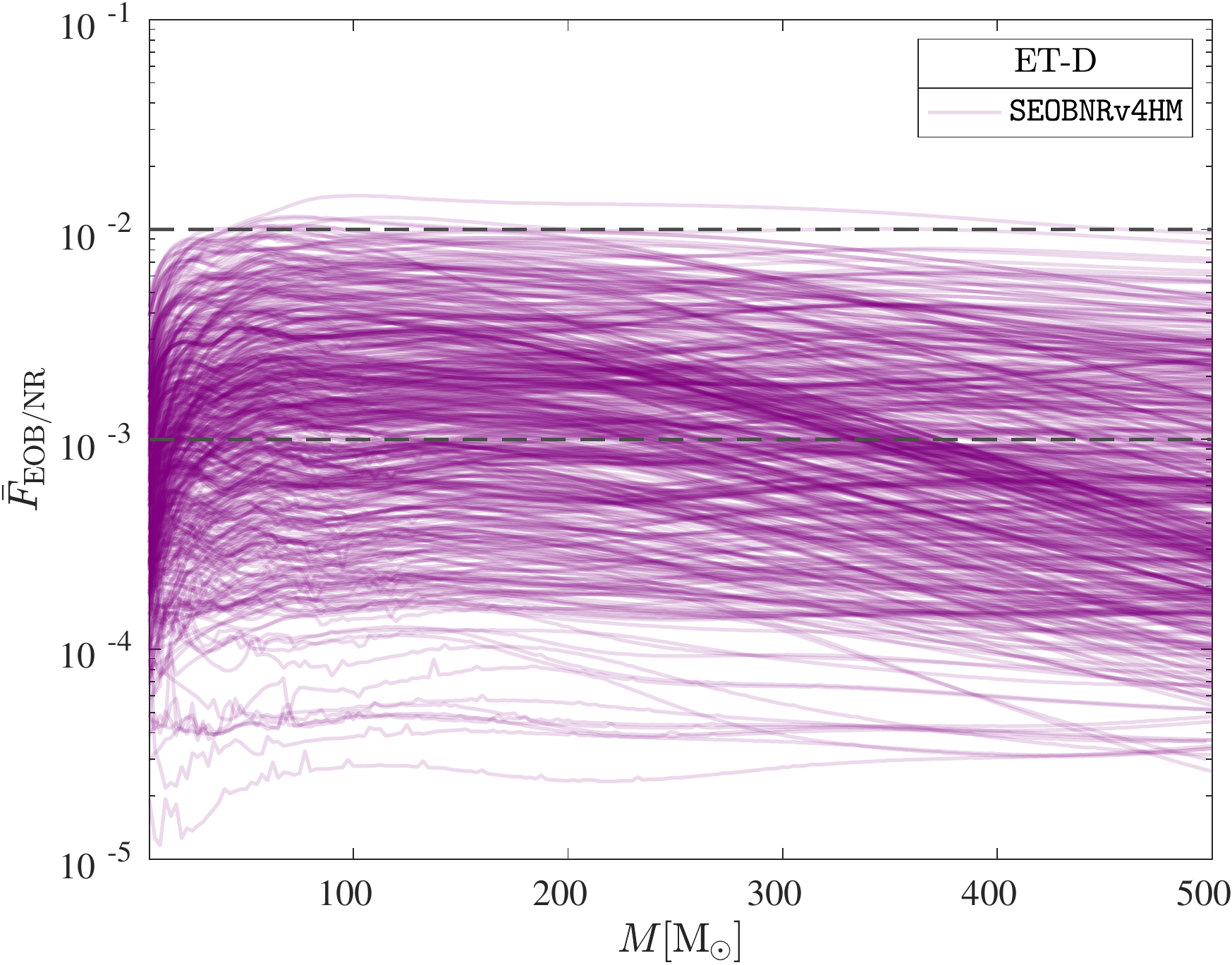} 
\includegraphics[width=0.32\textwidth]{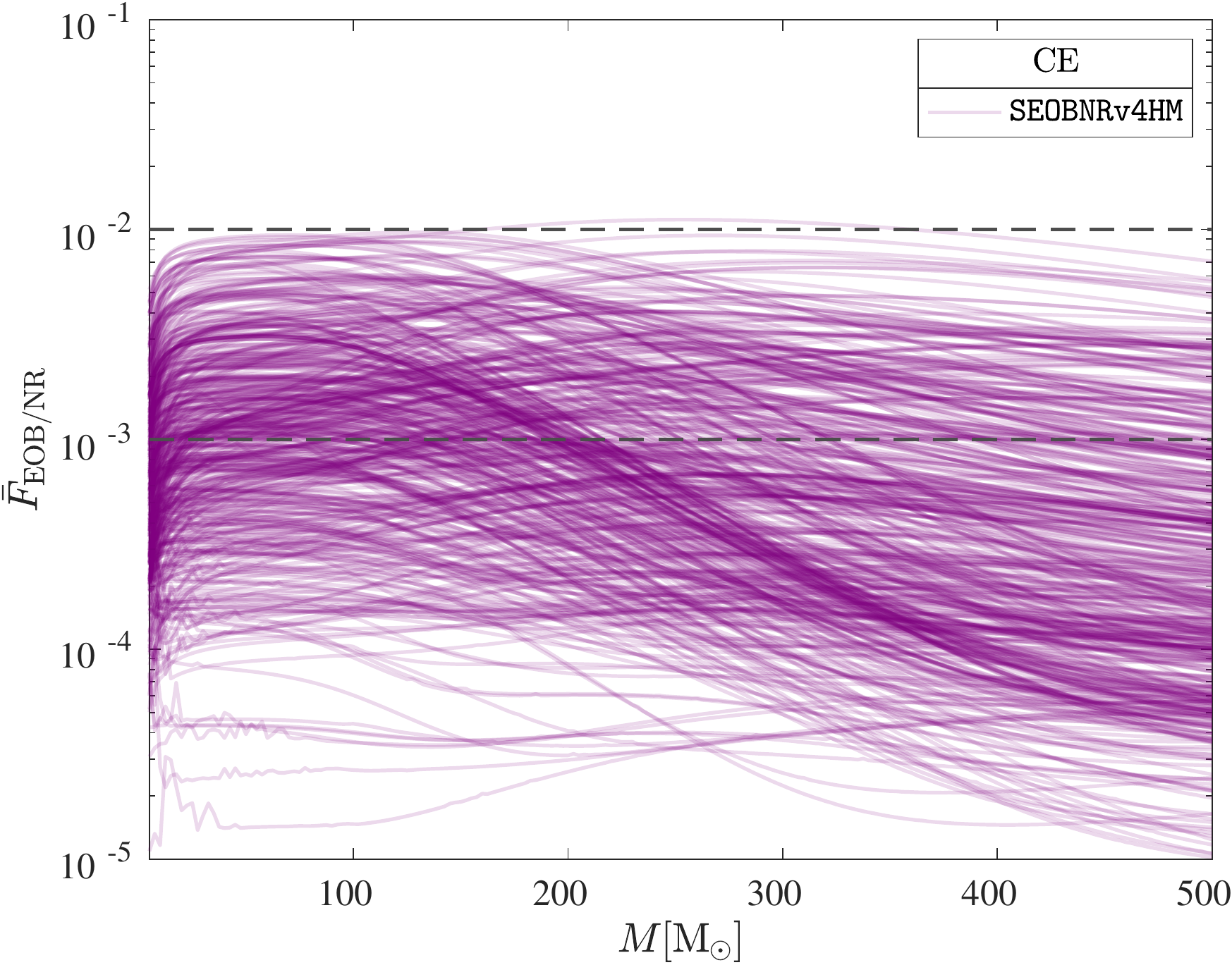} 
\caption{\label{fig:SEOBvsTEOB}Direct EOB/NR unfaithfulness comparison using the standard implementation of \TEOBResumS{} (top panels)
and {\SEOB} (bottom panels). Again, the unfaithfulness is evaluated 
for the sample of 534 nonprecessing quasicircular NR simulations of the SXS catalog 
(likewise Fig.~\ref{fig:barF}) using: (i) the zero-detuned, high-power noise spectral 
density of Advanced LIGO (first column), (ii) the expected PSD for Einstein Telescope (second column), (iii) the expected
PSD for Cosmic Explorer (third column). 
}
\end{center}
\end{figure*}

\subsection{Unfaithfulness}
\label{sec:barF_seobNRHM}
\begin{figure}[t]
\begin{center}
\includegraphics[width=0.45\textwidth]{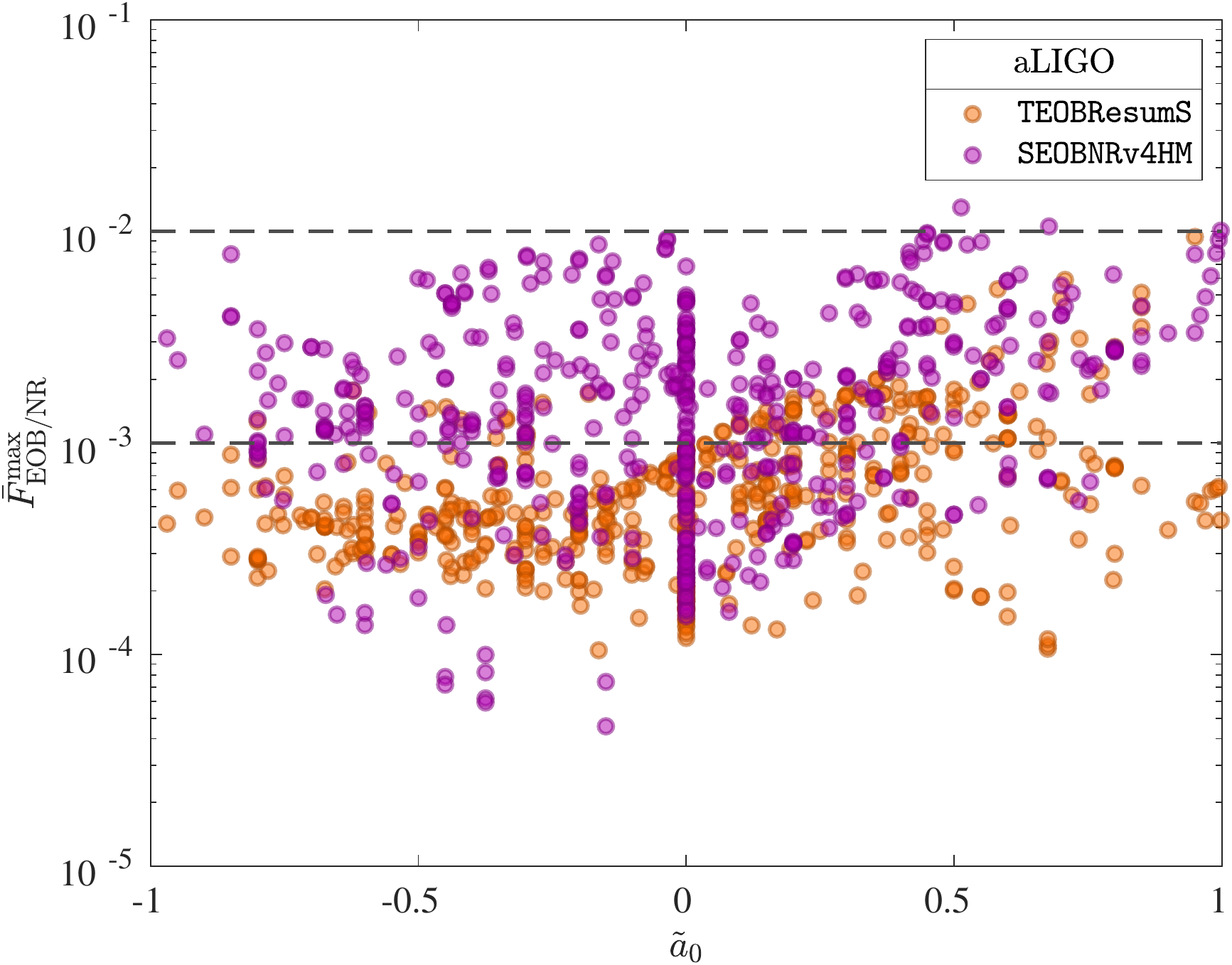}
\includegraphics[width=0.45\textwidth]{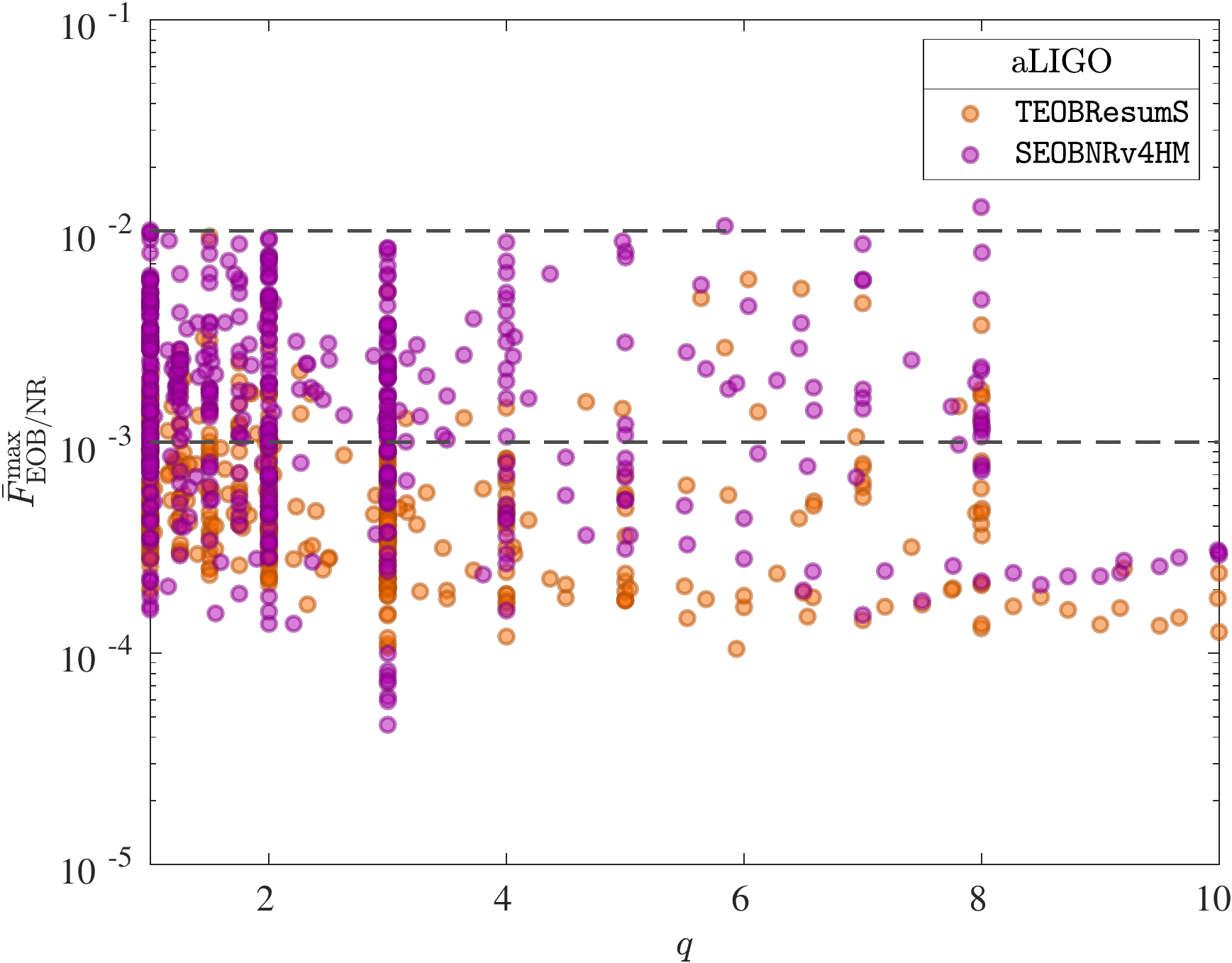}
\caption{\label{fig:max_SEOB_TEOB}Contrasting $\bar{F}^{\rm max}_{\rm EOB/NR}$
for \TEOBResumS{} and \SEOB{} versus $\tilde{a}_0$ and $\nu$. The values for \TEOBResumS{} 
are smaller than those of \SEOB{}, and also show a clear dependence on the effective spin,
indicating where the model may need further improvements.}
\end{center}
\end{figure}

Let us finally move to the calculation of the EOB/NR unfaithfulness using the \SEOB{} model. This calculation
is not new, since it was done for the first time in Ref.~\cite{Bohe:2016gbl} as test of the {\tt SEOBNRv4} 
model. However, from Ref.~\cite{Bohe:2016gbl} several {\it new} NR simulations offering a better covering of the
parameter space became available and the original $\bar{F}$ calculation was not updated since. In particular, 
updated comparisons don't seem to exist in Refs.~\cite{Cotesta:2018fcv,Ossokine:2020kjp}, nor 
in Ref.~\cite{Mihaylov:2021bpf}, that presents a faster version of the \SEOB{} model based on the 
application of the post-adiabatic approximation developed in Ref.~\cite{Nagar:2018gnk} 
(and notably already applied to the \SEOB{} Hamiltonian in Ref.~\cite{Rettegno:2019tzh}). 
To our knowledge, it seems that $\bar{F}_{\rm EOB/NR}$ has never been directly computed all over 
the 534 spin-aligned datasets currently available\footnote{The actual number of nonprecessing quasicircular datasets is larger
but we do not consider some problematic simulations.}. 
It should be mentioned, though, that there exists a comparison 
between \SEOB{} and the NR surrogate~\cite{Pratten:2020fqn}.
The purpose of this section is to complement the results of Ref.~\cite{Pratten:2020fqn} via a direct comparison with 
the SXS datasets. To put this analysis into the right context, we present these results by contrasting them with the corresponding ones 
obtained using the {\it standard}, publicly available, $C$ implementation
of \TEOBResumS{} already presented in Ref.~\cite{Riemenschneider:2021ppj}.
Since this model relies on fits for the NQC corrections, as detailed in Ref.~\cite{Riemenschneider:2021ppj},
its performance is slightly less good than the one 
we would obtain by using the (iterated) {\tt MATLAB} implementation
and similarly less good than what is theoretically achievable using \TEOBResumSlm{}.
Figure~\ref{fig:SEOBvsTEOB} directly compares $\bar{F}_{\rm EOB/NR}(M)$ from  \TEOBResumS{} (top panels) with the 
one from \SEOB{} (bottom panels). The calculation is done for Advanced LIGO (first column), ET-D (second column) and CE (third column). 
The bottom-left panel of Fig.~\ref{fig:SEOBvsTEOB} is the analogous of Fig.~2 of Ref.~\cite{Bohe:2016gbl}, but with the additional
SXS data that were not available at the time, and highlights the very different behavior of the two models
for low masses, where \SEOB{} grazes the $10^{-2}$ level for many configurations. 
This mirrors intrinsic structural differences, probably connected to the completely different  way 
of deforming the Hamiltonian of a point-particle around a Kerr black hole implemented in the two models~\cite{Rettegno:2019tzh}. 
If this is acceptable for Advanced LIGO (although it evidences that the \SEOB{} implementation is not accurate enough), 
 it is not acceptable for ET-D or CE1, where \SEOB{} grazes the $10^{-2}$ level for many configurations. 
Concerning the requirements for third generation detectors, Ref.~\cite{Purrer:2019jcp} concluded that 
current EOB models are not yet sufficiently accurate. Our analysis shows that 
things look better by at least one order of magnitude for \TEOBResumS{} or \TEOBResumSlm{},  that thus represent
more encouraging starting points for developing highly faithful waveform models.
Coming back to the Advanced LIGO design sensitivity curve, the results of the two left panels of Fig.~\ref{fig:SEOBvsTEOB}
are further summarized in  Fig.~\ref{fig:max_SEOB_TEOB}, that shows the corresponding $\bar{F}^{\rm max}_{\rm EOB/NR}$
either versus $\tilde{a}_0$ or versus $q$. Again, $\TEOBResumS{}$ is quite robust all over the parameter space,
although its performance worsens when the effective spin is increased. This clearly indicates where
the model needs to be improved further, coherently with the discussion made in the sections above.
By contrast, this structure is absent for \SEOB{} points, that look randomly distributed all over the parameter space.

\section{Conclusions}
\label{sec:conclusions}
We have presented an updated version of  the spin-aligned waveform model \TEOBResumS{} 
that differs from the previous ones for
(i) a more careful procedure to inform the spin sector of the model, including new choices for NR
simulations and a different functional form for the fit of the effective spin-orbit parameter $c_3$,
(ii) a specific effort to improve the behavior of the radiation reaction up to merger. In particular, our main 
achievement is to show that a careful inclusion of NQC corrections in the flux
typically allows to achieve a EOB/NR flux consistency below $1\%$ during the plunge. The consequent
recalibration of the spin-orbit sector eventually grants a model that shows a higher NR-faithfulness
all over the NR-covered parameter space. In addition we have provided the first ever detailed comparison
between \TEOBResumS{} and \SEOB{}.
Our results can be summarized as follows.
\begin{enumerate}
\item[(i)] We have presented a novel computation of the angular momentum flux from a selected sample of 
36 SXS datasets chosen so as to give a meaningful coverage of the full NR parameter space. 
Apparently, ours is the first computation of this kind from the early exploration of Ref.~\cite{Boyle:2008ge}.
We have introduced an efficient procedure to remove low-frequency oscillations that are present in the raw fluxes
obtained directly from the data. Such oscillations, if kept, would prevent us to perform quantitatively 
accurate EOB/NR comparisons when the fluxes are represented as functions of the frequency.

\item [(ii)] We have shown that the radiation reaction included in the standard implementation of 
                \TEOBResumS{}~\cite{Nagar:2020pcj,Riemenschneider:2021ppj}  already exhibits an 
                excellent consistency with the NR fluxes. However, this can be further improved by including
                NQC flux corrections in all $\ell=m$ modes up to $\ell=5$.

\item[(iii)] This modification to the radiation reaction effectively defined a {\it new} model, 
                called \TEOBResumSlm{}, that also required us to update the determination of the NR-informed
                effective spin-orbit parameter $c_3$. We did so by choosing a 
                {\it new} sample of SXS NR datasets, many of which have improved accuracy with respect to the ones used in previous work.   
                We evaluated the performance of this model all over the 534 spin-aligned SXS simulations
                available, using the Advanced LIGO PSD as well as the ones of ET and CE.
                To our knowledge, this is the first time an EOB model is being extensively tested for 3G detectors.
                We found that $\bar{F}_{\rm EOB/NR}$ is within $10^{-4}$ and  $10^{-3}$ for more than 
                 $80\%$ of the considered binaries. The outliers always occur for configurations 
                 with large, positive, spins, that are the most difficult to simulate numerically and to model 
                 analytically. Although we are still far from the expected 3G detector calibration error, 
                 between $\sim 10^{-4}$ and $\sim 10^{-5}$, our analysis shows that  (any version of) $\TEOBResumS{}$ 
                 can already be used for 3G-related studies provided the spin parameters are not too extreme. 
                 In our opinion, it might be possible that the increase in accuracy needed for 3G detectors 
                 advocated in Ref.~\cite{Purrer:2019jcp} will be less dramatic than suggested.
                
\item[(iv)] By contrast, when the same analyses are performed on the  {\tt SEOBNRv4HM} EOB waveform model, we find large
               differences between the analytical and numerical fluxes for a restricted sample of dataset for which, however, 
               \TEOBResumS{} is NR-consistent already in its native form. For the same configurations we also considered
               waveform and frequencies comparisons, underlining how the dynamics of \TEOBResumS{} is qualitatively consistent 
               with the expectations coming from test-particle limit calculations. 
               Similarly, we show that for the same configurations
               the dynamics of \SEOB{}, differently from the one of \TEOBResumS{}, is qualitatively inconsistent 
               with the expectations coming from test-particle limit calculations. 
               We finally fill the apparent gap in the literature of the calculation of 
               the EOB/NR unfaithfulness for the $\ell=m=2$ mode over all the 534 spin-aligned SXS NR simulations available,
               for Advanced LIGO, ET-D and CE detectors.
               The outcome of this calculation is directly contrasted with the corresponding one from the standard version of \TEOBResumS{},
               highlighting the different performance of the two models, especially during the inspiral. 
               This is worth noticing because \TEOBResumS{} and \SEOB{} were
               built using similar strategies and the same original PN information\footnote{Actually, \SEOB{} includes the exact spin-orbit sector 
               of a spinning test-body~\cite{Barausse:2009aa,Barausse:2009xi,Barausse:2011ys}, while it is only approximated within
               \TEOBResumS{}. It is however straightforward to build a \TEOBResumS{}-like Hamiltonian with the exact spinning test-body
               limit included~\cite{Rettegno:2019tzh}.}.
\end{enumerate}

The most important take-away message of our work is that \TEOBResumS{} can be improved (especially in the large-spin sector) 
only by means of minimal modifications to its structure and  a more careful choice of the NR simulations used to inform the model.
In this respect, it is worth mentioning that the available NR simulations could be better exploited to inform both $a_6^c$ and $c_3$. 
To maintain continuity with previous work, we did not change the function describing $a_6^c$ and we anchored the fit of $c_3$ 
to the equal-mass  case, using 16 equal-mass SXS dataset, while only additional 20 are used to determine the function up to $q=8$. 
This was motivated by the fact that in the past the SXS collaboration mainly focused on producing equal-mass binaries. 
Nowadays things have changed, and in particular there are many dataset available with $q=4$, since they were
needed to construct a NR waveform surrogate {\tt NRSur7dq4}~\cite{Varma:2019csw}. Since we are using only 2 dataset with $q\simeq 4$,
an improved model would be obtained by just anchoring the $c_3$ fit to more $q=4$ simulations, possibly with also an improved choice
of $a_6^c$ more carefully exploiting the nonspinning datasets. We expect that this will additionally improve the EOB/NR agreement, 
possibly pushing it below the $10^{-4}$ level for all binaries. This seems to be at reach given the simplicity and minimality of our 
procedures and will be tackled in future work.

\begin{figure}[t]
\includegraphics[width=0.45\textwidth]{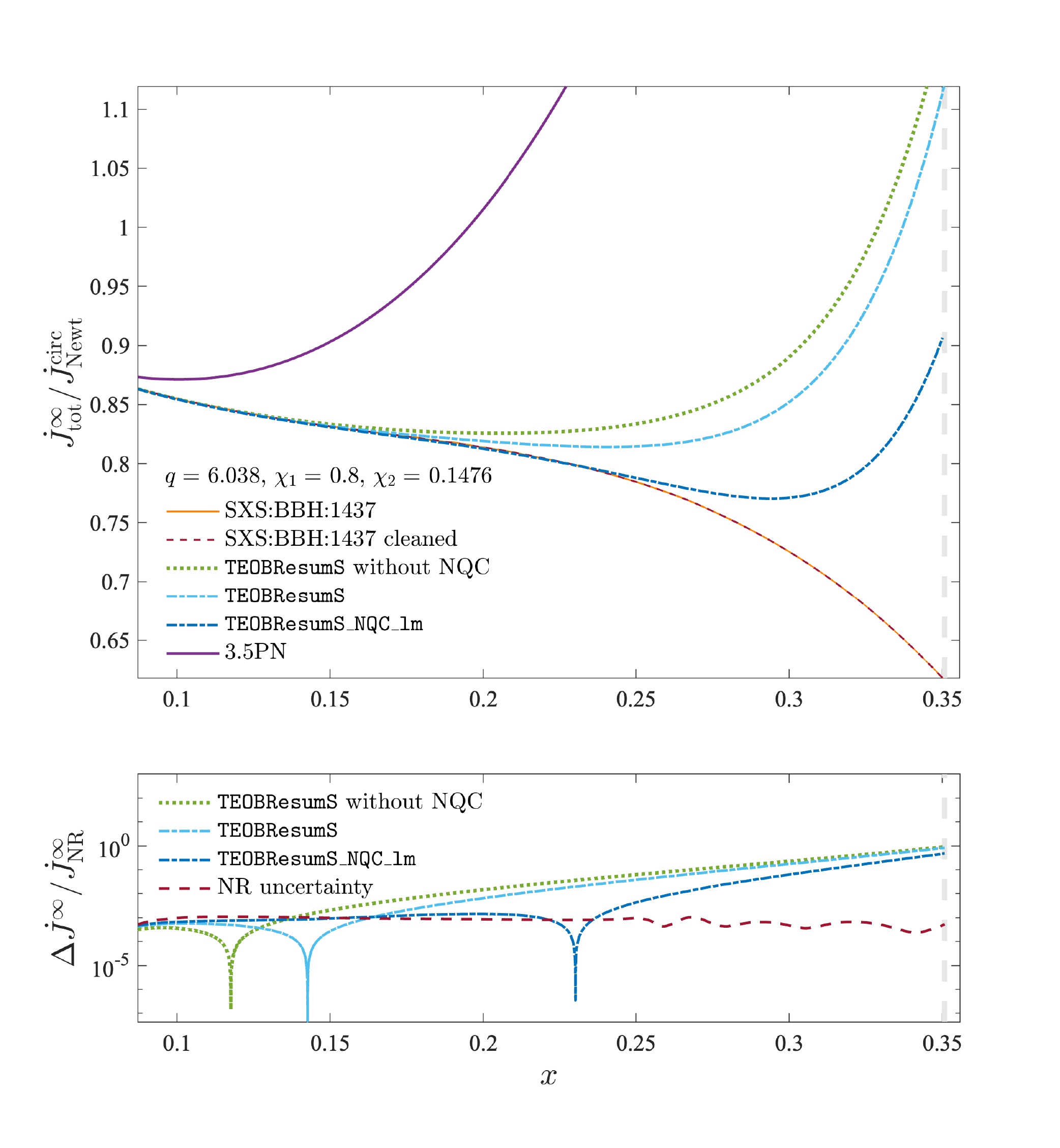} \\
\caption{\label{fig:1437} Contrasting EOB/NR total fluxes summed up to $\ell = 8$ 
using either \TEOBResumS{} or \TEOBResumSlm{}  for the dataset SXS:BBH:1437, 
with $(q, \chi_1, \chi_2) = (6.038, 0.8, 0.1476)$. 
The addition of NQC corrections increases the EOB/NR agreement, though it is not sufficient to completely remove the
growing behavior at the end of the evolution. As seen in Fig.~\ref{fig:new_multipoles}, for
 \TEOBResumSlm{} the multipoles up to $\ell = m = 5$ are consistent with the numerical flux,
meaning that the improvement is only needed for modes with $\ell \ge 6$. }
\end{figure}

\acknowledgements
A.A. has been supported by the fellowship Lumina Quaeruntur No.
LQ100032102 of the Czech Academy of Sciences.
We are grateful to M.~Breschi for a careful reading of the manuscript,
and to S.~Bernuzzi for daily discussions and for the music.
The {\tt TEOBResumS} code is publicly available at
\mbox{\url{https://bitbucket.org/eob_ihes/teobresums/}}.
The {\tt v2} version of the code, that implements the PA approximation and higher modes, 
is fully documented in Refs.~\cite{Nagar:2018gnk, Nagar:2018plt, Nagar:2019wds, Nagar:2020pcj, Riemenschneider:2021ppj}.
We recommend the above references to be cited by \TEOBResumS{}
users.

\appendix

\section{Issues in the NQC-corrected fluxes}
\label{sec:NQCissues} 

\begin{figure}[t]
\includegraphics[width=0.45\textwidth]{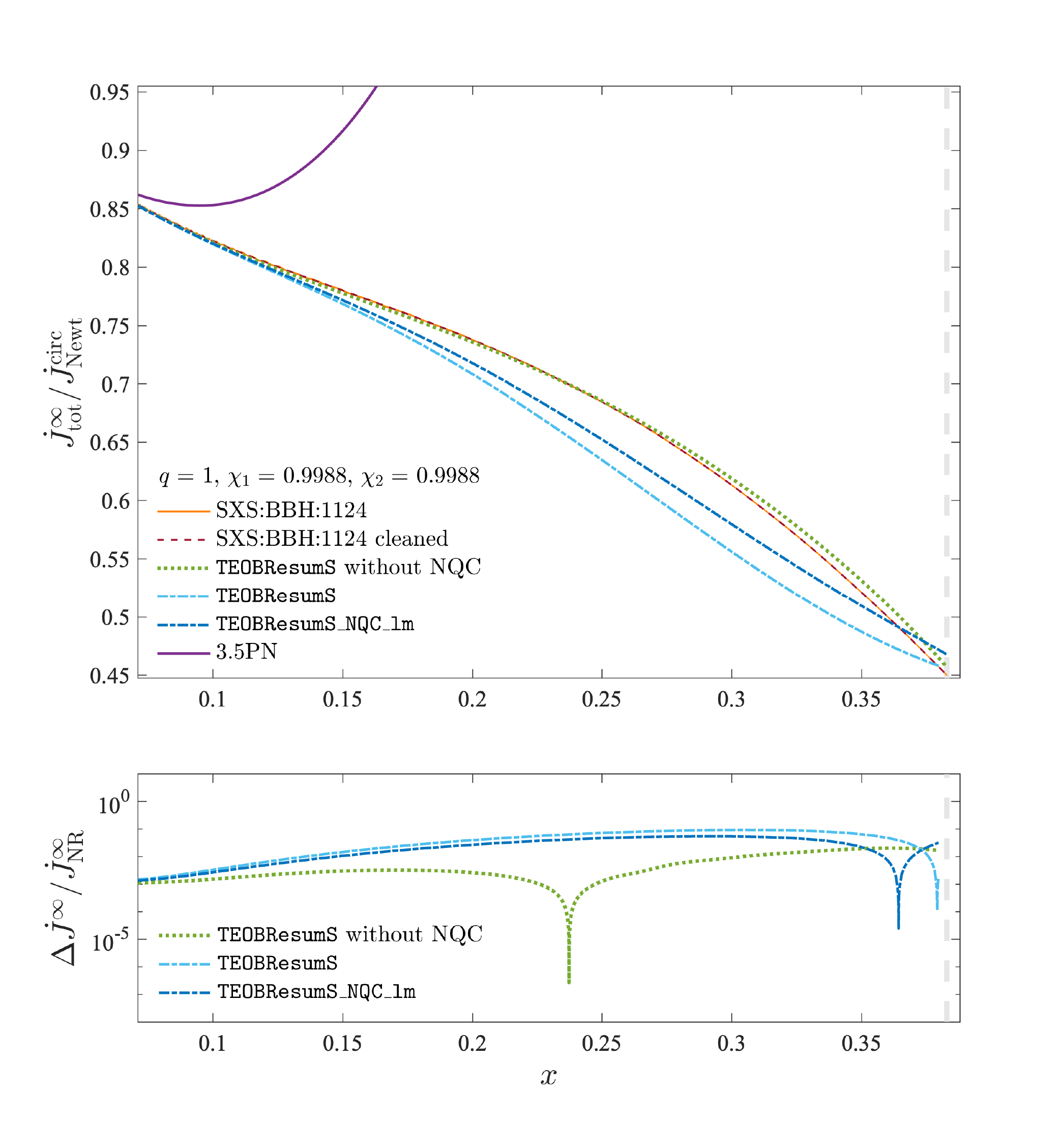} 
\caption{\label{fig:1124} Contrasting EOB/NR fluxes for the configuration
$(q, \chi_1, \chi_2) = (1, 0.9988, 0.9988)$, corresponding to dataset SXS:BBH:1124. 
The flux without NQC corrections is more consistent with the numerical one with respect
to the NQC-corrected ones, both from \TEOBResumS{} and from \TEOBResumSlm{}.}
\end{figure}

In this section we focus on some problematic EOB fluxes. 
Let us start by considering the dataset SXS:BBH:1437. As seen in Fig.~\ref{fig:1437},
for this configuration the additions of NQC corrections increases the agreement with NR but does not avoid the 
growing behavior at the end of the evolution. As pointed out in Fig.~\ref{fig:new_multipoles}, for \TEOBResumSlm{}
the EOB flux is consistent with the numerical one up to $\ell = m = 5$, so this behavior this is due to modes with $6 \le \ell \le 8$,
that only rely on analytical information and do not incorporate NQC corrections. 
Nevertheless, we underline that the EOB/NR relative difference for \TEOBResumSlm{} is of order $10^{-3}$
until $x \sim 0.24$, corresponding to $\sim 1.5$ orbits before merger. 

\begin{figure}[t]
\includegraphics[width=0.45\textwidth]{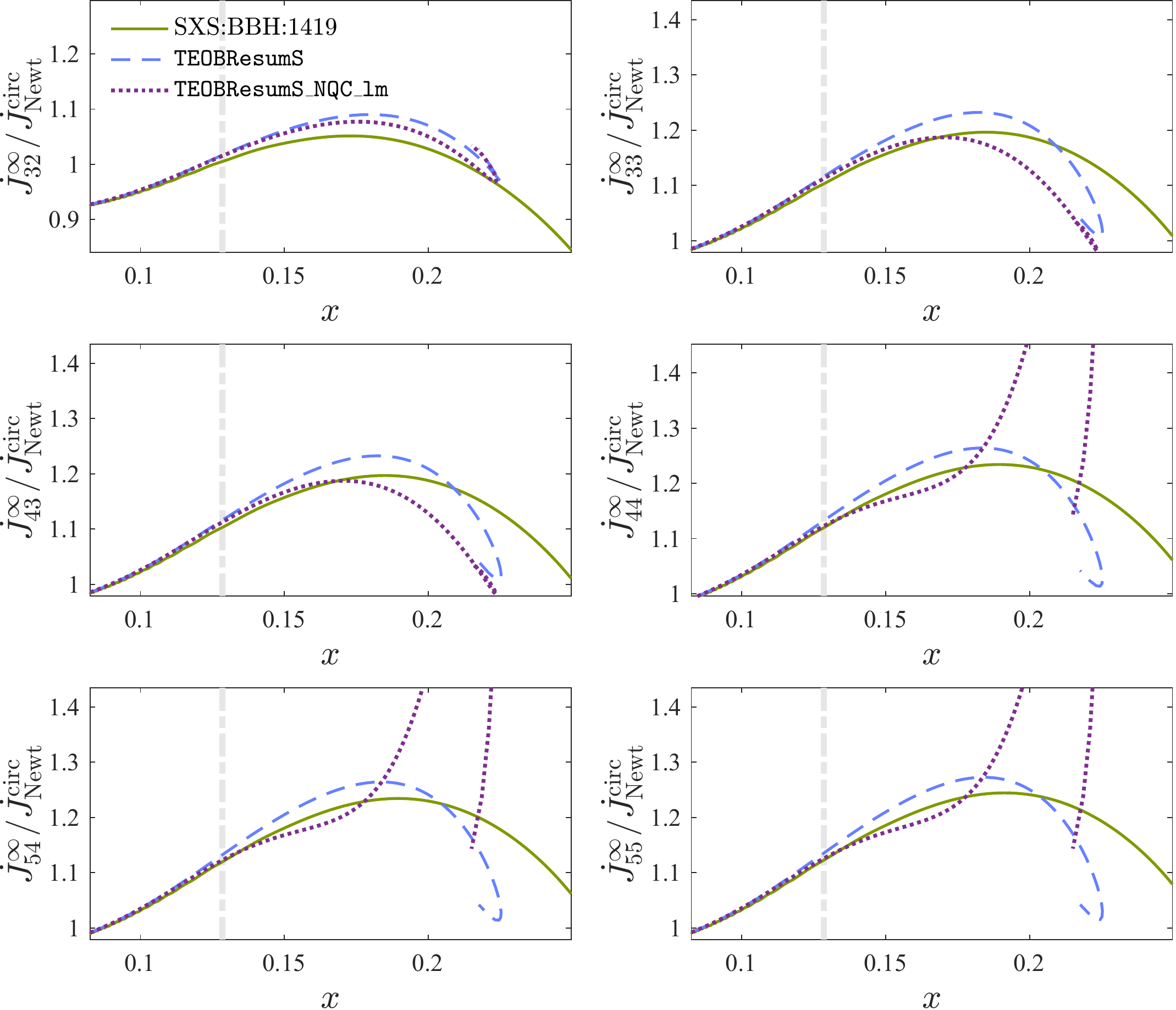} 
\caption{\label{fig:1419}  Comparison between EOB/NR fluxes for one of the
configurations excluded from Fig.~\ref{fig:flux_diff}, SXS:BBH:1419, with $(q, \chi_1, \chi_2) = (8, -0.8, -0.8)$. 
In this case the NQC correction factor in the $\ell=m=4$ mode becomes pathological and eventually 
\TEOBResumS{} yields a more NR-consistent flux.}
\end{figure}

Let us now consider dataset SXS:BBH:1124, corresponding to the extremely spinning
configuration $(q, \chi_1, \chi_2) = (1, 0.9988, 0.9988)$. In this case the purely
analytical flux is in excellent agreement with the NR one, keeping the fractional difference below $10^{-2}$
until merger, but surprisingly NQC corrections worsen the flux behavior all over the evolution. 
This may be 
attributed to two different facts: (i) the motion for a comparable mass binary
with such high spins is highly adiabatic, so that there is a reduced need of
non-circular correction factors; (ii) NQC corrections in the current model are added
from the beginning of the evolution, considering they are functions of the radial momentum
which is small but non-negligible during the inspiral, and its effect is progressively amplified. 
To avoid this issue it seems better to include the NQC factor only as a correction 
that is progressively switched on towards merger, similarly to what is currently implemented
in the version of \TEOBResumS{} valid for noncircular configurations~\cite{Nagar:2021gss, Nagar:2021xnh}.

Finally, we consider the two datasets excluded from the bottom panel of Fig.~\ref{fig:flux_diff},
namely SXS:BBH:1419 and SXS:BBH:1375, respectively corresponding to 
$(q, \chi_1, \chi_2) = (8, -0.80, -0.80)$ and $(q, \chi_1, \chi_2) = (8, -0.90, 0)$. 
The multipolar fluxes for the first configuration are
shown in Fig.~\ref{fig:1419},  from which we infer that the NQC correction factor is not correctly 
determined for the $\ell = m = 4$ and $\ell = m = 5$ modes, with the former multipole yielding 
the largest deviations. One notices, however, that up the $\ell=m=4$ mode excluded, NQC
corrections yield an agreement between the fluxes up to the LSO that is closer than the
standard case. The same happens for the dataset SXS:BBH:1375. 
We also found that for $(8,-0.80,-0.80)$ it is possible to fix the behavior of the $\ell=m=4$ mode 
by adjusting the $c_3$ value from the value predicted by the fit, $c_3^{\rm fit}=65.16$,
to $c_3=77$. On the contrary, this is not possible for $(8,-0.90,0)$, indicating that a more detailed
understanding of the determination of the NQC corrections is needed in this case.

\section{Improving the consistency between waveform and flux changing Newtonian prefactors}
\label{sec:hlm_Newt}

\begin{figure}[t]
\includegraphics[width=0.45\textwidth]{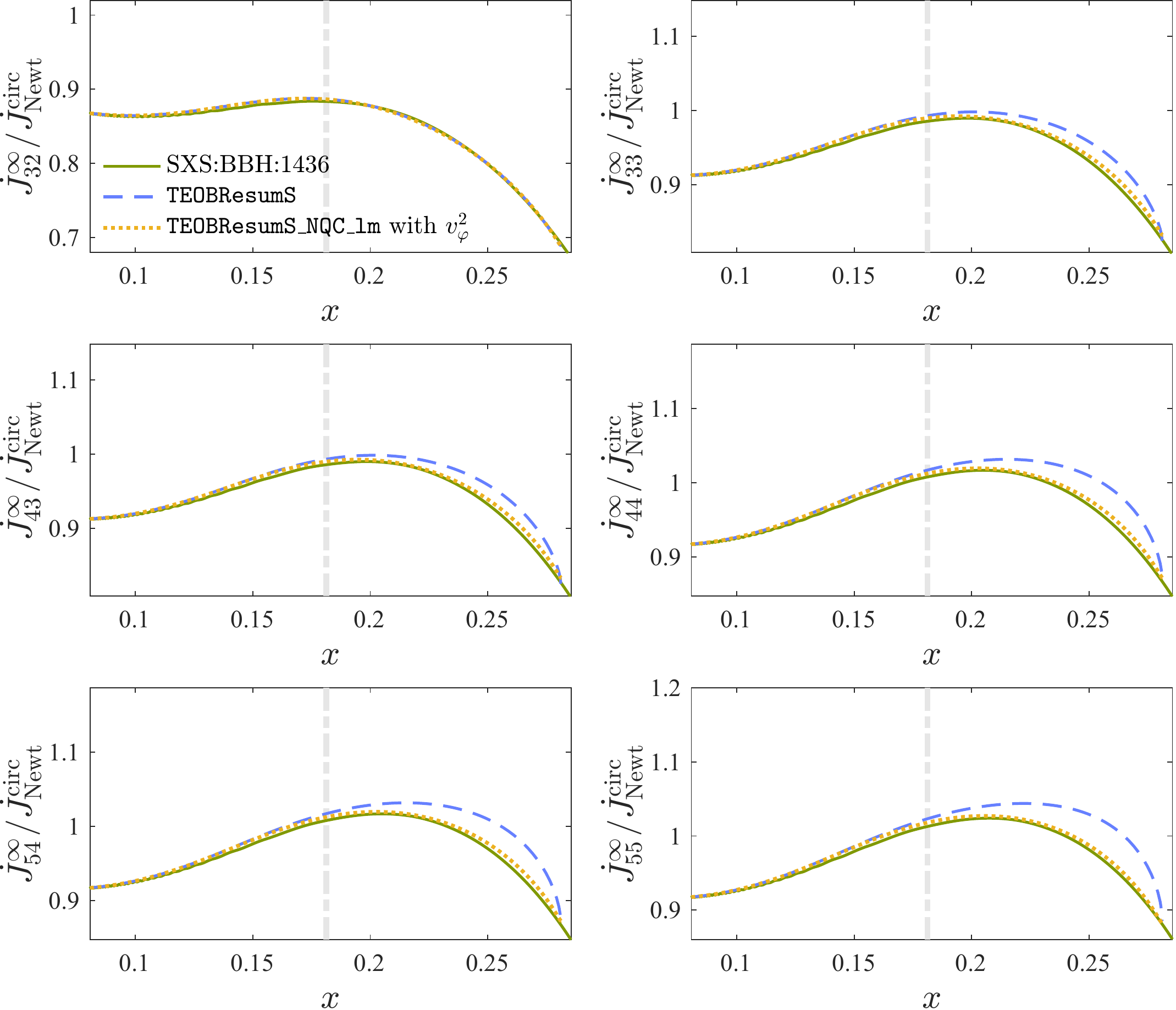}
\caption{\label{fig:1436vphi} Comparing EOB/NR multipolar fluxes for the dataset SXS:BBH:1436, 
using \TEOBResumSlm{} with the standard Newtonian prefactors in the waveform written as powers of $v_{\varphi}$ (see text).
For this configuration, this choice yields an excellent EOB/NR agreement up to merger.}
\end{figure}

Let us finally present an EOB/NR flux comparison using a model that has $\ell=m$ 
NQC corrections in the flux up to $\ell=5$ and uses {\it consistent} Newtonian prefactors
in the flux and in the waveform. In practice, this amounts at replacing $v_\Omega=\Omega^{1/3}$
with the standard $v_\varphi$ in Eqs.~(3.22)-(3.30) in Sec.~IIIC of Ref.~\cite{Nagar:2019wds}.
As can be seen in Fig.~\ref{fig:1436vphi}, this actually yields a more consistent flux for the configuration corresponding to
dataset SXS:BBH:1436 we analyzed previously. But, not surprisingly, this does not hold for the problematic
corner of the parameter space that motivated the different choice for the Newtonian
prefactors in the waveform, as evident when looking at the values in Table~\ref{tab:flux_diff}.
Moreover, we noticed there are some configurations, e.g. $(8, -0.8, -0.8)$, in which
even the $\ell = m = 2$ multipole is spoiled by the unsuccessful determination of NQC corrections.

For future developments, the EOB/NR flux agreement in Fig.~\ref{fig:1436vphi}
encourages us to look for different solutions to ensure the NQC determination works out
for high mass ratios and negative spins.

\begin{table}[t]
\caption{\label{tab:flux_diff} EOB/NR fractional flux differences at $x = 0.2$, both for 
\TEOBResumSlm{} and for its waveform/flux consistent version.}
 \begin{center}
 \begin{ruledtabular}
\begin{tabular}{c c c c}
   ID & $(q, \chi_1, \chi_2)$ & $ \Delta \dot{J}^{\texttt{NQC\_lm}}_{\rm EOBNR} $ & $ \Delta \dot{J}^{v_\varphi^2}_{\rm EOBNR} $  \\ 
\hline 
BBH:1155 & $(1,0,0)$ & $0.001957$ & 0.0022517\\ 
BBH:1222 & $(2,0,0)$ & $-0.0001001$ & 0.002563 \\ 
BBH:1179 & $(3,0,0)$ & $-0.00071698$ & 0.0040221\\ 
BBH:0190 & $(4.499,0,0)$ & $-0.0084789$ & $-0.0021009$\\ 
BBH:0192 & $(6.58,0,0)$ & $-0.0098679$ & $-0.0030977$\\ 
BBH:1107 & $(10,0,0)$ & $-0.011246$ & $-0.0096363$\\ 
\hline
BBH:1137 & $(1,-0.97,-0.97)$ & $0.099337$ & - \\ 
BBH:2084 & $(1,-0.90,0)$ & $0.072879$ & 0.038549 \\ 
BBH:2097 & $(1,+0.30,0)$ & $-0.0032907$ & $-0.0029234$ \\ 
BBH:2105 & $(1,+0.90,0)$ & $-0.01013$ & $-0.011484$ \\ 
BBH:1124 & $(1,+0.99,+0.99)$ & $-0.0272$  & $-0.027025$ \\ 
BBH:1146 & $(1.5,+0.95,+0.95)$ & $-0.035553$ & $-0.035553$\\ 
BBH:2111 & $(2,-0.60,+0.60)$ & $0.015314$ & $0.017929$ \\ 
BBH:2124 & $(2,+0.30,0)$ & $-0.0061851$ & $-0.0044956$ \\ 
BBH:2131 & $(2,+0.85,+0.85)$ & $-0.01893$ & $-0.018629$ \\ 
BBH:2132 & $(2,+0.87,0)$ & $-0.011947$ & $-0.010621$ \\ 
BBH:2133 & $(3,-0.73,+0.85)$ & $0.031988$ & 0.03876\\ 
BBH:2153 & $(3,+0.30,0)$ & $-0.0042875$ & $-0.0012004$\\ 
BBH:2162 & $(3,+0.60,+0.40)$ & $-0.011459$ & $-0.0097236$\\ 
BBH:1446 & $(3.154,-0.80,+0.78)$ & 0.04242 & $0.049673$\\ 
BBH:1936 & $(4,-0.80,-0.80)$ & 0.051971 & - \\ 
BBH:2040 & $(4,-0.80,-0.40)$ & 0.048942 & 0.063711 \\ 
BBH:1911 & $(4,0,-0.80)$ & 0.0019117  & 0.0083588 \\ 
BBH:2014 & $(4,+0.80,+0.40)$ & $-0.0089473$ & $-0.0078338$\\ 
BBH:1434 & $(4.368,+0.80,+0.80)$ & $-0.012975$ & $-0.01188$ \\ 
BBH:1463 & $(4.978,+0.61,+0.24)$ & $-0.0083662$ & $-0.0064106$ \\ 
BBH:0208 & $(5,-0.90,0)$ & $0.037164$ & $0.0016289$ \\ 
BBH:1428 & $(5.518,-0.80,-0.70)$ & $0.030081$ & -  \\ 
BBH:1437 & $(6.038,+0.80,+0.15)$ & $-0.001383$ & $-0.00033034$ \\ 
BBH:1436 & $(6.281,+0.009,-0.80)$ & $-0.002271$ & $0.0041482$ \\ 
BBH:1435 & $(6.588,-0.79,+0.7)$ & $0.0054703$ & - \\ 
BBH:1448 & $(6.944,-0.48,+0.52)$ & $0.0091571$ & $0.42486$ \\ 
BBH:1375 & $(8,-0.90, 0)$ &  $1.3589$ & - \\ 
BBH:1419 & $(8,-0.80,-0.80)$ & $0.19952$ & -  \\ 
BBH:1420 & $(8,-0.80,+0.80)$ & $-0.022786$ & - \\ 
BBH:1455 & $(8,-0.40, 0)$ & $0.0070483$ & $0.27028$ \\ 
\end{tabular}
\end{ruledtabular}
\end{center}
\end{table}

\section{Unfaithfulness with the ET-C noise}
\label{sec:ET-C}

\begin{figure}[t]
\center
\includegraphics[width=0.43\textwidth]{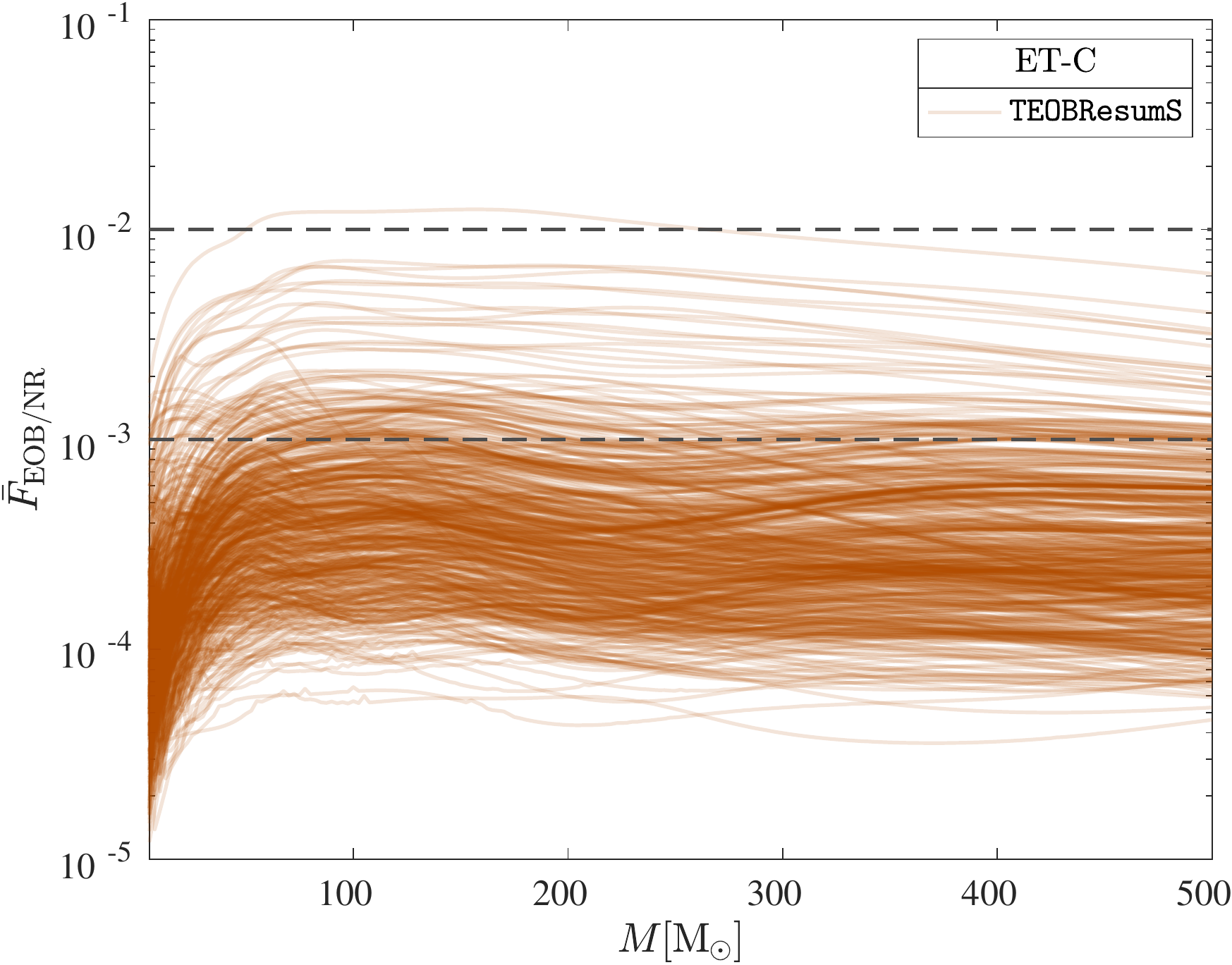} \\
\includegraphics[width=0.43\textwidth]{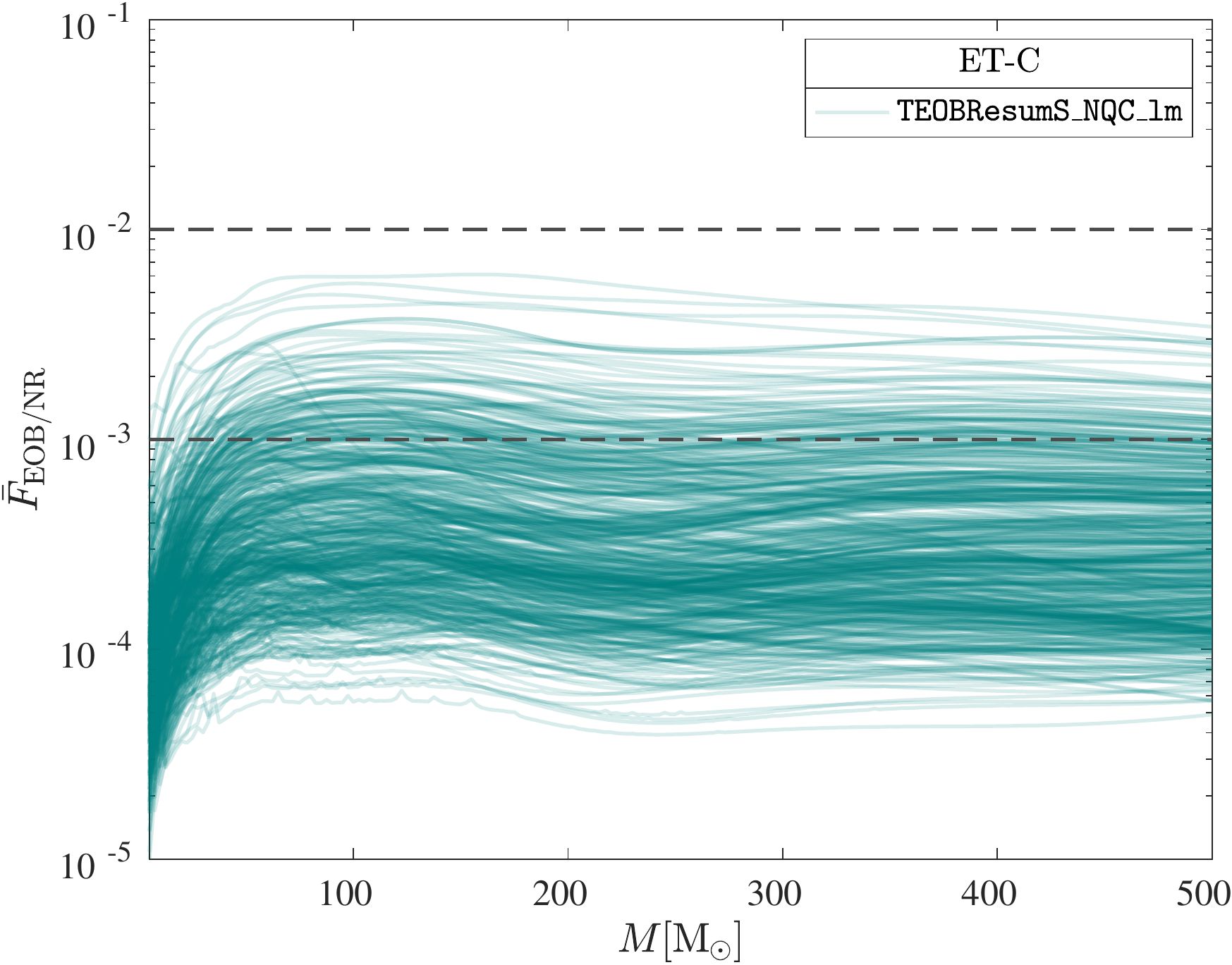} \\
\includegraphics[width=0.43\textwidth]{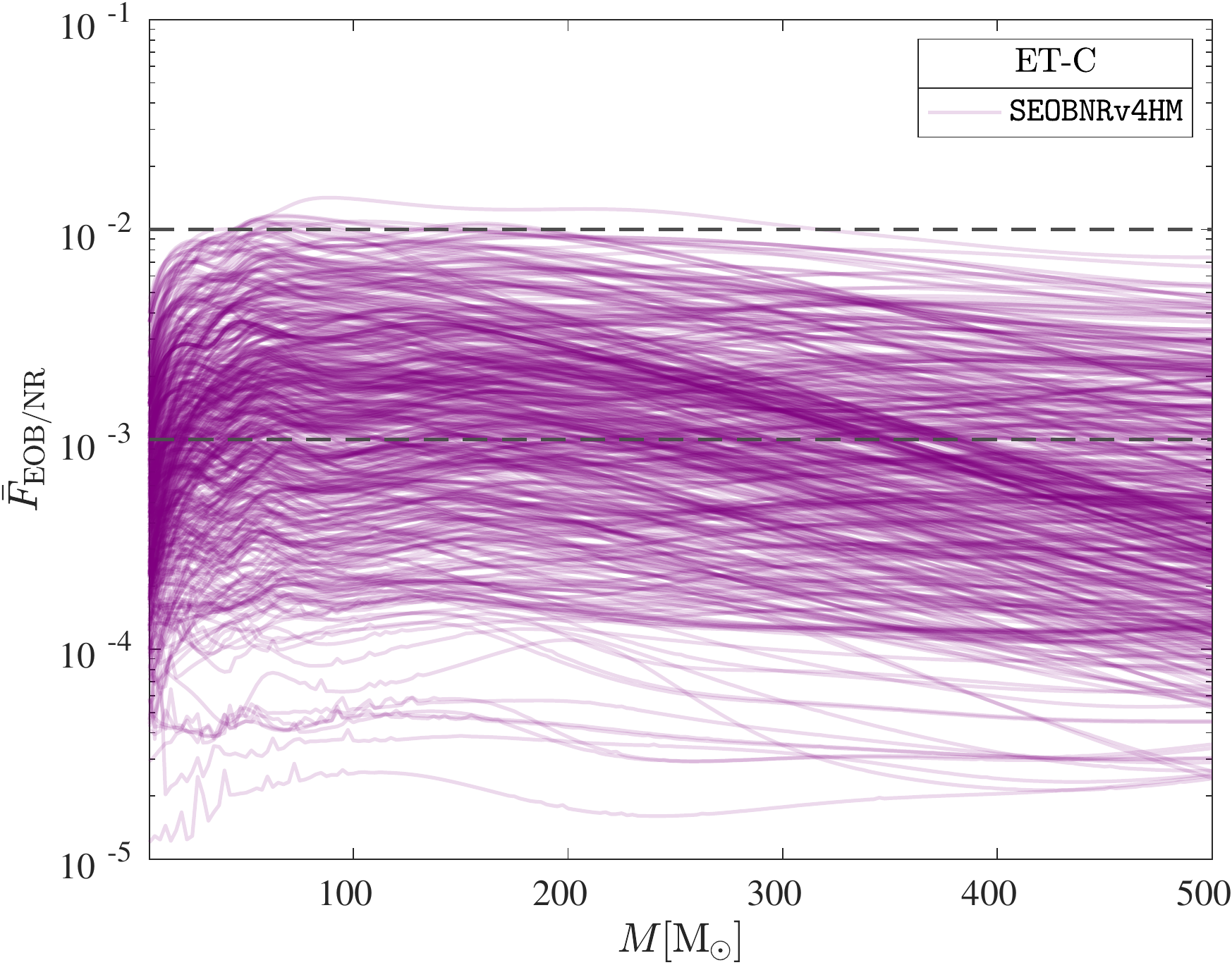}
\caption{\label{fig:barF_ET_C} EOB/NR unfaithfulness for \TEOBResumS{} (top),
\TEOBResumSlm{} (middle) and \SEOB{} (bottom), evaluated using 
 the ET-C version of the expected noise for Einstein Telescope~\cite{Hild:2009ns}.}
\end{figure}

We display in this section results for the EOB/NR unfaithfulness computation by using the less recent
PSD of Einstein Telescope, ET-C~\cite{Hild:2009ns}. As one can see in Fig.~\ref{fig:noises}, 
ET-D has a larger sensitivity with respect to ET-C for higher frequencies, where we expect both
EOB and NR waveforms to be less accurate\footnote{This is related as well to the choice of the extrapolation order 
and the ringdown modeling, as discussed above.}. Correspondently, the results shown in Fig.~\ref{fig:barF_ET_C}
and in Table~\ref{tab:maxFbar_ET_C} are slightly better than the ones we reported above for the latest PSD,
probably also owing to the fact that viceversa the ET-C version has a larger sensitivity at lower frequencies.

 \begin{table*}[t]
   \caption{\label{tab:maxFbar_ET_C} Analogous of Table~\ref{tab:maxFbar}, using the ET-C power spectral density.
   The central columns of the table display the fraction of datasets 
   whose maximum unfaithfulness $\bar{F}^{\rm max}_{\rm EOB/NR}$ is within the indicated limits for  \TEOBResumS{},
   \TEOBResumSlm{} or \SEOB{}. Again, as in Table~\ref{tab:maxFbar}, the last two columns display percentage numbers out of \textit{all} the mismatch values. 
   }
   \begin{center}
 \begin{ruledtabular}
   \begin{tabular}{l | c c c | c c}
       & $\bar{F}^{\rm max} < 10^{-3}$ & $10^{-3} < \bar{F}^{\rm max} < 10^{-2}$ & $\bar{F}^{\rm max}> 3\times 10^{-3}$ 
      & $10^{-4} < \bar{F}< 10^{-3}$ & $\bar{F} < 10^{-4}$\\
     \hline
     \TEOBResumS{}      & 85.0\% & 14.4\% & 2.6\% & 84.5\% & 5.3\% \\
      \TEOBResumSlm{} & 82.4\% & 18.0\% & 1.7\% & 81.1\% & 6.9\% \\
      \SEOB{}                  & 36.3\% & 38.0\% & 27.7\% & 48.1\% & 3.2\% \\
 \end{tabular}
 \end{ruledtabular}
 \end{center}
 \end{table*}
 
%
%
%
%
%
%
%

\bibliography{refs20211128.bib,local.bib}

\begin{thebibliography}{72}%
\makeatletter
\providecommand \@ifxundefined [1]{%
 \@ifx{#1\undefined}
}%
\providecommand \@ifnum [1]{%
 \ifnum #1\expandafter \@firstoftwo
 \else \expandafter \@secondoftwo
 \fi
}%
\providecommand \@ifx [1]{%
 \ifx #1\expandafter \@firstoftwo
 \else \expandafter \@secondoftwo
 \fi
}%
\providecommand \natexlab [1]{#1}%
\providecommand \enquote  [1]{``#1''}%
\providecommand \bibnamefont  [1]{#1}%
\providecommand \bibfnamefont [1]{#1}%
\providecommand \citenamefont [1]{#1}%
\providecommand \href@noop [0]{\@secondoftwo}%
\providecommand \href [0]{\begingroup \@sanitize@url \@href}%
\providecommand \@href[1]{\@@startlink{#1}\@@href}%
\providecommand \@@href[1]{\endgroup#1\@@endlink}%
\providecommand \@sanitize@url [0]{\catcode `\\12\catcode `\$12\catcode
  `\&12\catcode `\#12\catcode `\^12\catcode `\_12\catcode `\%12\relax}%
\providecommand \@@startlink[1]{}%
\providecommand \@@endlink[0]{}%
\providecommand \url  [0]{\begingroup\@sanitize@url \@url }%
\providecommand \@url [1]{\endgroup\@href {#1}{\urlprefix }}%
\providecommand \urlprefix  [0]{URL }%
\providecommand \Eprint [0]{\href }%
\providecommand \doibase [0]{http://dx.doi.org/}%
\providecommand \selectlanguage [0]{\@gobble}%
\providecommand \bibinfo  [0]{\@secondoftwo}%
\providecommand \bibfield  [0]{\@secondoftwo}%
\providecommand \translation [1]{[#1]}%
\providecommand \BibitemOpen [0]{}%
\providecommand \bibitemStop [0]{}%
\providecommand \bibitemNoStop [0]{.\EOS\space}%
\providecommand \EOS [0]{\spacefactor3000\relax}%
\providecommand \BibitemShut  [1]{\csname bibitem#1\endcsname}%
\let\auto@bib@innerbib\@empty
\bibitem [{\citenamefont {Acernese}\ \emph {et~al.}(2015)\citenamefont
  {Acernese} \emph {et~al.}}]{TheVirgo:2014hva}%
  \BibitemOpen
  \bibfield  {author} {\bibinfo {author} {\bibfnamefont {F.}~\bibnamefont
  {Acernese}} \emph {et~al.} (\bibinfo {collaboration} {VIRGO}),\ }\href
  {\doibase 10.1088/0264-9381/32/2/024001} {\bibfield  {journal} {\bibinfo
  {journal} {Class. Quant. Grav.}\ }\textbf {\bibinfo {volume} {32}},\ \bibinfo
  {pages} {024001} (\bibinfo {year} {2015})},\ \Eprint
  {http://arxiv.org/abs/1408.3978} {arXiv:1408.3978 [gr-qc]} \BibitemShut
  {NoStop}%
\bibitem [{\citenamefont {Aasi}\ \emph {et~al.}(2015)\citenamefont {Aasi} \emph
  {et~al.}}]{TheLIGOScientific:2014jea}%
  \BibitemOpen
  \bibfield  {author} {\bibinfo {author} {\bibfnamefont {J.}~\bibnamefont
  {Aasi}} \emph {et~al.} (\bibinfo {collaboration} {LIGO Scientific}),\ }\href
  {\doibase 10.1088/0264-9381/32/7/074001} {\bibfield  {journal} {\bibinfo
  {journal} {Class. Quant. Grav.}\ }\textbf {\bibinfo {volume} {32}},\ \bibinfo
  {pages} {074001} (\bibinfo {year} {2015})},\ \Eprint
  {http://arxiv.org/abs/1411.4547} {arXiv:1411.4547 [gr-qc]} \BibitemShut
  {NoStop}%
\bibitem [{\citenamefont {Abbott}\ \emph {et~al.}(2021)\citenamefont {Abbott}
  \emph {et~al.}}]{LIGOScientific:2020ibl}%
  \BibitemOpen
  \bibfield  {author} {\bibinfo {author} {\bibfnamefont {R.}~\bibnamefont
  {Abbott}} \emph {et~al.} (\bibinfo {collaboration} {LIGO Scientific,
  Virgo}),\ }\href {\doibase 10.1103/PhysRevX.11.021053} {\bibfield  {journal}
  {\bibinfo  {journal} {Phys. Rev. X}\ }\textbf {\bibinfo {volume} {11}},\
  \bibinfo {pages} {021053} (\bibinfo {year} {2021})},\ \Eprint
  {http://arxiv.org/abs/2010.14527} {arXiv:2010.14527 [gr-qc]} \BibitemShut
  {NoStop}%
\bibitem [{\citenamefont {Buonanno}\ and\ \citenamefont
  {Damour}(1999)}]{Buonanno:1998gg}%
  \BibitemOpen
  \bibfield  {author} {\bibinfo {author} {\bibfnamefont {A.}~\bibnamefont
  {Buonanno}}\ and\ \bibinfo {author} {\bibfnamefont {T.}~\bibnamefont
  {Damour}},\ }\href {\doibase 10.1103/PhysRevD.59.084006} {\bibfield
  {journal} {\bibinfo  {journal} {Phys. Rev.}\ }\textbf {\bibinfo {volume}
  {D59}},\ \bibinfo {pages} {084006} (\bibinfo {year} {1999})},\ \Eprint
  {http://arxiv.org/abs/gr-qc/9811091} {arXiv:gr-qc/9811091} \BibitemShut
  {NoStop}%
\bibitem [{\citenamefont {Buonanno}\ and\ \citenamefont
  {Damour}(2000)}]{Buonanno:2000ef}%
  \BibitemOpen
  \bibfield  {author} {\bibinfo {author} {\bibfnamefont {A.}~\bibnamefont
  {Buonanno}}\ and\ \bibinfo {author} {\bibfnamefont {T.}~\bibnamefont
  {Damour}},\ }\href {\doibase 10.1103/PhysRevD.62.064015} {\bibfield
  {journal} {\bibinfo  {journal} {Phys. Rev.}\ }\textbf {\bibinfo {volume}
  {D62}},\ \bibinfo {pages} {064015} (\bibinfo {year} {2000})},\ \Eprint
  {http://arxiv.org/abs/gr-qc/0001013} {arXiv:gr-qc/0001013} \BibitemShut
  {NoStop}%
\bibitem [{\citenamefont {Damour}\ \emph {et~al.}(2000)\citenamefont {Damour},
  \citenamefont {Jaranowski},\ and\ \citenamefont {Schaefer}}]{Damour:2000we}%
  \BibitemOpen
  \bibfield  {author} {\bibinfo {author} {\bibfnamefont {T.}~\bibnamefont
  {Damour}}, \bibinfo {author} {\bibfnamefont {P.}~\bibnamefont {Jaranowski}},
  \ and\ \bibinfo {author} {\bibfnamefont {G.}~\bibnamefont {Schaefer}},\
  }\href {\doibase 10.1103/PhysRevD.62.084011} {\bibfield  {journal} {\bibinfo
  {journal} {Phys. Rev.}\ }\textbf {\bibinfo {volume} {D62}},\ \bibinfo {pages}
  {084011} (\bibinfo {year} {2000})},\ \Eprint
  {http://arxiv.org/abs/gr-qc/0005034} {arXiv:gr-qc/0005034 [gr-qc]}
  \BibitemShut {NoStop}%
\bibitem [{\citenamefont {Damour}(2001)}]{Damour:2001tu}%
  \BibitemOpen
  \bibfield  {author} {\bibinfo {author} {\bibfnamefont {T.}~\bibnamefont
  {Damour}},\ }\href {\doibase 10.1103/PhysRevD.64.124013} {\bibfield
  {journal} {\bibinfo  {journal} {Phys. Rev.}\ }\textbf {\bibinfo {volume}
  {D64}},\ \bibinfo {pages} {124013} (\bibinfo {year} {2001})},\ \Eprint
  {http://arxiv.org/abs/gr-qc/0103018} {arXiv:gr-qc/0103018} \BibitemShut
  {NoStop}%
\bibitem [{\citenamefont {Damour}\ \emph {et~al.}(2015)\citenamefont {Damour},
  \citenamefont {Jaranowski},\ and\ \citenamefont {Schäfer}}]{Damour:2015isa}%
  \BibitemOpen
  \bibfield  {author} {\bibinfo {author} {\bibfnamefont {T.}~\bibnamefont
  {Damour}}, \bibinfo {author} {\bibfnamefont {P.}~\bibnamefont {Jaranowski}},
  \ and\ \bibinfo {author} {\bibfnamefont {G.}~\bibnamefont {Schäfer}},\
  }\href {\doibase 10.1103/PhysRevD.91.084024} {\bibfield  {journal} {\bibinfo
  {journal} {Phys. Rev.}\ }\textbf {\bibinfo {volume} {D91}},\ \bibinfo {pages}
  {084024} (\bibinfo {year} {2015})},\ \Eprint
  {http://arxiv.org/abs/1502.07245} {arXiv:1502.07245 [gr-qc]} \BibitemShut
  {NoStop}%
\bibitem [{\citenamefont {Nagar}\ and\ \citenamefont
  {Rettegno}(2019)}]{Nagar:2018gnk}%
  \BibitemOpen
  \bibfield  {author} {\bibinfo {author} {\bibfnamefont {A.}~\bibnamefont
  {Nagar}}\ and\ \bibinfo {author} {\bibfnamefont {P.}~\bibnamefont
  {Rettegno}},\ }\href {\doibase 10.1103/PhysRevD.99.021501} {\bibfield
  {journal} {\bibinfo  {journal} {Phys. Rev.}\ }\textbf {\bibinfo {volume}
  {D99}},\ \bibinfo {pages} {021501} (\bibinfo {year} {2019})},\ \Eprint
  {http://arxiv.org/abs/1805.03891} {arXiv:1805.03891 [gr-qc]} \BibitemShut
  {NoStop}%
\bibitem [{\citenamefont {Nagar}\ \emph {et~al.}(2018)\citenamefont {Nagar}
  \emph {et~al.}}]{Nagar:2018zoe}%
  \BibitemOpen
  \bibfield  {author} {\bibinfo {author} {\bibfnamefont {A.}~\bibnamefont
  {Nagar}} \emph {et~al.},\ }\href {\doibase 10.1103/PhysRevD.98.104052}
  {\bibfield  {journal} {\bibinfo  {journal} {Phys. Rev.}\ }\textbf {\bibinfo
  {volume} {D98}},\ \bibinfo {pages} {104052} (\bibinfo {year} {2018})},\
  \Eprint {http://arxiv.org/abs/1806.01772} {arXiv:1806.01772 [gr-qc]}
  \BibitemShut {NoStop}%
\bibitem [{\citenamefont {Cotesta}\ \emph {et~al.}(2018)\citenamefont
  {Cotesta}, \citenamefont {Buonanno}, \citenamefont {Boh\'e}, \citenamefont
  {Taracchini}, \citenamefont {Hinder},\ and\ \citenamefont
  {Ossokine}}]{Cotesta:2018fcv}%
  \BibitemOpen
  \bibfield  {author} {\bibinfo {author} {\bibfnamefont {R.}~\bibnamefont
  {Cotesta}}, \bibinfo {author} {\bibfnamefont {A.}~\bibnamefont {Buonanno}},
  \bibinfo {author} {\bibfnamefont {A.}~\bibnamefont {Boh\'e}}, \bibinfo
  {author} {\bibfnamefont {A.}~\bibnamefont {Taracchini}}, \bibinfo {author}
  {\bibfnamefont {I.}~\bibnamefont {Hinder}}, \ and\ \bibinfo {author}
  {\bibfnamefont {S.}~\bibnamefont {Ossokine}},\ }\href {\doibase
  10.1103/PhysRevD.98.084028} {\bibfield  {journal} {\bibinfo  {journal} {Phys.
  Rev.}\ }\textbf {\bibinfo {volume} {D98}},\ \bibinfo {pages} {084028}
  (\bibinfo {year} {2018})},\ \Eprint {http://arxiv.org/abs/1803.10701}
  {arXiv:1803.10701 [gr-qc]} \BibitemShut {NoStop}%
\bibitem [{\citenamefont {Nagar}\ \emph
  {et~al.}(2019{\natexlab{a}})\citenamefont {Nagar}, \citenamefont {Pratten},
  \citenamefont {Riemenschneider},\ and\ \citenamefont
  {Gamba}}]{Nagar:2019wds}%
  \BibitemOpen
  \bibfield  {author} {\bibinfo {author} {\bibfnamefont {A.}~\bibnamefont
  {Nagar}}, \bibinfo {author} {\bibfnamefont {G.}~\bibnamefont {Pratten}},
  \bibinfo {author} {\bibfnamefont {G.}~\bibnamefont {Riemenschneider}}, \ and\
  \bibinfo {author} {\bibfnamefont {R.}~\bibnamefont {Gamba}},\ }\href@noop {}
  {\  (\bibinfo {year} {2019}{\natexlab{a}})},\ \Eprint
  {http://arxiv.org/abs/1904.09550} {arXiv:1904.09550 [gr-qc]} \BibitemShut
  {NoStop}%
\bibitem [{\citenamefont {Nagar}\ \emph {et~al.}(2020)\citenamefont {Nagar},
  \citenamefont {Riemenschneider}, \citenamefont {Pratten}, \citenamefont
  {Rettegno},\ and\ \citenamefont {Messina}}]{Nagar:2020pcj}%
  \BibitemOpen
  \bibfield  {author} {\bibinfo {author} {\bibfnamefont {A.}~\bibnamefont
  {Nagar}}, \bibinfo {author} {\bibfnamefont {G.}~\bibnamefont
  {Riemenschneider}}, \bibinfo {author} {\bibfnamefont {G.}~\bibnamefont
  {Pratten}}, \bibinfo {author} {\bibfnamefont {P.}~\bibnamefont {Rettegno}}, \
  and\ \bibinfo {author} {\bibfnamefont {F.}~\bibnamefont {Messina}},\ }\href
  {\doibase 10.1103/PhysRevD.102.024077} {\bibfield  {journal} {\bibinfo
  {journal} {Phys. Rev. D}\ }\textbf {\bibinfo {volume} {102}},\ \bibinfo
  {pages} {024077} (\bibinfo {year} {2020})},\ \Eprint
  {http://arxiv.org/abs/2001.09082} {arXiv:2001.09082 [gr-qc]} \BibitemShut
  {NoStop}%
\bibitem [{\citenamefont {Ossokine}\ \emph {et~al.}(2020)\citenamefont
  {Ossokine} \emph {et~al.}}]{Ossokine:2020kjp}%
  \BibitemOpen
  \bibfield  {author} {\bibinfo {author} {\bibfnamefont {S.}~\bibnamefont
  {Ossokine}} \emph {et~al.},\ }\href {\doibase 10.1103/PhysRevD.102.044055}
  {\bibfield  {journal} {\bibinfo  {journal} {Phys. Rev. D}\ }\textbf {\bibinfo
  {volume} {102}},\ \bibinfo {pages} {044055} (\bibinfo {year} {2020})},\
  \Eprint {http://arxiv.org/abs/2004.09442} {arXiv:2004.09442 [gr-qc]}
  \BibitemShut {NoStop}%
\bibitem [{\citenamefont {Schmidt}\ \emph {et~al.}(2021)\citenamefont
  {Schmidt}, \citenamefont {Breschi}, \citenamefont {Gamba}, \citenamefont
  {Pagano}, \citenamefont {Rettegno}, \citenamefont {Riemenschneider},
  \citenamefont {Bernuzzi}, \citenamefont {Nagar},\ and\ \citenamefont
  {Del~Pozzo}}]{Schmidt:2020yuu}%
  \BibitemOpen
  \bibfield  {author} {\bibinfo {author} {\bibfnamefont {S.}~\bibnamefont
  {Schmidt}}, \bibinfo {author} {\bibfnamefont {M.}~\bibnamefont {Breschi}},
  \bibinfo {author} {\bibfnamefont {R.}~\bibnamefont {Gamba}}, \bibinfo
  {author} {\bibfnamefont {G.}~\bibnamefont {Pagano}}, \bibinfo {author}
  {\bibfnamefont {P.}~\bibnamefont {Rettegno}}, \bibinfo {author}
  {\bibfnamefont {G.}~\bibnamefont {Riemenschneider}}, \bibinfo {author}
  {\bibfnamefont {S.}~\bibnamefont {Bernuzzi}}, \bibinfo {author}
  {\bibfnamefont {A.}~\bibnamefont {Nagar}}, \ and\ \bibinfo {author}
  {\bibfnamefont {W.}~\bibnamefont {Del~Pozzo}},\ }\href {\doibase
  10.1103/PhysRevD.103.043020} {\bibfield  {journal} {\bibinfo  {journal}
  {Phys. Rev. D}\ }\textbf {\bibinfo {volume} {103}},\ \bibinfo {pages}
  {043020} (\bibinfo {year} {2021})},\ \Eprint
  {http://arxiv.org/abs/2011.01958} {arXiv:2011.01958 [gr-qc]} \BibitemShut
  {NoStop}%
\bibitem [{\citenamefont {Chiaramello}\ and\ \citenamefont
  {Nagar}(2020)}]{Chiaramello:2020ehz}%
  \BibitemOpen
  \bibfield  {author} {\bibinfo {author} {\bibfnamefont {D.}~\bibnamefont
  {Chiaramello}}\ and\ \bibinfo {author} {\bibfnamefont {A.}~\bibnamefont
  {Nagar}},\ }\href {\doibase 10.1103/PhysRevD.101.101501} {\bibfield
  {journal} {\bibinfo  {journal} {Phys. Rev. D}\ }\textbf {\bibinfo {volume}
  {101}},\ \bibinfo {pages} {101501} (\bibinfo {year} {2020})},\ \Eprint
  {http://arxiv.org/abs/2001.11736} {arXiv:2001.11736 [gr-qc]} \BibitemShut
  {NoStop}%
\bibitem [{\citenamefont {Nagar}\ \emph
  {et~al.}(2021{\natexlab{a}})\citenamefont {Nagar}, \citenamefont {Bonino},\
  and\ \citenamefont {Rettegno}}]{Nagar:2021gss}%
  \BibitemOpen
  \bibfield  {author} {\bibinfo {author} {\bibfnamefont {A.}~\bibnamefont
  {Nagar}}, \bibinfo {author} {\bibfnamefont {A.}~\bibnamefont {Bonino}}, \
  and\ \bibinfo {author} {\bibfnamefont {P.}~\bibnamefont {Rettegno}},\ }\href
  {\doibase 10.1103/PhysRevD.103.104021} {\bibfield  {journal} {\bibinfo
  {journal} {Phys. Rev. D}\ }\textbf {\bibinfo {volume} {103}},\ \bibinfo
  {pages} {104021} (\bibinfo {year} {2021}{\natexlab{a}})},\ \Eprint
  {http://arxiv.org/abs/2101.08624} {arXiv:2101.08624 [gr-qc]} \BibitemShut
  {NoStop}%
\bibitem [{\citenamefont {Nagar}\ and\ \citenamefont
  {Rettegno}(2021)}]{Nagar:2021xnh}%
  \BibitemOpen
  \bibfield  {author} {\bibinfo {author} {\bibfnamefont {A.}~\bibnamefont
  {Nagar}}\ and\ \bibinfo {author} {\bibfnamefont {P.}~\bibnamefont
  {Rettegno}},\ }\href@noop {} {\  (\bibinfo {year} {2021})},\ \Eprint
  {http://arxiv.org/abs/2108.02043} {arXiv:2108.02043 [gr-qc]} \BibitemShut
  {NoStop}%
\bibitem [{\citenamefont {Damour}\ \emph {et~al.}(2014)\citenamefont {Damour},
  \citenamefont {Guercilena}, \citenamefont {Hinder}, \citenamefont {Hopper},
  \citenamefont {Nagar} \emph {et~al.}}]{Damour:2014afa}%
  \BibitemOpen
  \bibfield  {author} {\bibinfo {author} {\bibfnamefont {T.}~\bibnamefont
  {Damour}}, \bibinfo {author} {\bibfnamefont {F.}~\bibnamefont {Guercilena}},
  \bibinfo {author} {\bibfnamefont {I.}~\bibnamefont {Hinder}}, \bibinfo
  {author} {\bibfnamefont {S.}~\bibnamefont {Hopper}}, \bibinfo {author}
  {\bibfnamefont {A.}~\bibnamefont {Nagar}},  \emph {et~al.},\ }\href@noop {}
  {\  (\bibinfo {year} {2014})},\ \Eprint {http://arxiv.org/abs/1402.7307}
  {arXiv:1402.7307 [gr-qc]} \BibitemShut {NoStop}%
\bibitem [{\citenamefont {Nagar}\ \emph
  {et~al.}(2021{\natexlab{b}})\citenamefont {Nagar}, \citenamefont {Rettegno},
  \citenamefont {Gamba},\ and\ \citenamefont {Bernuzzi}}]{Nagar:2020xsk}%
  \BibitemOpen
  \bibfield  {author} {\bibinfo {author} {\bibfnamefont {A.}~\bibnamefont
  {Nagar}}, \bibinfo {author} {\bibfnamefont {P.}~\bibnamefont {Rettegno}},
  \bibinfo {author} {\bibfnamefont {R.}~\bibnamefont {Gamba}}, \ and\ \bibinfo
  {author} {\bibfnamefont {S.}~\bibnamefont {Bernuzzi}},\ }\href {\doibase
  10.1103/PhysRevD.103.064013} {\bibfield  {journal} {\bibinfo  {journal}
  {Phys. Rev. D}\ }\textbf {\bibinfo {volume} {103}},\ \bibinfo {pages}
  {064013} (\bibinfo {year} {2021}{\natexlab{b}})},\ \Eprint
  {http://arxiv.org/abs/2009.12857} {arXiv:2009.12857 [gr-qc]} \BibitemShut
  {NoStop}%
\bibitem [{\citenamefont {Gamba}\ \emph
  {et~al.}(2021{\natexlab{a}})\citenamefont {Gamba}, \citenamefont {Breschi},
  \citenamefont {Carullo}, \citenamefont {Rettegno}, \citenamefont {Albanesi},
  \citenamefont {Bernuzzi},\ and\ \citenamefont {Nagar}}]{Gamba:2021gap}%
  \BibitemOpen
  \bibfield  {author} {\bibinfo {author} {\bibfnamefont {R.}~\bibnamefont
  {Gamba}}, \bibinfo {author} {\bibfnamefont {M.}~\bibnamefont {Breschi}},
  \bibinfo {author} {\bibfnamefont {G.}~\bibnamefont {Carullo}}, \bibinfo
  {author} {\bibfnamefont {P.}~\bibnamefont {Rettegno}}, \bibinfo {author}
  {\bibfnamefont {S.}~\bibnamefont {Albanesi}}, \bibinfo {author}
  {\bibfnamefont {S.}~\bibnamefont {Bernuzzi}}, \ and\ \bibinfo {author}
  {\bibfnamefont {A.}~\bibnamefont {Nagar}},\ }\href@noop {} {\bibfield
  {journal} {\bibinfo  {journal} {Submitted to Nature Astronomy}\ } (\bibinfo
  {year} {2021}{\natexlab{a}})},\ \Eprint {http://arxiv.org/abs/2106.05575}
  {arXiv:2106.05575 [gr-qc]} \BibitemShut {NoStop}%
\bibitem [{\citenamefont {Pratten}\ \emph {et~al.}(2020)\citenamefont
  {Pratten}, \citenamefont {Husa}, \citenamefont {Garcia-Quiros}, \citenamefont
  {Colleoni}, \citenamefont {Ramos-Buades}, \citenamefont {Estelles},\ and\
  \citenamefont {Jaume}}]{Pratten:2020fqn}%
  \BibitemOpen
  \bibfield  {author} {\bibinfo {author} {\bibfnamefont {G.}~\bibnamefont
  {Pratten}}, \bibinfo {author} {\bibfnamefont {S.}~\bibnamefont {Husa}},
  \bibinfo {author} {\bibfnamefont {C.}~\bibnamefont {Garcia-Quiros}}, \bibinfo
  {author} {\bibfnamefont {M.}~\bibnamefont {Colleoni}}, \bibinfo {author}
  {\bibfnamefont {A.}~\bibnamefont {Ramos-Buades}}, \bibinfo {author}
  {\bibfnamefont {H.}~\bibnamefont {Estelles}}, \ and\ \bibinfo {author}
  {\bibfnamefont {R.}~\bibnamefont {Jaume}},\ }\href {\doibase
  10.1103/PhysRevD.102.064001} {\bibfield  {journal} {\bibinfo  {journal}
  {Phys. Rev. D}\ }\textbf {\bibinfo {volume} {102}},\ \bibinfo {pages}
  {064001} (\bibinfo {year} {2020})},\ \Eprint
  {http://arxiv.org/abs/2001.11412} {arXiv:2001.11412 [gr-qc]} \BibitemShut
  {NoStop}%
\bibitem [{\citenamefont {Garc\'\i{}a-Quir\'os}\ \emph
  {et~al.}(2020)\citenamefont {Garc\'\i{}a-Quir\'os}, \citenamefont {Colleoni},
  \citenamefont {Husa}, \citenamefont {Estell\'es}, \citenamefont {Pratten},
  \citenamefont {Ramos-Buades}, \citenamefont {Mateu-Lucena},\ and\
  \citenamefont {Jaume}}]{Garcia-Quiros:2020qpx}%
  \BibitemOpen
  \bibfield  {author} {\bibinfo {author} {\bibfnamefont {C.}~\bibnamefont
  {Garc\'\i{}a-Quir\'os}}, \bibinfo {author} {\bibfnamefont {M.}~\bibnamefont
  {Colleoni}}, \bibinfo {author} {\bibfnamefont {S.}~\bibnamefont {Husa}},
  \bibinfo {author} {\bibfnamefont {H.}~\bibnamefont {Estell\'es}}, \bibinfo
  {author} {\bibfnamefont {G.}~\bibnamefont {Pratten}}, \bibinfo {author}
  {\bibfnamefont {A.}~\bibnamefont {Ramos-Buades}}, \bibinfo {author}
  {\bibfnamefont {M.}~\bibnamefont {Mateu-Lucena}}, \ and\ \bibinfo {author}
  {\bibfnamefont {R.}~\bibnamefont {Jaume}},\ }\href {\doibase
  10.1103/PhysRevD.102.064002} {\bibfield  {journal} {\bibinfo  {journal}
  {Phys. Rev. D}\ }\textbf {\bibinfo {volume} {102}},\ \bibinfo {pages}
  {064002} (\bibinfo {year} {2020})},\ \Eprint
  {http://arxiv.org/abs/2001.10914} {arXiv:2001.10914 [gr-qc]} \BibitemShut
  {NoStop}%
\bibitem [{\citenamefont {Pratten}\ \emph {et~al.}(2021)\citenamefont {Pratten}
  \emph {et~al.}}]{Pratten:2020ceb}%
  \BibitemOpen
  \bibfield  {author} {\bibinfo {author} {\bibfnamefont {G.}~\bibnamefont
  {Pratten}} \emph {et~al.},\ }\href {\doibase 10.1103/PhysRevD.103.104056}
  {\bibfield  {journal} {\bibinfo  {journal} {Phys. Rev. D}\ }\textbf {\bibinfo
  {volume} {103}},\ \bibinfo {pages} {104056} (\bibinfo {year} {2021})},\
  \Eprint {http://arxiv.org/abs/2004.06503} {arXiv:2004.06503 [gr-qc]}
  \BibitemShut {NoStop}%
\bibitem [{\citenamefont {Akcay}\ \emph {et~al.}(2021)\citenamefont {Akcay},
  \citenamefont {Gamba},\ and\ \citenamefont {Bernuzzi}}]{Akcay:2020qrj}%
  \BibitemOpen
  \bibfield  {author} {\bibinfo {author} {\bibfnamefont {S.}~\bibnamefont
  {Akcay}}, \bibinfo {author} {\bibfnamefont {R.}~\bibnamefont {Gamba}}, \ and\
  \bibinfo {author} {\bibfnamefont {S.}~\bibnamefont {Bernuzzi}},\ }\href
  {\doibase 10.1103/PhysRevD.103.024014} {\bibfield  {journal} {\bibinfo
  {journal} {Phys. Rev. D}\ }\textbf {\bibinfo {volume} {103}},\ \bibinfo
  {pages} {024014} (\bibinfo {year} {2021})},\ \Eprint
  {http://arxiv.org/abs/2005.05338} {arXiv:2005.05338 [gr-qc]} \BibitemShut
  {NoStop}%
\bibitem [{\citenamefont {Gamba}\ \emph
  {et~al.}(2021{\natexlab{b}})\citenamefont {Gamba}, \citenamefont {Ak\c{c}ay},
  \citenamefont {Bernuzzi},\ and\ \citenamefont {Williams}}]{Gamba:2021ydi}%
  \BibitemOpen
  \bibfield  {author} {\bibinfo {author} {\bibfnamefont {R.}~\bibnamefont
  {Gamba}}, \bibinfo {author} {\bibfnamefont {S.}~\bibnamefont {Ak\c{c}ay}},
  \bibinfo {author} {\bibfnamefont {S.}~\bibnamefont {Bernuzzi}}, \ and\
  \bibinfo {author} {\bibfnamefont {J.}~\bibnamefont {Williams}},\ }\href@noop
  {} {\  (\bibinfo {year} {2021}{\natexlab{b}})},\ \Eprint
  {http://arxiv.org/abs/2111.03675} {arXiv:2111.03675 [gr-qc]} \BibitemShut
  {NoStop}%
\bibitem [{\citenamefont {Rettegno}\ \emph {et~al.}(2019)\citenamefont
  {Rettegno}, \citenamefont {Martinetti}, \citenamefont {Nagar}, \citenamefont
  {Bini}, \citenamefont {Riemenschneider},\ and\ \citenamefont
  {Damour}}]{Rettegno:2019tzh}%
  \BibitemOpen
  \bibfield  {author} {\bibinfo {author} {\bibfnamefont {P.}~\bibnamefont
  {Rettegno}}, \bibinfo {author} {\bibfnamefont {F.}~\bibnamefont
  {Martinetti}}, \bibinfo {author} {\bibfnamefont {A.}~\bibnamefont {Nagar}},
  \bibinfo {author} {\bibfnamefont {D.}~\bibnamefont {Bini}}, \bibinfo {author}
  {\bibfnamefont {G.}~\bibnamefont {Riemenschneider}}, \ and\ \bibinfo {author}
  {\bibfnamefont {T.}~\bibnamefont {Damour}},\ }\href@noop {} {\  (\bibinfo
  {year} {2019})},\ \Eprint {http://arxiv.org/abs/1911.10818} {arXiv:1911.10818
  [gr-qc]} \BibitemShut {NoStop}%
\bibitem [{\citenamefont {Damour}\ \emph {et~al.}(2012)\citenamefont {Damour},
  \citenamefont {Nagar}, \citenamefont {Pollney},\ and\ \citenamefont
  {Reisswig}}]{Damour:2011fu}%
  \BibitemOpen
  \bibfield  {author} {\bibinfo {author} {\bibfnamefont {T.}~\bibnamefont
  {Damour}}, \bibinfo {author} {\bibfnamefont {A.}~\bibnamefont {Nagar}},
  \bibinfo {author} {\bibfnamefont {D.}~\bibnamefont {Pollney}}, \ and\
  \bibinfo {author} {\bibfnamefont {C.}~\bibnamefont {Reisswig}},\ }\href
  {\doibase 10.1103/PhysRevLett.108.131101} {\bibfield  {journal} {\bibinfo
  {journal} {Phys.Rev.Lett.}\ }\textbf {\bibinfo {volume} {108}},\ \bibinfo
  {pages} {131101} (\bibinfo {year} {2012})},\ \Eprint
  {http://arxiv.org/abs/1110.2938} {arXiv:1110.2938 [gr-qc]} \BibitemShut
  {NoStop}%
\bibitem [{\citenamefont {Nagar}\ \emph {et~al.}(2016)\citenamefont {Nagar},
  \citenamefont {Damour}, \citenamefont {Reisswig},\ and\ \citenamefont
  {Pollney}}]{Nagar:2015xqa}%
  \BibitemOpen
  \bibfield  {author} {\bibinfo {author} {\bibfnamefont {A.}~\bibnamefont
  {Nagar}}, \bibinfo {author} {\bibfnamefont {T.}~\bibnamefont {Damour}},
  \bibinfo {author} {\bibfnamefont {C.}~\bibnamefont {Reisswig}}, \ and\
  \bibinfo {author} {\bibfnamefont {D.}~\bibnamefont {Pollney}},\ }\href
  {\doibase 10.1103/PhysRevD.93.044046} {\bibfield  {journal} {\bibinfo
  {journal} {Phys. Rev.}\ }\textbf {\bibinfo {volume} {D93}},\ \bibinfo {pages}
  {044046} (\bibinfo {year} {2016})},\ \Eprint
  {http://arxiv.org/abs/1506.08457} {arXiv:1506.08457 [gr-qc]} \BibitemShut
  {NoStop}%
\bibitem [{\citenamefont {Ossokine}\ \emph {et~al.}(2018)\citenamefont
  {Ossokine}, \citenamefont {Dietrich}, \citenamefont {Foley}, \citenamefont
  {Katebi},\ and\ \citenamefont {Lovelace}}]{Ossokine:2017dge}%
  \BibitemOpen
  \bibfield  {author} {\bibinfo {author} {\bibfnamefont {S.}~\bibnamefont
  {Ossokine}}, \bibinfo {author} {\bibfnamefont {T.}~\bibnamefont {Dietrich}},
  \bibinfo {author} {\bibfnamefont {E.}~\bibnamefont {Foley}}, \bibinfo
  {author} {\bibfnamefont {R.}~\bibnamefont {Katebi}}, \ and\ \bibinfo {author}
  {\bibfnamefont {G.}~\bibnamefont {Lovelace}},\ }\href {\doibase
  10.1103/PhysRevD.98.104057} {\bibfield  {journal} {\bibinfo  {journal} {Phys.
  Rev.}\ }\textbf {\bibinfo {volume} {D98}},\ \bibinfo {pages} {104057}
  (\bibinfo {year} {2018})},\ \Eprint {http://arxiv.org/abs/1712.06533}
  {arXiv:1712.06533 [gr-qc]} \BibitemShut {NoStop}%
\bibitem [{\citenamefont {Le~Tiec}\ \emph {et~al.}(2011)\citenamefont
  {Le~Tiec}, \citenamefont {Mroue}, \citenamefont {Barack}, \citenamefont
  {Buonanno}, \citenamefont {Pfeiffer}, \citenamefont {Sago},\ and\
  \citenamefont {Taracchini}}]{LeTiec:2011bk}%
  \BibitemOpen
  \bibfield  {author} {\bibinfo {author} {\bibfnamefont {A.}~\bibnamefont
  {Le~Tiec}}, \bibinfo {author} {\bibfnamefont {A.~H.}\ \bibnamefont {Mroue}},
  \bibinfo {author} {\bibfnamefont {L.}~\bibnamefont {Barack}}, \bibinfo
  {author} {\bibfnamefont {A.}~\bibnamefont {Buonanno}}, \bibinfo {author}
  {\bibfnamefont {H.~P.}\ \bibnamefont {Pfeiffer}}, \bibinfo {author}
  {\bibfnamefont {N.}~\bibnamefont {Sago}}, \ and\ \bibinfo {author}
  {\bibfnamefont {A.}~\bibnamefont {Taracchini}},\ }\href {\doibase
  10.1103/PhysRevLett.107.141101} {\bibfield  {journal} {\bibinfo  {journal}
  {Phys. Rev. Lett.}\ }\textbf {\bibinfo {volume} {107}},\ \bibinfo {pages}
  {141101} (\bibinfo {year} {2011})},\ \Eprint {http://arxiv.org/abs/1106.3278}
  {arXiv:1106.3278 [gr-qc]} \BibitemShut {NoStop}%
\bibitem [{\citenamefont {Le~Tiec}\ \emph {et~al.}(2013)\citenamefont {Le~Tiec}
  \emph {et~al.}}]{LeTiec:2013uey}%
  \BibitemOpen
  \bibfield  {author} {\bibinfo {author} {\bibfnamefont {A.}~\bibnamefont
  {Le~Tiec}} \emph {et~al.},\ }\href {\doibase 10.1103/PhysRevD.88.124027}
  {\bibfield  {journal} {\bibinfo  {journal} {Phys. Rev. D}\ }\textbf {\bibinfo
  {volume} {88}},\ \bibinfo {pages} {124027} (\bibinfo {year} {2013})},\
  \Eprint {http://arxiv.org/abs/1309.0541} {arXiv:1309.0541 [gr-qc]}
  \BibitemShut {NoStop}%
\bibitem [{\citenamefont {Hinderer}\ \emph {et~al.}(2013)\citenamefont
  {Hinderer} \emph {et~al.}}]{Hinderer:2013uwa}%
  \BibitemOpen
  \bibfield  {author} {\bibinfo {author} {\bibfnamefont {T.}~\bibnamefont
  {Hinderer}} \emph {et~al.},\ }\href {\doibase 10.1103/PhysRevD.88.084005}
  {\bibfield  {journal} {\bibinfo  {journal} {Phys. Rev.}\ }\textbf {\bibinfo
  {volume} {D88}},\ \bibinfo {pages} {084005} (\bibinfo {year} {2013})},\
  \Eprint {http://arxiv.org/abs/1309.0544} {arXiv:1309.0544 [gr-qc]}
  \BibitemShut {NoStop}%
\bibitem [{\citenamefont {Boyle}\ \emph {et~al.}(2008)\citenamefont {Boyle},
  \citenamefont {Buonanno}, \citenamefont {Kidder}, \citenamefont {Mroue},
  \citenamefont {Pan} \emph {et~al.}}]{Boyle:2008ge}%
  \BibitemOpen
  \bibfield  {author} {\bibinfo {author} {\bibfnamefont {M.}~\bibnamefont
  {Boyle}}, \bibinfo {author} {\bibfnamefont {A.}~\bibnamefont {Buonanno}},
  \bibinfo {author} {\bibfnamefont {L.~E.}\ \bibnamefont {Kidder}}, \bibinfo
  {author} {\bibfnamefont {A.~H.}\ \bibnamefont {Mroue}}, \bibinfo {author}
  {\bibfnamefont {Y.}~\bibnamefont {Pan}},  \emph {et~al.},\ }\href {\doibase
  10.1103/PhysRevD.78.104020} {\bibfield  {journal} {\bibinfo  {journal}
  {Phys.Rev.}\ }\textbf {\bibinfo {volume} {D78}},\ \bibinfo {pages} {104020}
  (\bibinfo {year} {2008})},\ \Eprint {http://arxiv.org/abs/0804.4184}
  {arXiv:0804.4184 [gr-qc]} \BibitemShut {NoStop}%
\bibitem [{\citenamefont {Boyle}\ \emph {et~al.}(2019)\citenamefont {Boyle}
  \emph {et~al.}}]{Boyle:2019kee}%
  \BibitemOpen
  \bibfield  {author} {\bibinfo {author} {\bibfnamefont {M.}~\bibnamefont
  {Boyle}} \emph {et~al.},\ }\href {\doibase 10.1088/1361-6382/ab34e2}
  {\bibfield  {journal} {\bibinfo  {journal} {Class. Quant. Grav.}\ }\textbf
  {\bibinfo {volume} {36}},\ \bibinfo {pages} {195006} (\bibinfo {year}
  {2019})},\ \Eprint {http://arxiv.org/abs/1904.04831} {arXiv:1904.04831
  [gr-qc]} \BibitemShut {NoStop}%
\bibitem [{\citenamefont {Reitze}\ \emph {et~al.}(2021)\citenamefont {Reitze}
  \emph {et~al.}}]{Reitze:2021gzo}%
  \BibitemOpen
  \bibfield  {author} {\bibinfo {author} {\bibfnamefont {D.}~\bibnamefont
  {Reitze}} \emph {et~al.},\ }\href@noop {} {\  (\bibinfo {year} {2021})},\
  \Eprint {http://arxiv.org/abs/2111.06986} {arXiv:2111.06986 [gr-qc]}
  \BibitemShut {NoStop}%
\bibitem [{\citenamefont {Couvares}\ \emph {et~al.}(2021)\citenamefont
  {Couvares} \emph {et~al.}}]{Couvares:2021ajn}%
  \BibitemOpen
  \bibfield  {author} {\bibinfo {author} {\bibfnamefont {P.}~\bibnamefont
  {Couvares}} \emph {et~al.},\ }\href@noop {} {\  (\bibinfo {year} {2021})},\
  \Eprint {http://arxiv.org/abs/2111.06987} {arXiv:2111.06987 [gr-qc]}
  \BibitemShut {NoStop}%
\bibitem [{\citenamefont {Punturo}\ \emph {et~al.}(2021)\citenamefont {Punturo}
  \emph {et~al.}}]{Punturo:2021ryo}%
  \BibitemOpen
  \bibfield  {author} {\bibinfo {author} {\bibfnamefont {M.}~\bibnamefont
  {Punturo}} \emph {et~al.},\ }\href@noop {} {\  (\bibinfo {year} {2021})},\
  \Eprint {http://arxiv.org/abs/2111.06988} {arXiv:2111.06988 [gr-qc]}
  \BibitemShut {NoStop}%
\bibitem [{\citenamefont {Katsanevas}\ \emph {et~al.}(2021)\citenamefont
  {Katsanevas} \emph {et~al.}}]{Katsanevas:2021fzj}%
  \BibitemOpen
  \bibfield  {author} {\bibinfo {author} {\bibfnamefont {S.}~\bibnamefont
  {Katsanevas}} \emph {et~al.},\ }\href@noop {} {\  (\bibinfo {year} {2021})},\
  \Eprint {http://arxiv.org/abs/2111.06989} {arXiv:2111.06989 [gr-qc]}
  \BibitemShut {NoStop}%
\bibitem [{\citenamefont {Kalogera}\ \emph {et~al.}(2021)\citenamefont
  {Kalogera} \emph {et~al.}}]{Kalogera:2021bya}%
  \BibitemOpen
  \bibfield  {author} {\bibinfo {author} {\bibfnamefont {V.}~\bibnamefont
  {Kalogera}} \emph {et~al.},\ }\href@noop {} {\  (\bibinfo {year} {2021})},\
  \Eprint {http://arxiv.org/abs/2111.06990} {arXiv:2111.06990 [gr-qc]}
  \BibitemShut {NoStop}%
\bibitem [{\citenamefont {McClelland}\ \emph {et~al.}(2021)\citenamefont
  {McClelland} \emph {et~al.}}]{McClelland:2021wqy}%
  \BibitemOpen
  \bibfield  {author} {\bibinfo {author} {\bibfnamefont {D.}~\bibnamefont
  {McClelland}} \emph {et~al.},\ }\href@noop {} {\  (\bibinfo {year} {2021})},\
  \Eprint {http://arxiv.org/abs/2111.06991} {arXiv:2111.06991 [gr-qc]}
  \BibitemShut {NoStop}%
\bibitem [{\citenamefont {Damour}\ and\ \citenamefont
  {Nagar}(2014)}]{Damour:2014sva}%
  \BibitemOpen
  \bibfield  {author} {\bibinfo {author} {\bibfnamefont {T.}~\bibnamefont
  {Damour}}\ and\ \bibinfo {author} {\bibfnamefont {A.}~\bibnamefont {Nagar}},\
  }\href {\doibase 10.1103/PhysRevD.90.044018} {\bibfield  {journal} {\bibinfo
  {journal} {Phys.Rev.}\ }\textbf {\bibinfo {volume} {D90}},\ \bibinfo {pages}
  {044018} (\bibinfo {year} {2014})},\ \Eprint {http://arxiv.org/abs/1406.6913}
  {arXiv:1406.6913 [gr-qc]} \BibitemShut {NoStop}%
\bibitem [{aLI()}]{aLIGODesign_PSD}%
  \BibitemOpen
  \href@noop {} {\enquote {\bibinfo {title} {{Updated Advanced LIGO sensitivity
  design curve}},}\ }\bibinfo {howpublished}
  {\url{https://dcc.ligo.org/LIGO-T1800044/public}}\BibitemShut {NoStop}%
\bibitem [{\citenamefont {Hild}\ \emph {et~al.}(2010)\citenamefont {Hild},
  \citenamefont {Chelkowski}, \citenamefont {Freise}, \citenamefont {Franc},
  \citenamefont {Morgado}, \citenamefont {Flaminio},\ and\ \citenamefont
  {DeSalvo}}]{Hild:2009ns}%
  \BibitemOpen
  \bibfield  {author} {\bibinfo {author} {\bibfnamefont {S.}~\bibnamefont
  {Hild}}, \bibinfo {author} {\bibfnamefont {S.}~\bibnamefont {Chelkowski}},
  \bibinfo {author} {\bibfnamefont {A.}~\bibnamefont {Freise}}, \bibinfo
  {author} {\bibfnamefont {J.}~\bibnamefont {Franc}}, \bibinfo {author}
  {\bibfnamefont {N.}~\bibnamefont {Morgado}}, \bibinfo {author} {\bibfnamefont
  {R.}~\bibnamefont {Flaminio}}, \ and\ \bibinfo {author} {\bibfnamefont
  {R.}~\bibnamefont {DeSalvo}},\ }\href {\doibase
  10.1088/0264-9381/27/1/015003} {\bibfield  {journal} {\bibinfo  {journal}
  {Class. Quant. Grav.}\ }\textbf {\bibinfo {volume} {27}},\ \bibinfo {pages}
  {015003} (\bibinfo {year} {2010})},\ \Eprint {http://arxiv.org/abs/0906.2655}
  {arXiv:0906.2655 [gr-qc]} \BibitemShut {NoStop}%
\bibitem [{\citenamefont {Hild}\ \emph {et~al.}(2011)\citenamefont {Hild} \emph
  {et~al.}}]{Hild:2010id}%
  \BibitemOpen
  \bibfield  {author} {\bibinfo {author} {\bibfnamefont {S.}~\bibnamefont
  {Hild}} \emph {et~al.},\ }\href {\doibase 10.1088/0264-9381/28/9/094013}
  {\bibfield  {journal} {\bibinfo  {journal} {Class. Quant. Grav.}\ }\textbf
  {\bibinfo {volume} {28}},\ \bibinfo {pages} {094013} (\bibinfo {year}
  {2011})},\ \Eprint {http://arxiv.org/abs/1012.0908} {arXiv:1012.0908 [gr-qc]}
  \BibitemShut {NoStop}%
\bibitem [{\citenamefont {Evans}\ \emph {et~al.}(2021)\citenamefont {Evans}
  \emph {et~al.}}]{Evans:2021gyd}%
  \BibitemOpen
  \bibfield  {author} {\bibinfo {author} {\bibfnamefont {M.}~\bibnamefont
  {Evans}} \emph {et~al.},\ }\href@noop {} {\  (\bibinfo {year} {2021})},\
  \Eprint {http://arxiv.org/abs/2109.09882} {arXiv:2109.09882 [astro-ph.IM]}
  \BibitemShut {NoStop}%
\bibitem [{\citenamefont {Boh{\'e}}\ \emph {et~al.}(2017)\citenamefont
  {Boh{\'e}} \emph {et~al.}}]{Bohe:2016gbl}%
  \BibitemOpen
  \bibfield  {author} {\bibinfo {author} {\bibfnamefont {A.}~\bibnamefont
  {Boh{\'e}}} \emph {et~al.},\ }\href {\doibase 10.1103/PhysRevD.95.044028}
  {\bibfield  {journal} {\bibinfo  {journal} {Phys. Rev.}\ }\textbf {\bibinfo
  {volume} {D95}},\ \bibinfo {pages} {044028} (\bibinfo {year} {2017})},\
  \Eprint {http://arxiv.org/abs/1611.03703} {arXiv:1611.03703 [gr-qc]}
  \BibitemShut {NoStop}%
\bibitem [{\citenamefont {Cotesta}\ \emph {et~al.}(2020)\citenamefont
  {Cotesta}, \citenamefont {Marsat},\ and\ \citenamefont
  {P\"urrer}}]{Cotesta:2020qhw}%
  \BibitemOpen
  \bibfield  {author} {\bibinfo {author} {\bibfnamefont {R.}~\bibnamefont
  {Cotesta}}, \bibinfo {author} {\bibfnamefont {S.}~\bibnamefont {Marsat}}, \
  and\ \bibinfo {author} {\bibfnamefont {M.}~\bibnamefont {P\"urrer}},\ }\href
  {\doibase 10.1103/PhysRevD.101.124040} {\bibfield  {journal} {\bibinfo
  {journal} {Phys. Rev. D}\ }\textbf {\bibinfo {volume} {101}},\ \bibinfo
  {pages} {124040} (\bibinfo {year} {2020})},\ \Eprint
  {http://arxiv.org/abs/2003.12079} {arXiv:2003.12079 [gr-qc]} \BibitemShut
  {NoStop}%
\bibitem [{\citenamefont {Mitman}\ \emph {et~al.}(2021)\citenamefont {Mitman}
  \emph {et~al.}}]{Mitman:2021xkq}%
  \BibitemOpen
  \bibfield  {author} {\bibinfo {author} {\bibfnamefont {K.}~\bibnamefont
  {Mitman}} \emph {et~al.},\ }\href {\doibase 10.1103/PhysRevD.104.024051}
  {\bibfield  {journal} {\bibinfo  {journal} {Phys. Rev. D}\ }\textbf {\bibinfo
  {volume} {104}},\ \bibinfo {pages} {024051} (\bibinfo {year} {2021})},\
  \Eprint {http://arxiv.org/abs/2105.02300} {arXiv:2105.02300 [gr-qc]}
  \BibitemShut {NoStop}%
\bibitem [{\citenamefont {Campanelli}\ \emph {et~al.}(2006)\citenamefont
  {Campanelli}, \citenamefont {Lousto},\ and\ \citenamefont
  {Zlochower}}]{Campanelli:2006uy}%
  \BibitemOpen
  \bibfield  {author} {\bibinfo {author} {\bibfnamefont {M.}~\bibnamefont
  {Campanelli}}, \bibinfo {author} {\bibfnamefont {C.}~\bibnamefont {Lousto}},
  \ and\ \bibinfo {author} {\bibfnamefont {Y.}~\bibnamefont {Zlochower}},\
  }\href {\doibase 10.1103/PhysRevD.74.041501} {\bibfield  {journal} {\bibinfo
  {journal} {Phys.Rev.}\ }\textbf {\bibinfo {volume} {D74}},\ \bibinfo {pages}
  {041501} (\bibinfo {year} {2006})},\ \Eprint
  {http://arxiv.org/abs/gr-qc/0604012} {arXiv:gr-qc/0604012 [gr-qc]}
  \BibitemShut {NoStop}%
\bibitem [{\citenamefont {Bini}\ and\ \citenamefont
  {Damour}(2012)}]{Bini:2012ji}%
  \BibitemOpen
  \bibfield  {author} {\bibinfo {author} {\bibfnamefont {D.}~\bibnamefont
  {Bini}}\ and\ \bibinfo {author} {\bibfnamefont {T.}~\bibnamefont {Damour}},\
  }\href {\doibase 10.1103/PhysRevD.86.124012} {\bibfield  {journal} {\bibinfo
  {journal} {Phys.Rev.}\ }\textbf {\bibinfo {volume} {D86}},\ \bibinfo {pages}
  {124012} (\bibinfo {year} {2012})},\ \Eprint {http://arxiv.org/abs/1210.2834}
  {arXiv:1210.2834 [gr-qc]} \BibitemShut {NoStop}%
\bibitem [{\citenamefont {Damour}\ and\ \citenamefont
  {Gopakumar}(2006)}]{Damour:2006tr}%
  \BibitemOpen
  \bibfield  {author} {\bibinfo {author} {\bibfnamefont {T.}~\bibnamefont
  {Damour}}\ and\ \bibinfo {author} {\bibfnamefont {A.}~\bibnamefont
  {Gopakumar}},\ }\href {\doibase 10.1103/PhysRevD.73.124006} {\bibfield
  {journal} {\bibinfo  {journal} {Phys. Rev.}\ }\textbf {\bibinfo {volume}
  {D73}},\ \bibinfo {pages} {124006} (\bibinfo {year} {2006})},\ \Eprint
  {http://arxiv.org/abs/gr-qc/0602117} {arXiv:gr-qc/0602117} \BibitemShut
  {NoStop}%
\bibitem [{\citenamefont {Damour}\ and\ \citenamefont
  {Nagar}(2007)}]{Damour:2007xr}%
  \BibitemOpen
  \bibfield  {author} {\bibinfo {author} {\bibfnamefont {T.}~\bibnamefont
  {Damour}}\ and\ \bibinfo {author} {\bibfnamefont {A.}~\bibnamefont {Nagar}},\
  }\href {\doibase 10.1103/PhysRevD.76.064028} {\bibfield  {journal} {\bibinfo
  {journal} {Phys. Rev.}\ }\textbf {\bibinfo {volume} {D76}},\ \bibinfo {pages}
  {064028} (\bibinfo {year} {2007})},\ \Eprint {http://arxiv.org/abs/0705.2519}
  {arXiv:0705.2519 [gr-qc]} \BibitemShut {NoStop}%
\bibitem [{\citenamefont {Damour}\ \emph {et~al.}(2013)\citenamefont {Damour},
  \citenamefont {Nagar},\ and\ \citenamefont {Bernuzzi}}]{Damour:2012ky}%
  \BibitemOpen
  \bibfield  {author} {\bibinfo {author} {\bibfnamefont {T.}~\bibnamefont
  {Damour}}, \bibinfo {author} {\bibfnamefont {A.}~\bibnamefont {Nagar}}, \
  and\ \bibinfo {author} {\bibfnamefont {S.}~\bibnamefont {Bernuzzi}},\ }\href
  {\doibase 10.1103/PhysRevD.87.084035} {\bibfield  {journal} {\bibinfo
  {journal} {Phys.Rev.}\ }\textbf {\bibinfo {volume} {D87}},\ \bibinfo {pages}
  {084035} (\bibinfo {year} {2013})},\ \Eprint {http://arxiv.org/abs/1212.4357}
  {arXiv:1212.4357 [gr-qc]} \BibitemShut {NoStop}%
\bibitem [{\citenamefont {Damour}\ \emph {et~al.}(2009)\citenamefont {Damour},
  \citenamefont {Iyer},\ and\ \citenamefont {Nagar}}]{Damour:2008gu}%
  \BibitemOpen
  \bibfield  {author} {\bibinfo {author} {\bibfnamefont {T.}~\bibnamefont
  {Damour}}, \bibinfo {author} {\bibfnamefont {B.~R.}\ \bibnamefont {Iyer}}, \
  and\ \bibinfo {author} {\bibfnamefont {A.}~\bibnamefont {Nagar}},\ }\href
  {\doibase 10.1103/PhysRevD.79.064004} {\bibfield  {journal} {\bibinfo
  {journal} {Phys. Rev.}\ }\textbf {\bibinfo {volume} {D79}},\ \bibinfo {pages}
  {064004} (\bibinfo {year} {2009})},\ \Eprint {http://arxiv.org/abs/0811.2069}
  {arXiv:0811.2069 [gr-qc]} \BibitemShut {NoStop}%
\bibitem [{\citenamefont {Nagar}\ \emph {et~al.}(2017)\citenamefont {Nagar},
  \citenamefont {Riemenschneider},\ and\ \citenamefont
  {Pratten}}]{Nagar:2017jdw}%
  \BibitemOpen
  \bibfield  {author} {\bibinfo {author} {\bibfnamefont {A.}~\bibnamefont
  {Nagar}}, \bibinfo {author} {\bibfnamefont {G.}~\bibnamefont
  {Riemenschneider}}, \ and\ \bibinfo {author} {\bibfnamefont {G.}~\bibnamefont
  {Pratten}},\ }\href {\doibase 10.1103/PhysRevD.96.084045} {\bibfield
  {journal} {\bibinfo  {journal} {Phys. Rev.}\ }\textbf {\bibinfo {volume}
  {D96}},\ \bibinfo {pages} {084045} (\bibinfo {year} {2017})},\ \Eprint
  {http://arxiv.org/abs/1703.06814} {arXiv:1703.06814 [gr-qc]} \BibitemShut
  {NoStop}%
\bibitem [{\citenamefont {Riemenschneider}\ \emph {et~al.}(2021)\citenamefont
  {Riemenschneider}, \citenamefont {Rettegno}, \citenamefont {Breschi},
  \citenamefont {Albertini}, \citenamefont {Gamba}, \citenamefont {Bernuzzi},\
  and\ \citenamefont {Nagar}}]{Riemenschneider:2021ppj}%
  \BibitemOpen
  \bibfield  {author} {\bibinfo {author} {\bibfnamefont {G.}~\bibnamefont
  {Riemenschneider}}, \bibinfo {author} {\bibfnamefont {P.}~\bibnamefont
  {Rettegno}}, \bibinfo {author} {\bibfnamefont {M.}~\bibnamefont {Breschi}},
  \bibinfo {author} {\bibfnamefont {A.}~\bibnamefont {Albertini}}, \bibinfo
  {author} {\bibfnamefont {R.}~\bibnamefont {Gamba}}, \bibinfo {author}
  {\bibfnamefont {S.}~\bibnamefont {Bernuzzi}}, \ and\ \bibinfo {author}
  {\bibfnamefont {A.}~\bibnamefont {Nagar}},\ }\href {\doibase
  10.1103/PhysRevD.104.104045} {\bibfield  {journal} {\bibinfo  {journal}
  {Phys. Rev. D}\ }\textbf {\bibinfo {volume} {104}},\ \bibinfo {pages}
  {104045} (\bibinfo {year} {2021})},\ \Eprint
  {http://arxiv.org/abs/2104.07533} {arXiv:2104.07533 [gr-qc]} \BibitemShut
  {NoStop}%
\bibitem [{\citenamefont {Damour}\ and\ \citenamefont
  {Nagar}(2009)}]{Damour:2009kr}%
  \BibitemOpen
  \bibfield  {author} {\bibinfo {author} {\bibfnamefont {T.}~\bibnamefont
  {Damour}}\ and\ \bibinfo {author} {\bibfnamefont {A.}~\bibnamefont {Nagar}},\
  }\href {\doibase 10.1103/PhysRevD.79.081503} {\bibfield  {journal} {\bibinfo
  {journal} {Phys. Rev.}\ }\textbf {\bibinfo {volume} {D79}},\ \bibinfo {pages}
  {081503} (\bibinfo {year} {2009})},\ \Eprint {http://arxiv.org/abs/0902.0136}
  {arXiv:0902.0136 [gr-qc]} \BibitemShut {NoStop}%
\bibitem [{\citenamefont {P\"urrer}\ and\ \citenamefont
  {Haster}(2020)}]{Purrer:2019jcp}%
  \BibitemOpen
  \bibfield  {author} {\bibinfo {author} {\bibfnamefont {M.}~\bibnamefont
  {P\"urrer}}\ and\ \bibinfo {author} {\bibfnamefont {C.-J.}\ \bibnamefont
  {Haster}},\ }\href {\doibase 10.1103/PhysRevResearch.2.023151} {\bibfield
  {journal} {\bibinfo  {journal} {Phys. Rev. Res.}\ }\textbf {\bibinfo {volume}
  {2}},\ \bibinfo {pages} {023151} (\bibinfo {year} {2020})},\ \Eprint
  {http://arxiv.org/abs/1912.10055} {arXiv:1912.10055 [gr-qc]} \BibitemShut
  {NoStop}%
\bibitem [{\citenamefont {Moxon}\ \emph {et~al.}(2021)\citenamefont {Moxon},
  \citenamefont {Scheel}, \citenamefont {Teukolsky}, \citenamefont {Deppe},
  \citenamefont {Fischer}, \citenamefont {H\'ebert}, \citenamefont {Kidder},\
  and\ \citenamefont {Throwe}}]{Moxon:2021gbv}%
  \BibitemOpen
  \bibfield  {author} {\bibinfo {author} {\bibfnamefont {J.}~\bibnamefont
  {Moxon}}, \bibinfo {author} {\bibfnamefont {M.~A.}\ \bibnamefont {Scheel}},
  \bibinfo {author} {\bibfnamefont {S.~A.}\ \bibnamefont {Teukolsky}}, \bibinfo
  {author} {\bibfnamefont {N.}~\bibnamefont {Deppe}}, \bibinfo {author}
  {\bibfnamefont {N.}~\bibnamefont {Fischer}}, \bibinfo {author} {\bibfnamefont
  {F.}~\bibnamefont {H\'ebert}}, \bibinfo {author} {\bibfnamefont {L.~E.}\
  \bibnamefont {Kidder}}, \ and\ \bibinfo {author} {\bibfnamefont
  {W.}~\bibnamefont {Throwe}},\ }\href@noop {} {\  (\bibinfo {year} {2021})},\
  \Eprint {http://arxiv.org/abs/2110.08635} {arXiv:2110.08635 [gr-qc]}
  \BibitemShut {NoStop}%
\bibitem [{\citenamefont {Fischer}\ \emph {et~al.}(2021)\citenamefont {Fischer}
  \emph {et~al.}}]{Fischer:2021qbh}%
  \BibitemOpen
  \bibfield  {author} {\bibinfo {author} {\bibfnamefont {N.~L.}\ \bibnamefont
  {Fischer}} \emph {et~al.},\ }\href@noop {} {\  (\bibinfo {year} {2021})},\
  \Eprint {http://arxiv.org/abs/2111.06767} {arXiv:2111.06767 [gr-qc]}
  \BibitemShut {NoStop}%
\bibitem [{\citenamefont {Zertuche}\ \emph {et~al.}(2021)\citenamefont
  {Zertuche} \emph {et~al.}}]{Zertuche:2021xkb}%
  \BibitemOpen
  \bibfield  {author} {\bibinfo {author} {\bibfnamefont {L.~M.~n.}\
  \bibnamefont {Zertuche}} \emph {et~al.},\ }\href@noop {} {\  (\bibinfo {year}
  {2021})},\ \Eprint {http://arxiv.org/abs/2110.15922} {arXiv:2110.15922
  [gr-qc]} \BibitemShut {NoStop}%
\bibitem [{\citenamefont {Albanesi}\ \emph {et~al.}(2021)\citenamefont
  {Albanesi}, \citenamefont {Nagar},\ and\ \citenamefont
  {Bernuzzi}}]{Albanesi:2021rby}%
  \BibitemOpen
  \bibfield  {author} {\bibinfo {author} {\bibfnamefont {S.}~\bibnamefont
  {Albanesi}}, \bibinfo {author} {\bibfnamefont {A.}~\bibnamefont {Nagar}}, \
  and\ \bibinfo {author} {\bibfnamefont {S.}~\bibnamefont {Bernuzzi}},\ }\href
  {\doibase 10.1103/PhysRevD.104.024067} {\bibfield  {journal} {\bibinfo
  {journal} {Phys. Rev. D}\ }\textbf {\bibinfo {volume} {104}},\ \bibinfo
  {pages} {024067} (\bibinfo {year} {2021})},\ \Eprint
  {http://arxiv.org/abs/2104.10559} {arXiv:2104.10559 [gr-qc]} \BibitemShut
  {NoStop}%
\bibitem [{\citenamefont {Harms}\ \emph {et~al.}(2014)\citenamefont {Harms},
  \citenamefont {Bernuzzi}, \citenamefont {Nagar},\ and\ \citenamefont
  {Zenginoglu}}]{Harms:2014dqa}%
  \BibitemOpen
  \bibfield  {author} {\bibinfo {author} {\bibfnamefont {E.}~\bibnamefont
  {Harms}}, \bibinfo {author} {\bibfnamefont {S.}~\bibnamefont {Bernuzzi}},
  \bibinfo {author} {\bibfnamefont {A.}~\bibnamefont {Nagar}}, \ and\ \bibinfo
  {author} {\bibfnamefont {A.}~\bibnamefont {Zenginoglu}},\ }\href {\doibase
  10.1088/0264-9381/31/24/245004} {\bibfield  {journal} {\bibinfo  {journal}
  {Class.Quant.Grav.}\ }\textbf {\bibinfo {volume} {31}},\ \bibinfo {pages}
  {245004} (\bibinfo {year} {2014})},\ \Eprint {http://arxiv.org/abs/1406.5983}
  {arXiv:1406.5983 [gr-qc]} \BibitemShut {NoStop}%
\bibitem [{\citenamefont {Barausse}\ \emph {et~al.}(2012)\citenamefont
  {Barausse}, \citenamefont {Buonanno}, \citenamefont {Hughes}, \citenamefont
  {Khanna}, \citenamefont {O'Sullivan} \emph {et~al.}}]{Barausse:2011kb}%
  \BibitemOpen
  \bibfield  {author} {\bibinfo {author} {\bibfnamefont {E.}~\bibnamefont
  {Barausse}}, \bibinfo {author} {\bibfnamefont {A.}~\bibnamefont {Buonanno}},
  \bibinfo {author} {\bibfnamefont {S.~A.}\ \bibnamefont {Hughes}}, \bibinfo
  {author} {\bibfnamefont {G.}~\bibnamefont {Khanna}}, \bibinfo {author}
  {\bibfnamefont {S.}~\bibnamefont {O'Sullivan}},  \emph {et~al.},\ }\href
  {\doibase 10.1103/PhysRevD.85.024046} {\bibfield  {journal} {\bibinfo
  {journal} {Phys.Rev.}\ }\textbf {\bibinfo {volume} {D85}},\ \bibinfo {pages}
  {024046} (\bibinfo {year} {2012})},\ \Eprint {http://arxiv.org/abs/1110.3081}
  {arXiv:1110.3081 [gr-qc]} \BibitemShut {NoStop}%
\bibitem [{\citenamefont {Taracchini}\ \emph {et~al.}(2013)\citenamefont
  {Taracchini}, \citenamefont {Buonanno}, \citenamefont {Hughes},\ and\
  \citenamefont {Khanna}}]{Taracchini:2013wfa}%
  \BibitemOpen
  \bibfield  {author} {\bibinfo {author} {\bibfnamefont {A.}~\bibnamefont
  {Taracchini}}, \bibinfo {author} {\bibfnamefont {A.}~\bibnamefont
  {Buonanno}}, \bibinfo {author} {\bibfnamefont {S.~A.}\ \bibnamefont
  {Hughes}}, \ and\ \bibinfo {author} {\bibfnamefont {G.}~\bibnamefont
  {Khanna}},\ }\href {\doibase 10.1103/PhysRevD.88.044001} {\bibfield
  {journal} {\bibinfo  {journal} {Phys.Rev.}\ }\textbf {\bibinfo {volume}
  {D88}},\ \bibinfo {pages} {044001} (\bibinfo {year} {2013})},\ \Eprint
  {http://arxiv.org/abs/1305.2184} {arXiv:1305.2184 [gr-qc]} \BibitemShut
  {NoStop}%
\bibitem [{\citenamefont {Mihaylov}\ \emph {et~al.}(2021)\citenamefont
  {Mihaylov}, \citenamefont {Ossokine}, \citenamefont {Buonanno},\ and\
  \citenamefont {Ghosh}}]{Mihaylov:2021bpf}%
  \BibitemOpen
  \bibfield  {author} {\bibinfo {author} {\bibfnamefont {D.~P.}\ \bibnamefont
  {Mihaylov}}, \bibinfo {author} {\bibfnamefont {S.}~\bibnamefont {Ossokine}},
  \bibinfo {author} {\bibfnamefont {A.}~\bibnamefont {Buonanno}}, \ and\
  \bibinfo {author} {\bibfnamefont {A.}~\bibnamefont {Ghosh}},\ }\href@noop {}
  {\  (\bibinfo {year} {2021})},\ \Eprint {http://arxiv.org/abs/2105.06983}
  {arXiv:2105.06983 [gr-qc]} \BibitemShut {NoStop}%
\bibitem [{\citenamefont {Barausse}\ \emph {et~al.}(2009)\citenamefont
  {Barausse}, \citenamefont {Racine},\ and\ \citenamefont
  {Buonanno}}]{Barausse:2009aa}%
  \BibitemOpen
  \bibfield  {author} {\bibinfo {author} {\bibfnamefont {E.}~\bibnamefont
  {Barausse}}, \bibinfo {author} {\bibfnamefont {E.}~\bibnamefont {Racine}}, \
  and\ \bibinfo {author} {\bibfnamefont {A.}~\bibnamefont {Buonanno}},\ }\href
  {\doibase 10.1103/PhysRevD.80.104025} {\bibfield  {journal} {\bibinfo
  {journal} {Phys. Rev.}\ }\textbf {\bibinfo {volume} {D80}},\ \bibinfo {pages}
  {104025} (\bibinfo {year} {2009})},\ \Eprint {http://arxiv.org/abs/0907.4745}
  {arXiv:0907.4745 [gr-qc]} \BibitemShut {NoStop}%
\bibitem [{\citenamefont {Barausse}\ and\ \citenamefont
  {Buonanno}(2010)}]{Barausse:2009xi}%
  \BibitemOpen
  \bibfield  {author} {\bibinfo {author} {\bibfnamefont {E.}~\bibnamefont
  {Barausse}}\ and\ \bibinfo {author} {\bibfnamefont {A.}~\bibnamefont
  {Buonanno}},\ }\href {\doibase 10.1103/PhysRevD.81.084024} {\bibfield
  {journal} {\bibinfo  {journal} {Phys.Rev.}\ }\textbf {\bibinfo {volume}
  {D81}},\ \bibinfo {pages} {084024} (\bibinfo {year} {2010})},\ \Eprint
  {http://arxiv.org/abs/0912.3517} {arXiv:0912.3517 [gr-qc]} \BibitemShut
  {NoStop}%
\bibitem [{\citenamefont {Barausse}\ and\ \citenamefont
  {Buonanno}(2011)}]{Barausse:2011ys}%
  \BibitemOpen
  \bibfield  {author} {\bibinfo {author} {\bibfnamefont {E.}~\bibnamefont
  {Barausse}}\ and\ \bibinfo {author} {\bibfnamefont {A.}~\bibnamefont
  {Buonanno}},\ }\href {\doibase 10.1103/PhysRevD.84.104027} {\bibfield
  {journal} {\bibinfo  {journal} {Phys.Rev.}\ }\textbf {\bibinfo {volume}
  {D84}},\ \bibinfo {pages} {104027} (\bibinfo {year} {2011})},\ \Eprint
  {http://arxiv.org/abs/1107.2904} {arXiv:1107.2904 [gr-qc]} \BibitemShut
  {NoStop}%
\bibitem [{\citenamefont {Varma}\ \emph {et~al.}(2019)\citenamefont {Varma},
  \citenamefont {Field}, \citenamefont {Scheel}, \citenamefont {Blackman},
  \citenamefont {Gerosa}, \citenamefont {Stein}, \citenamefont {Kidder},\ and\
  \citenamefont {Pfeiffer}}]{Varma:2019csw}%
  \BibitemOpen
  \bibfield  {author} {\bibinfo {author} {\bibfnamefont {V.}~\bibnamefont
  {Varma}}, \bibinfo {author} {\bibfnamefont {S.~E.}\ \bibnamefont {Field}},
  \bibinfo {author} {\bibfnamefont {M.~A.}\ \bibnamefont {Scheel}}, \bibinfo
  {author} {\bibfnamefont {J.}~\bibnamefont {Blackman}}, \bibinfo {author}
  {\bibfnamefont {D.}~\bibnamefont {Gerosa}}, \bibinfo {author} {\bibfnamefont
  {L.~C.}\ \bibnamefont {Stein}}, \bibinfo {author} {\bibfnamefont {L.~E.}\
  \bibnamefont {Kidder}}, \ and\ \bibinfo {author} {\bibfnamefont {H.~P.}\
  \bibnamefont {Pfeiffer}},\ }\href {\doibase 10.1103/PhysRevResearch.1.033015}
  {\bibfield  {journal} {\bibinfo  {journal} {Phys. Rev. Research.}\ }\textbf
  {\bibinfo {volume} {1}},\ \bibinfo {pages} {033015} (\bibinfo {year}
  {2019})},\ \Eprint {http://arxiv.org/abs/1905.09300} {arXiv:1905.09300
  [gr-qc]} \BibitemShut {NoStop}%
\bibitem [{\citenamefont {Nagar}\ \emph
  {et~al.}(2019{\natexlab{b}})\citenamefont {Nagar}, \citenamefont {Messina},
  \citenamefont {Rettegno}, \citenamefont {Bini}, \citenamefont {Damour},
  \citenamefont {Geralico}, \citenamefont {Akcay},\ and\ \citenamefont
  {Bernuzzi}}]{Nagar:2018plt}%
  \BibitemOpen
  \bibfield  {author} {\bibinfo {author} {\bibfnamefont {A.}~\bibnamefont
  {Nagar}}, \bibinfo {author} {\bibfnamefont {F.}~\bibnamefont {Messina}},
  \bibinfo {author} {\bibfnamefont {P.}~\bibnamefont {Rettegno}}, \bibinfo
  {author} {\bibfnamefont {D.}~\bibnamefont {Bini}}, \bibinfo {author}
  {\bibfnamefont {T.}~\bibnamefont {Damour}}, \bibinfo {author} {\bibfnamefont
  {A.}~\bibnamefont {Geralico}}, \bibinfo {author} {\bibfnamefont
  {S.}~\bibnamefont {Akcay}}, \ and\ \bibinfo {author} {\bibfnamefont
  {S.}~\bibnamefont {Bernuzzi}},\ }\href {\doibase 10.1103/PhysRevD.99.044007}
  {\bibfield  {journal} {\bibinfo  {journal} {Phys. Rev.}\ }\textbf {\bibinfo
  {volume} {D99}},\ \bibinfo {pages} {044007} (\bibinfo {year}
  {2019}{\natexlab{b}})},\ \Eprint {http://arxiv.org/abs/1812.07923}
  {arXiv:1812.07923 [gr-qc]} \BibitemShut {NoStop}%
\end{thebibliography}%

\end{document}